\documentclass[12pt]{article}
\pdfoutput=1
\usepackage{amssymb,amsmath}
\usepackage{epsfig}
\usepackage{epstopdf}
\usepackage{latexsym}
\usepackage{graphicx}
\usepackage{subfigure}
\usepackage{booktabs}
\usepackage{bbm}
\usepackage[margin=20pt,small]{caption}

\usepackage[toc]{appendix}

\usepackage{color}
\usepackage{datetime}

\ifpdf
\fi

\renewcommand{\theequation}{\arabic{section}.\arabic{equation}}

\def\cN{{\cal N}}
\def\cO{{\cal O}}
\def\cP{{\cal P}}

\def\cL{{\cal L}}

\def\cS{{\cal S}}

\def\cL{{\cal L}}
\def\cA{{\cal A}}

\def\cV{{\cal V}}
\def\cH{{\cal H}}
\def\cC{{\cal C}}
\def\cD{{\cal D}}

\definecolor{cardinal}{rgb}{0.6,0,0}
\definecolor{darkgreen}{rgb}{0,0.5,0}
\definecolor{golden}{rgb}{0.92, 0.7, 0}
\definecolor{midnight}{rgb}{0, 0, 0.5}
\definecolor{darkblue}{rgb}{0.2, 0, 0.8}

\begin{document}

\begin{titlepage}

\begin{flushright}
UTTG-05-12 \\
NSF-KITP-12-060
\end{flushright}

\bigskip
\bigskip
\bigskip
\centerline{\Large \bf Holographic Thermalization with Chemical Potential}
\bigskip
\bigskip
\centerline{{\bf Elena Caceres$^{1,2}$ and Arnab Kundu$^{2, 3}$}}
\bigskip

\centerline{$^1$ Facultad de Ciencias}
\centerline{Universidad de Colima} 
\centerline{Bernal Diaz del Castillo 340, Colima, Mexico.}
\bigskip
\centerline{$^2$ Theory Group, Department of Physics}
\centerline{University of Texas at Austin} 
\centerline{Austin, TX 78712, USA.}
\bigskip
\centerline{${}^{3}$ Kavli Institute for Theoretical Physics}
\centerline{University of California}
\centerline{Santa Barbara CA 93106-4030, USA.}
\bigskip
\centerline{elenac@zippy.ph.utexas.edu, arnab@physics.utexas.edu}
\bigskip
\bigskip

\begin{abstract}

\noindent We study the thermalization of a strongly coupled quantum field theory in the presence of a chemical potential. More precisely, using the holographic prescription, we calculate non-local operators such as two point function, Wilson loop and entanglement entropy in a time-dependent background that interpolates between AdS$_{d+1}$ and AdS$_{d+1}$-Reissner-Nordstr\"om for $d=3,4$. We find that  it is the entanglement entropy that thermalizes the latest and thus sets a time-scale for equilibration in the field theory. We study the  dependence of the thermalization time on the probe length and the chemical potential. We find an interesting non-monotonic behavior. For a fixed small value of $T \ell$ and small values of  $\mu/T$ the thermalization time  decreases  as we increase $\mu/T$, thus the plasma thermalizes faster. For large values of $\mu/T$ the dependence changes and the thermalization time increases with increasing $\mu/T$. On the other hand, if we increase the value of $(T \ell)$ this non-monotonic behavior becomes less pronounced and eventually disappears indicating two different regimes for the physics of thermalization: non-monotonic dependence of the thermalization time on the chemical potential  for $T\ell << 1$ and monotonic for $T \ell >>1$. 

\end{abstract}

\newpage

\tableofcontents

\end{titlepage}

\newpage

\section{Introduction}

A standard method to study near-equilibrium physics is to slightly perturb the  system away from the equilibrium state and study the linear response of the system to these perturbations. One of the main virtues of this method  is that one can compute observables by calculating correlation functions in the equilibrium ensemble; The response functions are given in terms of Green's function evaluated in the equilibrium state. However, calculating Green's functions in a strongly coupled theory is not an easy task. Fortunately, the gauge/gravity duality is a powerful tool to compute correlation functions for a large class of strongly coupled theories. It maps the problem to a classical gravity calculation in higher dimensional curved  spacetime that in the case of conformal theories  is asymptotically anti--de-Sitter (AdS). The gauge/gravity duality has proven to be very useful in the study of near-equilibrium physics  (see \cite{Son:2007vk} and \cite{Hubeny:2010ry} and references therein). However, out-of-equilibrium processes cannot be understood using linear response theory. The study of out-of-equilibrium processes in strongly coupled field theories involves non-trivial temporal dynamics and  is an extremely challenging problem. The gauge/gravity duality translates this problem into a less daunting though equally fascinating one: the study of time-dependent gravitational dynamics, and in particular, black hole formation.

Non-equilibrium states in the gravity dual can be created by  turning  on time-dependent background fields at the boundary.  For example if  a  field is given a temporal dependence of compact support $(0, \delta t)$  at the boundary it  creates a wave that  propagates into the bulk. It was shown in \cite{Bhattacharyya:2009uu} that,  under suitable conditions, the gravitational collapse of this wave will form a black brane in the bulk. This process of black hole formation via gravitational collapse is expected to be the dual description of the thermalization process in a strongly coupled field theory.

The goal of the present work is to use the  holographic setup explained above to study the thermalization  in a strongly coupled theory in the presence of chemical potential. 
 A practical motivation for this work comes from the fact that the  equilibration time  observed at the Relativistic-Heavy-Ion-Collider (RHIC) is  much  shorter than predicted {\it via} perturbative approaches to thermalization. This indicates that the strong coupling is crucial in understanding the thermalization process of the Quark Gluon Plasma (QGP). Hence, this is a natural arena where gauge/gravity duality can be useful. Indeed, in \cite{Balasubramanian:2010ce, Balasubramanian:2011ur} the authors studied thermalization probes in  a time-dependent background  with vanishing chemical potential. They found that, unlike what is expected preturbatively, the thermalization proceeds top-down {\it i.e.} UV modes thermalize first, IR modes thermalize later. They also found that among the probes studied it is the entanglement entropy that thermalizes the latest and thus sets the time-scale for thermalization. The thermalization time scales  as $\tau_{\rm crit} \sim  \ell/2$, where $\ell$ is the typical length of the probe. A numerical estimate  yields $\tau_{\rm crit} \sim  0.3 fm/c$. Since the RHIC and the Large Hadron Collider (LHC) processes  occur at different values of the chemical potential, a natural question to ask is how do these results change when we consider a theory with a non-vanishing chemical potential. However tempting, our intention is  not to  make precise numerical predictions based on a toy model  but to  learn  qualitative features related to the effect of the chemical potential  in the thermalization process of  a strongly coupled theory.

Beyond the possible connection to RHIC and LHC physics, the study of non-local probes in a time-dependent background is an interesting problem in its own right and it can teach us important lessons about out-of-equilibrium physics.

To explore the thermalization of a strongly coupled field theory with chemical potential we consider a  time-dependent spacetime that interpolates between AdS$_{d+1} $ and AdS-Reissner-Nordst\"om (AdS-RN) background of the same bulk dimensions. This $(d+1)$-dim interpolating background, also known as the AdS-RN-Vaidya spacetime, can be understood as describing the gravitational collapse of a thin shell of charged null dust. We probe the theory with several non-local observables: two point functions, Wilson loop and entanglement entropy. We find that, as in the case of vanishing chemical potential, the thermalization is top-down and it is the entanglement entropy that thermalizes the latest and thus, sets the time scale for thermalization in the field theory.

Undoubtedly,  our most interesting result is that the  behavior of  the thermalization time as a function of the chemical potential seems to separate the physics in two regimes:   $T \ell \ll 1 $ and   $T \ell \gg 1 $.  For a fixed  small value of $T \ell$ and small values of $\mu/T$ the thermalization  time decreases as we increase $\mu/T$, thus the plasma thermalizes faster. For large values of $\mu/T$ the dependence changes and the thermalization time increases with increasing $\mu/T$. On the other hand, if we increase the value of $T \ell$ this non-monotonic behavior becomes less pronounced and eventually disappears altogether. Hence, we observe  two different regimes for the physics
of thermalization: non-monotonic dependence of the thermalization time on the chemical potential for $T \ell \ll 1$ and monotonic for $T \ell \gg 1$. The characteristics of the  non-monotonic behavior depend on the spacetime dimension, and is suppressed in lower dimensions.

The fact that the physics arranges itself\footnote{Or, should it be {\it herself}?} to manifest different qualitative features in the above two regimes tempts us to label them as follows: one ``classical" and the other ``quantum". For a system in thermal equilibrium, $T\ell \gg 1$ corresponds to a classical regime and $T\ell \ll 1$ to a quantum regime. Somewhere in the middle these two behaviors connect smoothly demonstrating a classical-to-quantum transition. We do observe qualitatively different physics as far as the behavior of the thermalization time is considered in these two regimes. However, we are dealing with a system truly away from equilibrium and it is not clear to us whether a ``classical" or a ``quantum" regime would stand as a precise notion for such processes, and if so, how the intricacies of such regimes interplay.

It should be emphasized that the background we are working with is {\it ab initio} ``phenomenologically motivated" rather than one obtained from rigorously solving Einstein equations to study the formation of a black hole in AdS-spacetime. Another approach towards analyzing such questions is to start with an out-of-equilibrium initial field configuration and study the thermalization process by solving for the gravitational background itself. Valiant efforts have been made along such directions in {\it e.g.} \cite{Grumiller:2008va}-\cite{CaronHuot:2011dr} by considering colliding gravitational shock waves. It remains to be seen whether a generalization of \cite{Bhattacharyya:2009uu} in the presence of a charged matter yields the desired form of the AdS-RN-Vaidya background that we work with.

This paper is organized as follows: In Section 2 we introduce the holographic setup {\it i.e.} we present the bulk action and the background metric of AdS-Reissner-Nordstr\"om in $(d+1)$ dimensions. Next, in Section 3, we study the equilibrium behavior of the  non-local observables mentioned above in $d=3,4$. Sections 4 and 5  contain our main results. In section 4 we present the generalization of the Vaidya metric to AdS-RN backgrounds by using a charged null dust to form the black hole. We then explore the non-equilibrium behavior of the non-local probes and determine the thermalization time for each probe. In Section 5 we summarize our results and elaborate on possible future directions. A detailed analysis of minimal surfaces and apparent horizons  in AdS-RN-Vaydia is found in Appendix A. Appendix B gathers some standard facts related to the embedding of 4 and 5 dimensional charged AdS black holes in 10 or 11-dim supergravity.

{\bf Note Added}: While this paper was close to completion, we became aware of ref. \cite{GalanteSchvellinger} which partially overlaps with our work.

\section{The bulk action and the background}

We will begin with the AdS-RN background and analyze equilibrium properties of the dual CFT by looking at two-point correlation function, Wilson loops and entanglement entropy. The holographic prescription for computing such observables is to compute minimal surfaces of appropriate dimensions in AdS-space. For generality we will analyze these observables in an AdS$_{d+1}$-background, where $d=3, 4$. As far as the background is concerned, our approach is seemingly ``phenomenological" or the so called ``bottom up" where we simply obtain these backgrounds as solutions to some effective gravity action in the corresponding dimension. However, for $d=3$ and $d=4$ --- which are the ones that we will study here --- one can embed the corresponding solutions within ($11$ or $10$-dimensional type IIB) supergravity. Also note that within supergravity one usually obtains a more general multi-charged black hole solution. The backgrounds that we consider here corresponds to setting all these charges equal.\footnote{The $S^5$ reduction of type IIB supergravity yields an $\cN=8$ gauged supergravity in $d=4$ with $SO(6)$ gauge group. This admits a consistent $\cN=2$ truncation coupled to two Abelian vector multiplets with an $U(1)^3$ Cartan subgroup of the full $SO(6)$. In general this admits a three charge black hole solution. Similarly, the $S^7$-reduction of $11$-dimensional supergravity admits an $\cN=2$ consistent truncation with $U(1)^4$ gauge group, which in general admits a four charge black hole solution. See {\it e.g.} \cite{Behrndt:1998jd} for a discussion of such solutions. Some features of such solutions are also discussed in Appendix A.} It is also noteworthy that one can obtain such charged black hole solutions in $d=6$ as well which comes from the $S^4$-reduction of $11$-dimensional supergravity; we will, however, not consider this case.

AdS-RN backgrounds are solutions of Einstein-Hilbert action with a -ve cosmological constant coupled to a Maxwell field. Let us start with the effective action of the form
\begin{eqnarray} \label{action1}
S_0 = \frac{1}{8\pi G_N^{(d+1)}}\left(\frac{1}{2} \int d^{d+1} x \sqrt{-g} \left(R - 2 \Lambda \right) - \frac{1}{4} \int d^{d+1}x \sqrt{-g} F_{\mu\nu} F^{\mu\nu} \right) \ ,
\end{eqnarray}
which gives the following equations of motion
\begin{eqnarray} 
&& R_{\mu\nu} - \frac{1}{2} \left(R- 2 \Lambda \right) g_{\mu\nu} = g^{\alpha\rho} F_{\rho\mu} F_{\alpha\nu} - \frac{1}{4} g_{\mu\nu} \left(F^{\alpha\beta} F_{\alpha \beta}\right) \ , \label{eom1} \\
&& \partial_\rho \left[ \sqrt{-g} g^{\mu\rho} g^{\nu\sigma} F_{\mu\nu} \right] = 0 \ . \label{eom2}
\end{eqnarray}
The solutions to (\ref{eom1}) and (\ref{eom2}) for a general $d(\ge 3)$ are explicitly given below:
\begin{eqnarray} \label{RN}
&& ds^2 = \frac{L^2}{z^2} \left(- f(z) dt^2 + \frac{dz^2}{f(z)} + d\vec{x}^2 \right) \ , \quad \Lambda = -\frac{d(d-1)}{2 L^2} \ , \nonumber\\
&& f(z) = 1- M z^d + \frac{(d-2)}{(d-1) L^2}Q^2 z^{2(d-1)} \ , \quad A_t = Q (z_H^{d-2} - z^{d-2})  \ .
\end{eqnarray}
Here $\vec{x}$ is a $(d-1)$-dimensional vector, $M$ is the mass of the black hole and $Q$ is the charge. The constant $z_H$ denotes the location of the horizon which is obtained by solving $f(z)=0$. In order for the one-form $A_t$ to be well-defined at the horizon, we have arranged $A_t(z_H) = 0$. The case of $d=2$ is special and is not captured by the solution given above. In this case, the background takes the following form
\begin{eqnarray} \label{RN2}
&& ds^2 = \frac{L^2}{z^2} \left(- f(z) dt^2 + \frac{dz^2}{f(z)} + dx^2 \right) \ , \nonumber\\
&& f(z) = 1-  M z^2 + \frac{Q^2 z^2}{L^2} \log \left( z/ L\right) \ , \quad A_t = Q \log\left(z_H/z\right) \ , \quad \Lambda = -\frac{1}{L^2} \ .
\end{eqnarray}

In the coordinate system used above the boundary is located at $z=0$. Having a non-zero electric field corresponds to having a chemical potential in the dual field theory and we simply read off this chemical potential from the vector field $A_t$ as
\begin{eqnarray}
\mu = \lim_{z\to 0} A_t(z) \ .
\end{eqnarray}
Note that the above definition does not work for $d=2$.\footnote{The case of $d=2$ is rather special. In this case, the identification of the source and the VEV are subtle and the chemical potential should be identified as the sub-leading term as $z\to 0$: $\mu \equiv Q \log \left( z_H / L\right)$. For more details on this issue, see \cite{Jensen:2010em}.}

\begin{figure}[!ht]
\begin{center}
\subfigure[] {\includegraphics[angle=0,
width=0.45\textwidth]{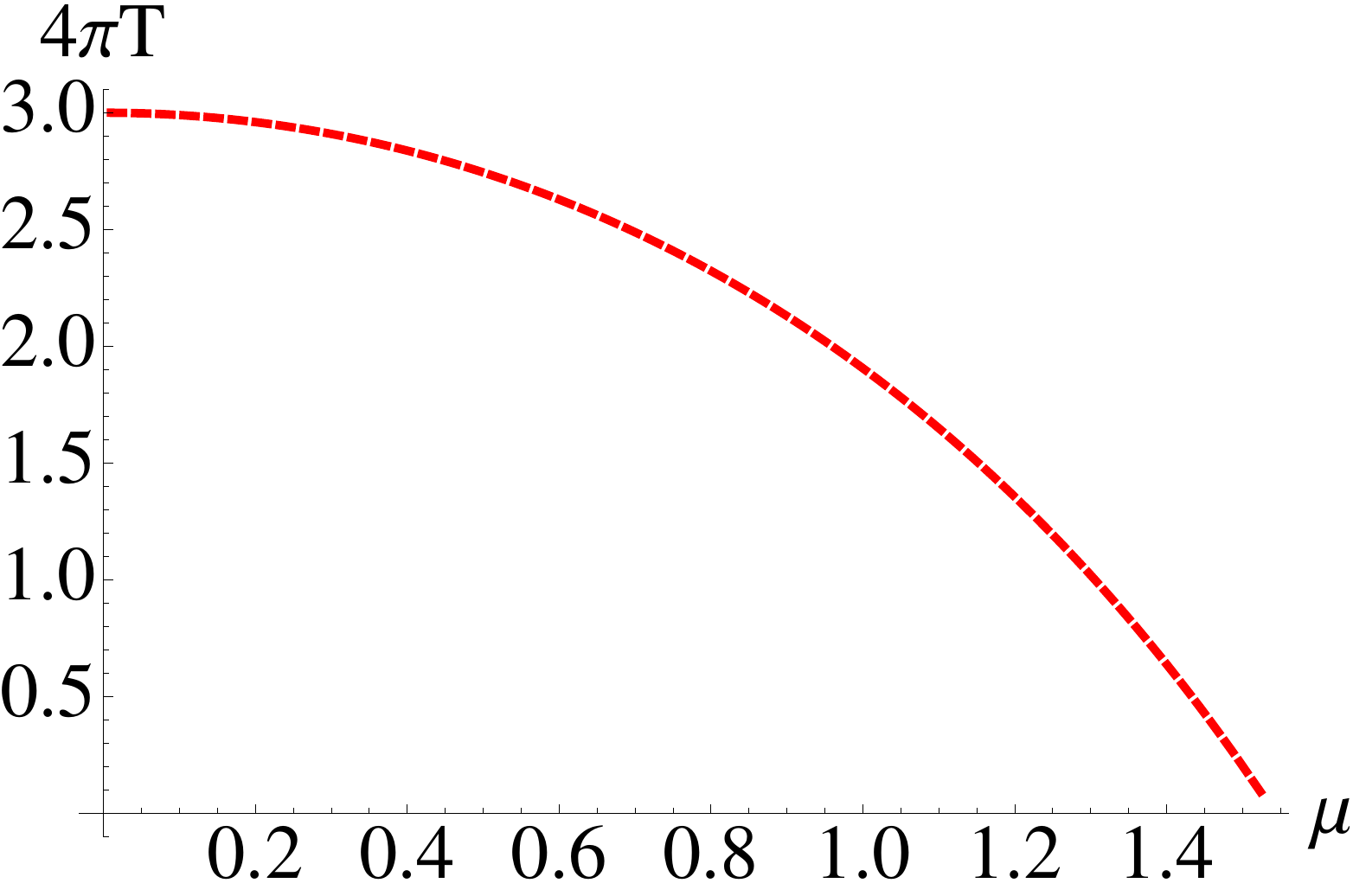} }
 \subfigure[] {\includegraphics[angle=0,
width=0.45\textwidth]{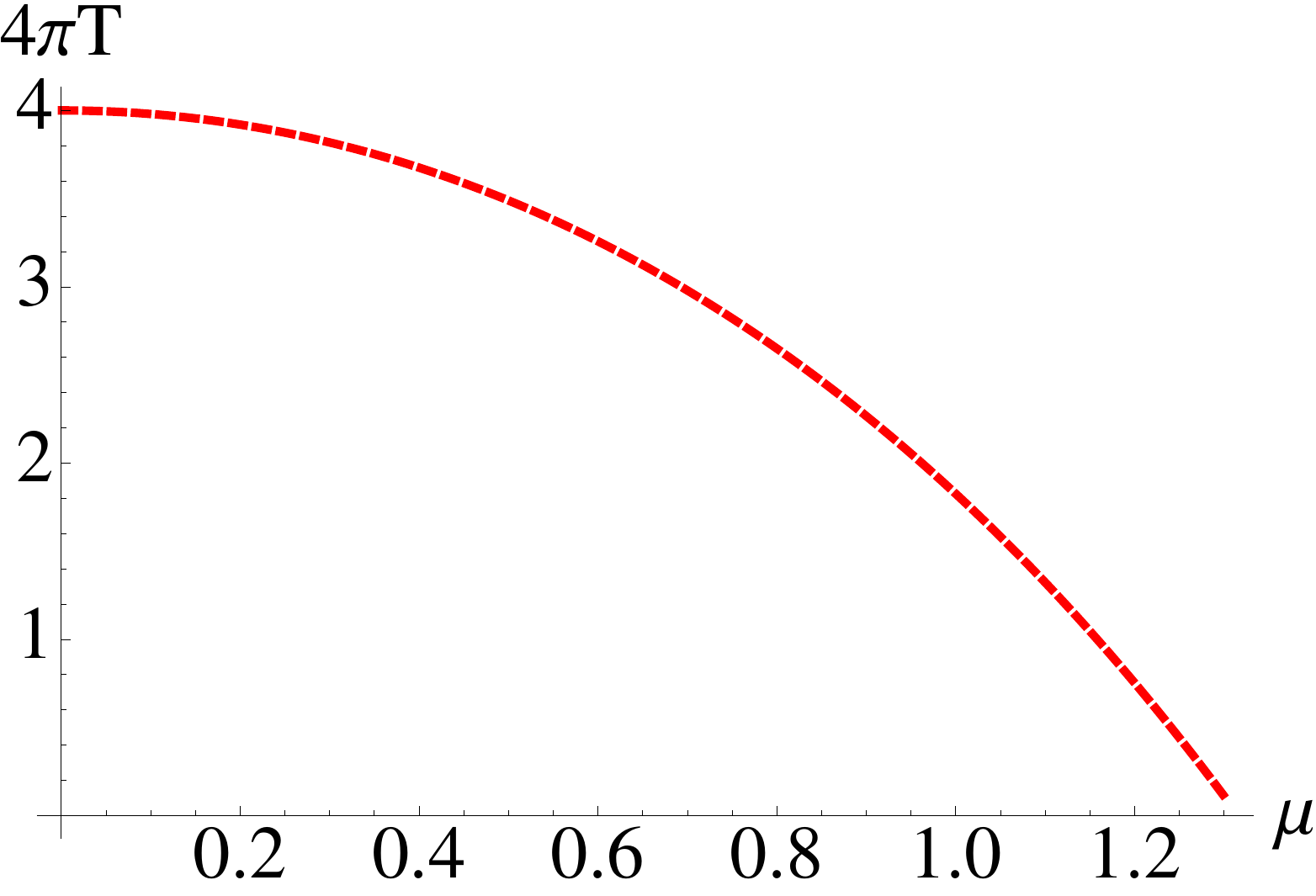} }
\caption{\small The dependence of temperature on the chemical potential in (a) $d=3$ and (b) $d=4$. We are measuring the temperature and the chemical potential in units of the AdS-radius. We have also set $M=1$.}
\label{Tvsq}
\end{center}
\end{figure}

On the other hand, the temperature of the black hole is given by
\begin{eqnarray}
T = - \left.  \frac{1}{4\pi} \frac{d}{dz} f(z) \right |_{z_H} \ .
\end{eqnarray}
We will briefly investigate the behavior of the temperature as the chemical potential of the system is cranked up. This will be a numerical endeavor and for that purpose we will set $L=1$, which implies we will be measuring the temperature and the chemical potential in units of $L$. Typically, for a fixed value of $M$, $Q$ has an upper bound beyond which there is no real positive solution of $f(z)=0$.\footnote{This is true for all cases except $d=2$ where $Q$ can be as large as possible.} As this upper bound is reached, the background temperature approaches zero and we approach an extremal solution with zero temperature but finite horizon radius (giving rise to finite entropy). The corresponding supergravity solutions (in $d=3, 4$) are the BPS black holes discussed in \cite{Behrndt:1998ns}. These extremal solutions possess a naked singularity. The behavior of the temperature with the chemical potential (for fixed $M=1$) is shown in fig.~\ref{Tvsq}. 
It is noteworthy to comment here that the physics of AdS-RN background has many interesting properties investigated earlier in {\it e.g.} \cite{Cvetic:1999ne, Chamblin:1999tk} and more recently from a more application-towards-condensed matter theory point of view reviewed in {\it e.g.} \cite{Hartnoll:2009sz}.

Before we proceed further, note that with the following change of coordinate
\begin{eqnarray} \label{EF}
dt = dv + \frac{dz}{f} \ ,
\end{eqnarray}
the above AdS-RN metrics given in equations (\ref{RN}) can be brought in the following form\footnote{Note that introducing the Eddington-Finkelstein coordinate in (\ref{EF}) yields a gauge field of the form: $A_v = A(z)$, $A_z = A(z)/f(z)$, where $A(z)$ is given in (\ref{RN}). We can make a gauge choice to recast all physical information by choosing a gauge $A_v(z)$ only.}
\begin{eqnarray} \label{metvaidya}
ds^2 = \frac{L^2}{z^2} \left( - f(z) dv^2 - 2 dv dz + d\vec{x}^2  \right) \ , \quad A_v = A_v(z) \ ,
\end{eqnarray}
where $\vec{x}$ is a $(d-1)$-dimensional vector. The form of the background in (\ref{metvaidya}) is suitable for expressing the Vaidya metric, which we will discuss later. Note that the form above is generic for any $d$, where the information of the dimensionality is entirely carried by the function $f(z)$ and the electric field $A_v(z)$. For our purposes, the explicit expression of $A_v(z)$ will not matter at all since ultimately we will study minimal surfaces which do not couple to this vector field.

\section{Non-local observables in equilibrium}

Before proceeding to discuss aspects of the thermalization, we begin by exploring the equilibrium behavior of the non-local observables in the presence of a chemical potential. The non-local observables we will study here are the two-point functions (for operators with large dimensions), expectation values for Wilson loops and entanglement entropy. To study this we take the background in the form presented in (\ref{metvaidya}).

\subsection{Two point function: the geodesic approximation}

The idea here is to study the thermal Wightman function of an operator with large conformal dimension. As shown in \cite{Balasubramanian:1999zv}, in this limit the equal-time two point function is given by
\begin{eqnarray}
\langle \cO(t, \vec{x}) \cO(t, \vec{x}') \rangle \sim e^{- \Delta \cL_{\rm thermal}}   \ ,
\end{eqnarray}
where $\Delta$ is the conformal dimension and $\cL_{\rm thermal}$ is the renormalized geodesic length. The above expression makes use of a saddle point approximation.

We want to consider space-like geodesic connecting the two boundary points: $(t, x_1) = (t_0, - \ell/2)$ and $(t', x_1') = (t_0, \ell/2)$, where (whenever necessary) all other spatial directions are identical at the two end points. Such a geodesic is parametrized by $v= v(x)$ and $z=z(x)$ where $x_1 \equiv x$. The boundary conditions satisfied by this geodesic is
\begin{eqnarray}
z(-\ell/2) = z_0 = z(\ell/2) \ , \quad v(-\ell/2) = t_0 = v(\ell/2) \ .
\end{eqnarray}
Here $z_0$ is the IR radial cut-off near the boundary.

The geodesic length is
\begin{eqnarray}
\cL = \int_{-\ell/2}^{\ell/2} dx \frac{L}{z} \left[ 1 - f v'^2 - 2 v' z' \right]^{1/2} \ , \quad ' \equiv \frac{d}{dx} \ .
\end{eqnarray}
There are two conservation equations: one because the Lagrangian is independent of $x$ and the other because the Lagrangian is independent of $v$. The first one gives
\begin{eqnarray} \label{con}
1 - 2 z' v' - f(z) v'^2 = \frac{z_*^2}{z^2} \ ,
\end{eqnarray}
where $z_*$ is the value of $z(x)$ at the midpoint. Using the definition of $v$ we get
\begin{eqnarray} \label{vdef}
v = t_0 - \int_{z_0}^z \frac{dz}{f(z)} \quad \implies \quad \frac{dv}{dx} = - \frac{1}{f(z)} \frac{dz}{dx} \ .
\end{eqnarray}
we can substitute $v'$ in favour of $z'$ in (\ref{con}) and obtain
\begin{eqnarray}
\frac{dz}{dx} = \pm \sqrt{f(z)} \left[ \frac{z_*^2}{z^2} - 1 \right]^{1/2} \ ,
\end{eqnarray}
where the positive sign is taken for $x>0$ and the negative sign is taken for $x<0$. The boundary separation length is then read off as
\begin{eqnarray} \label{ell2}
\frac{\ell}{2} = \int_{z_0}^{z_*} \frac{1} {\sqrt{f(z)}} \left[ \frac{z_*^2}{z^2} - 1 \right]^{-1/2} \ ,
\end{eqnarray}
where $z_0$ is the IR radial cut-off.\footnote{The length integral is perfectly convergent, therefore we can take $z_0 = 0$ without any issue.} The corresponding integral can be analytically carried out in purely AdS-background to yield
\begin{eqnarray} \label{ellads1}
\ell_{\rm AdS} = 2 z_* \ .
\end{eqnarray}

On the other hand, using the conservation equation, the geodesic length can be obtained to be
\begin{eqnarray} \label{Lthermal}
\cL = 2 L \int_{z_0}^{z_*} \frac{dz}{z} \frac{1}{\sqrt{f(z) \left( 1 - \frac{z^2}{z_*^2} \right)}} \ .
\end{eqnarray}
Note that both formulae in (\ref{ell2}) and (\ref{Lthermal}) apply for any $d$. The geodesic length is a divergent quantity; we can regularize this length by subtracting off the divergent piece in pure AdS-space. In pure AdS-space, this length is given by
\begin{eqnarray}
\cL_{\rm AdS} & = & 2 L \int_{z_0}^{z_*} \frac{dz}{z} \frac{1}{\sqrt{ \left( 1 - \frac{z^2}{z_*^2} \right)}} = - 2 L \log \left[ \frac{z_0}{z_* + \sqrt{z_*^2 - z_0^2}} \right] \nonumber\\
& = & - 2 L \log \left(\frac{z_0}{L}\right) + 2 L \log\left(\frac{L}{2 \ell_{\rm AdS}}\right) \ ,
\end{eqnarray}
where we have used (\ref{ellads1}) to substitute for $z_*$ in the above expression. We are interested in the finite part of this geodesic length.  We therefore consider the following renormalized length\footnote{Note that in \cite{Balasubramanian:2011ur}, the authors include a finite piece $2\log 2$ in the subtracted part. This should not change the qualitative behaviour since it's a constant.} 
\begin{eqnarray}
\cL_{\rm thermal} = 2 L \int_{z_0}^{z_*} \frac{dz}{z}\frac{1} { \sqrt{\left( 1 - \frac{z^2}{z_*^2} \right)}}  \frac{1}{\sqrt{f(z)}} +  2 L \log\left(\frac{z_0}{L}\right) \ .
\end{eqnarray}
Now we can numerically study how $\cL_{\rm thermal}$ behaves in various dimensions as we tune the chemical potential. Recall that our underlying theory is conformal, therefore the only relevant parameter that we can vary is a dimensionless ratio constructed from the temperature and the chemical potential. In $(d+1)$ bulk space-time dimensions (dual to a theory in $d$ boundary space-time dimensions)
\begin{eqnarray}
\left[ T \right] \sim \frac{1}{\rm length} \ , \quad \left[ \mu \right] \sim \frac{1}{\rm length} \ .
\end{eqnarray}
Thus we can consider 
\begin{eqnarray} \label{ratio}
\chi_{(d)} = \frac{1}{4\pi} \left(\frac{\mu}{T} \right)
\end{eqnarray}
to be the relevant parameter that we will vary. In practice we can vary $\chi_{(d)}$ by first fixing $M=1$ and then just varying the parameter $Q$ till it reaches its maximum value beyond which naked singularities appear in the gravitational background. It is straightforward to check that for this range of values for $Q$, $\chi_{(d)} \in [0,\infty]$.

\begin{figure}[!ht]
\begin{center}
\subfigure[] {\includegraphics[angle=0,
width=0.45\textwidth]{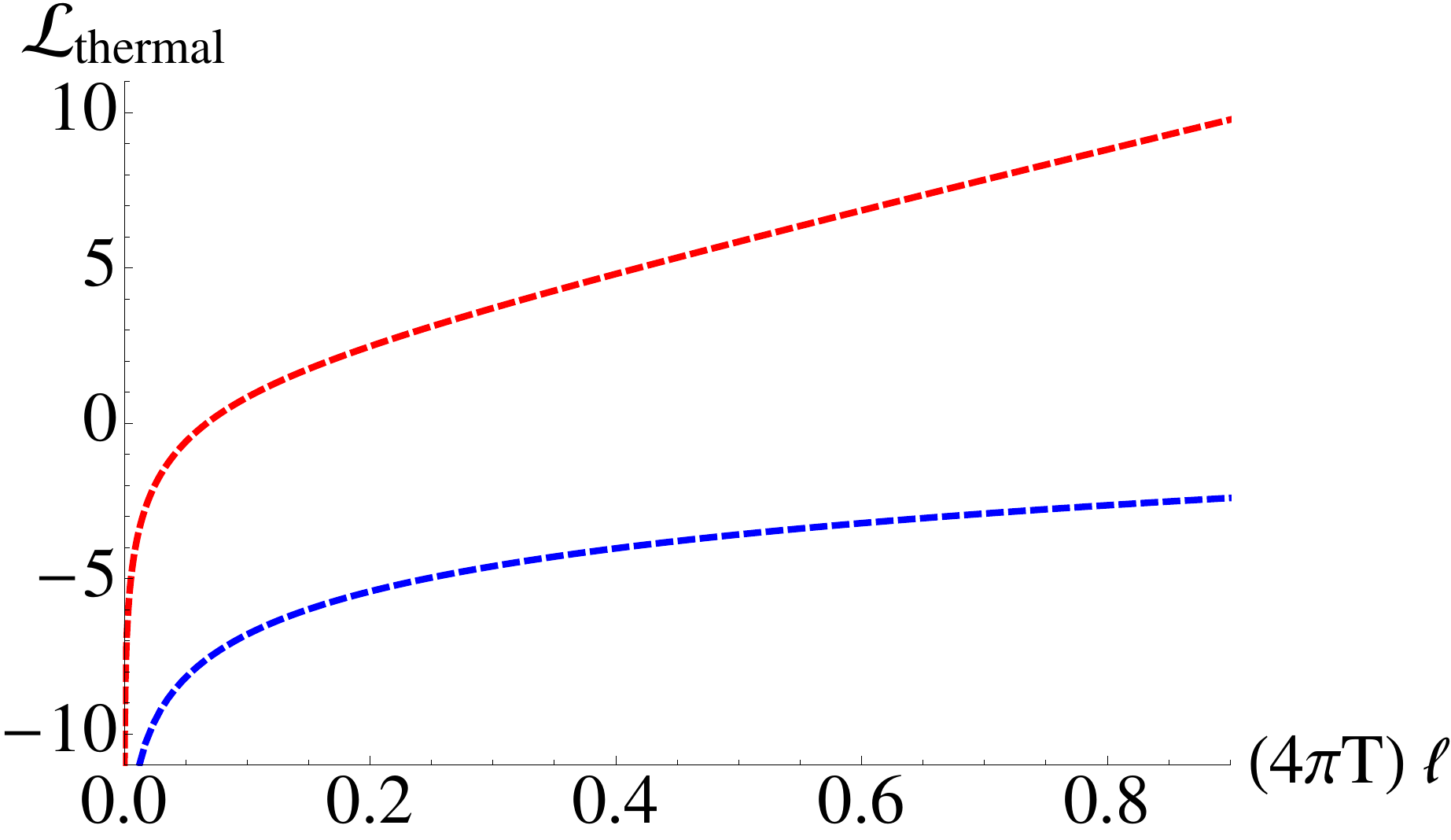} } 
\subfigure[] {\includegraphics[angle=0,
width=0.45\textwidth]{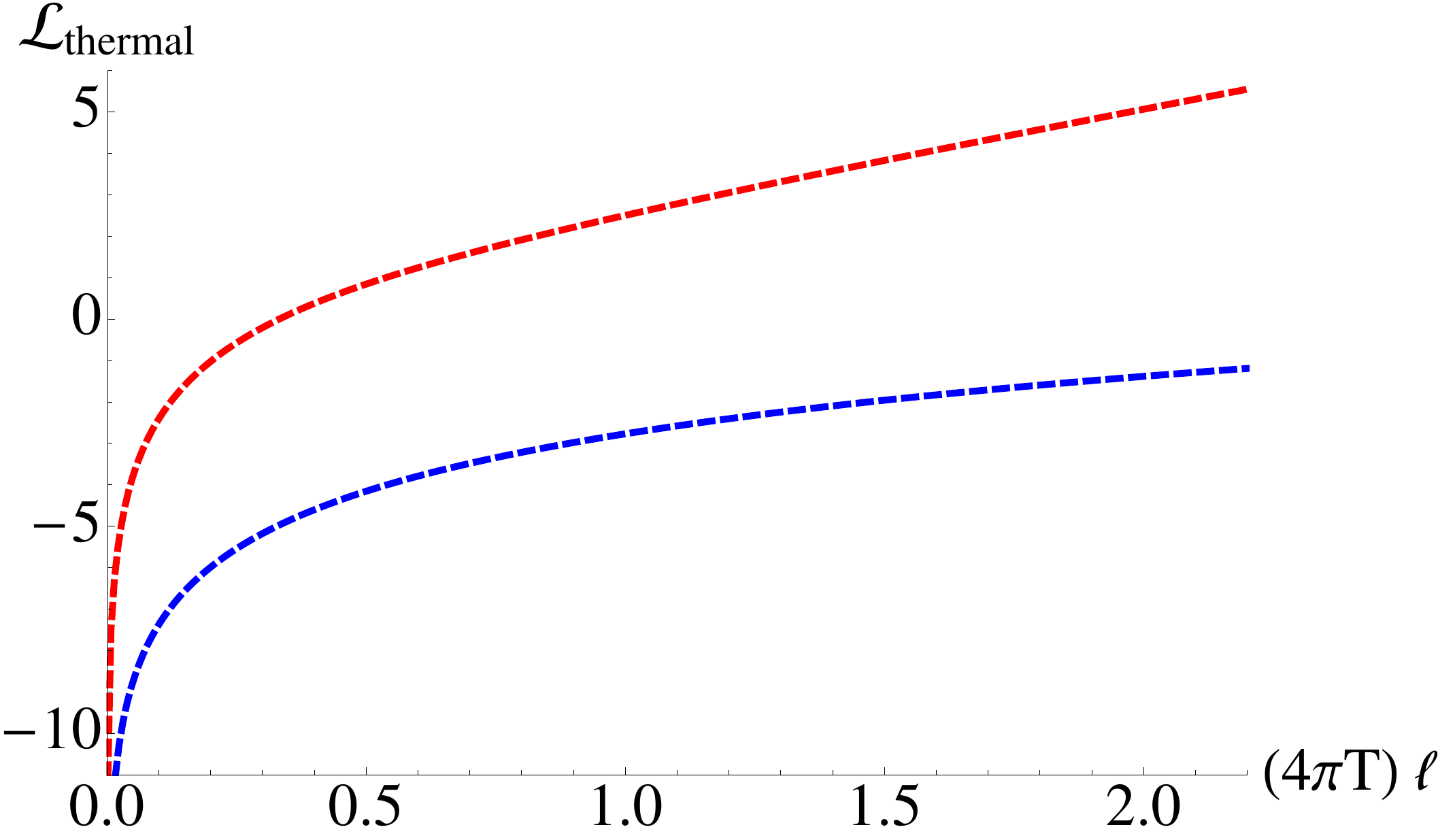} }
\subfigure[] {\includegraphics[angle=0,
width=0.55\textwidth]{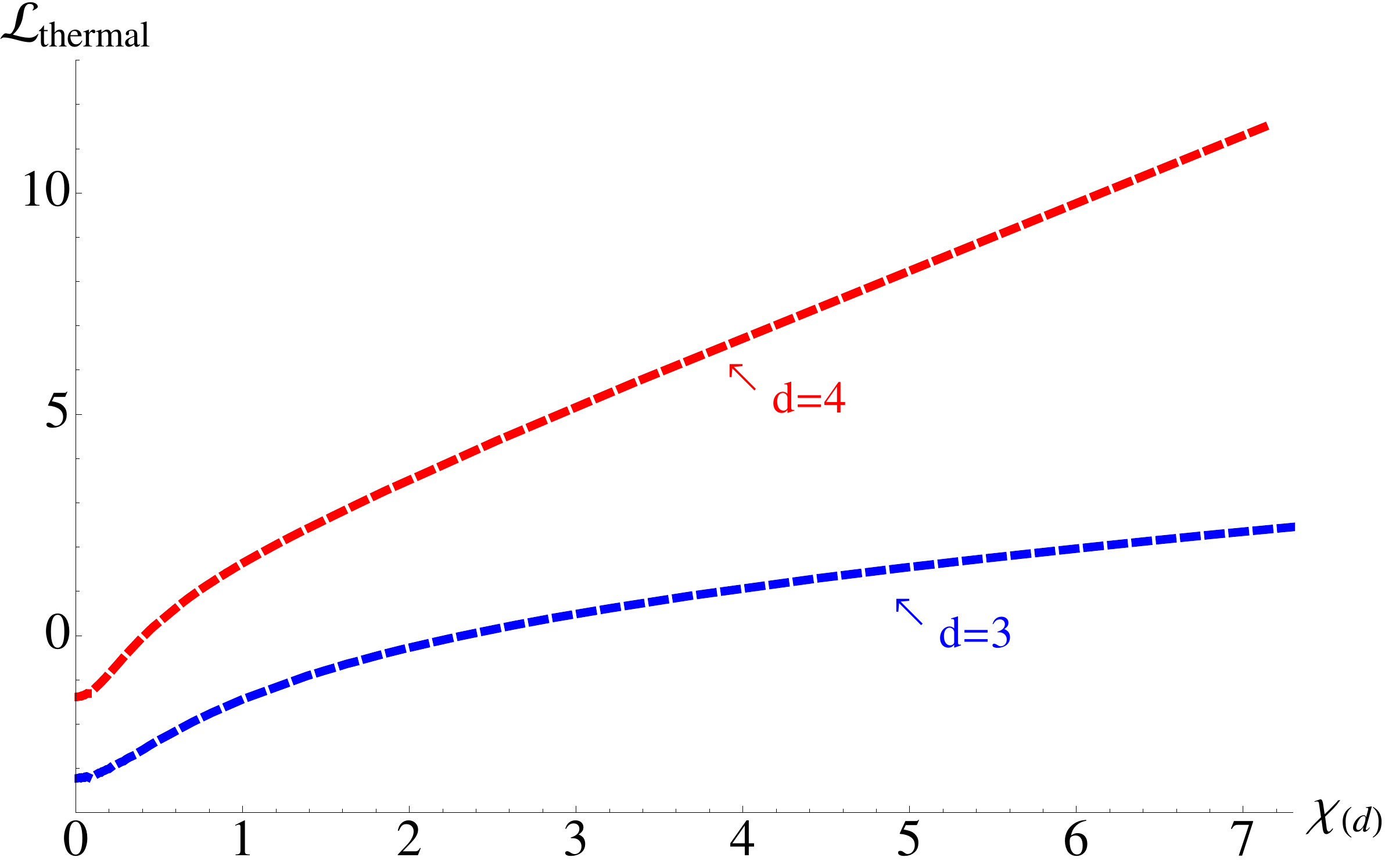} } 
\caption{\small Left panel (a): The case $d=3$. The blue curve corresponds to $\chi_{(3)} \approx 0.003$ and the red curve corresponds to $\chi_{(3)} \approx 22.4$. Right panel (b): The case $d=4$. The blue curve corresponds to $\chi_{(4)} \approx 0.002$ and the red curve corresponds to $\chi_{(4)} \approx 18.3$. In (c), we have shown how $\cL_{\rm thermal}$ scales with $\chi_{(d)}$ for both $d=3$ with $(4\pi T) \ell = 0.6$ and $d=4$ with $(4\pi T)\ell = 2$. Here $\cL_{\rm thermal}$ is measured in units of the AdS-radius, $L$.}
\label{figd34RN}
\end{center}
\end{figure}

For a given value of $\chi_{(d)}$, we can now study the behavior of $\cL_{\rm thermal}$ as a function of the boundary separation length. To generate these curves we do the following: To begin with we set $L=1$, which means the dimensionful quantity $\cL_{\rm thermal}$ (and later the area or the volume corresponding to the Wilson loop or the entanglement entropy calculations) is measured in units of the AdS-radius. Now we fix $M=1$ and for a given value of $Q$ start with a $z_*$ very close to the horizon. Next we keep changing $z_*$ till we reach $z_*\to 0$.  Each choice of $z_*$ generates an unique value of $\cL_{\rm thermal}$ and $\ell$ through equations (\ref{Lthermal}) and (\ref{ell2}) respectively. Finally we plot this result. The behaviour of $\cL_{\rm thermal}$ is shown in fig.~\ref{figd34RN}. The boundary length $\ell$ will be expressed in terms of the boundary temperature, $T$. It is clear from fig.~\ref{figd34RN}, $\cL_{\rm thermal} \sim (T \ell )$ for large $T \ell $.

The general observation we see from these plots is increasing $\chi_{(d)}$ monotonically increases the value of $\cL_{\rm thermal}$ for a given boundary separation $\ell$ in both $d=3, 4$. The dependence of $\cL_{\rm thermal}$ on the dimensionless ratio $\chi_{(d)}$ is shown in fig.~\ref{figd34RN}(c): the monotonically increasing function is generally non-linear --- however --- for $\chi_{(d)}\gg 1$, $\cL_{\rm thermal} \sim \chi_{(d)}$ with a slope that depends on $d$. Non-linearities appear only for small values of $\chi_{(d)}$. Also, we have checked explicitly that in the linear regime the slope of $\cL_{\rm thermal}$ vs $\chi_{(d)}$ curve depends on the fixed value of $(T\ell)$. This will be a generic feature in all the observables we will consider here.

\subsection{Space-like Wilson loops}

Wilson loop is another gauge invariant non-local observable that can probe thermal properties of a field theory, {\it e.g.} the expectation value of the Wilson loop can detect confinement/deconfinement transition in theories like QCD. Here we will study the thermalization of space-like Wilson loops from a holographic approach.

\begin{figure}[!ht]
\begin{center}
\subfigure[] {\includegraphics[angle=0,
width=0.45\textwidth]{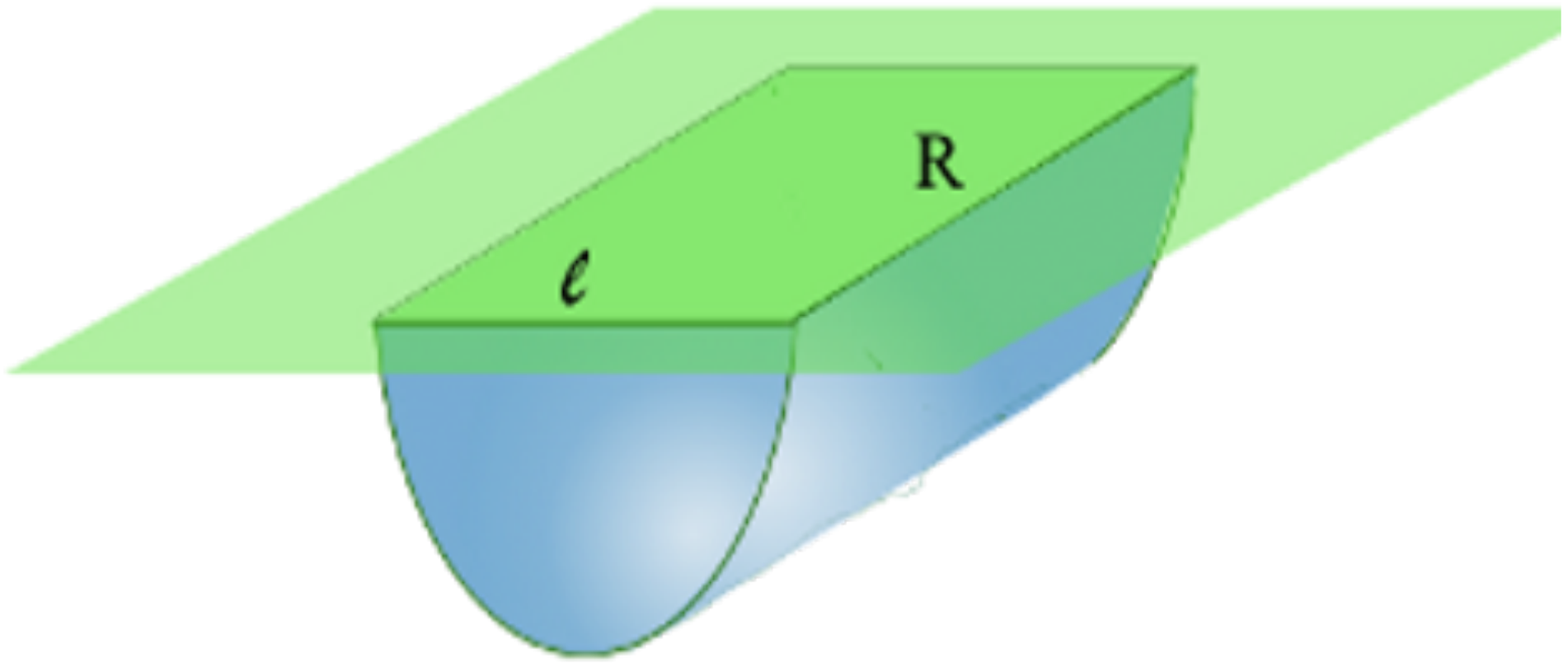} }
 \subfigure[] {\includegraphics[angle=0,
width=0.45\textwidth]{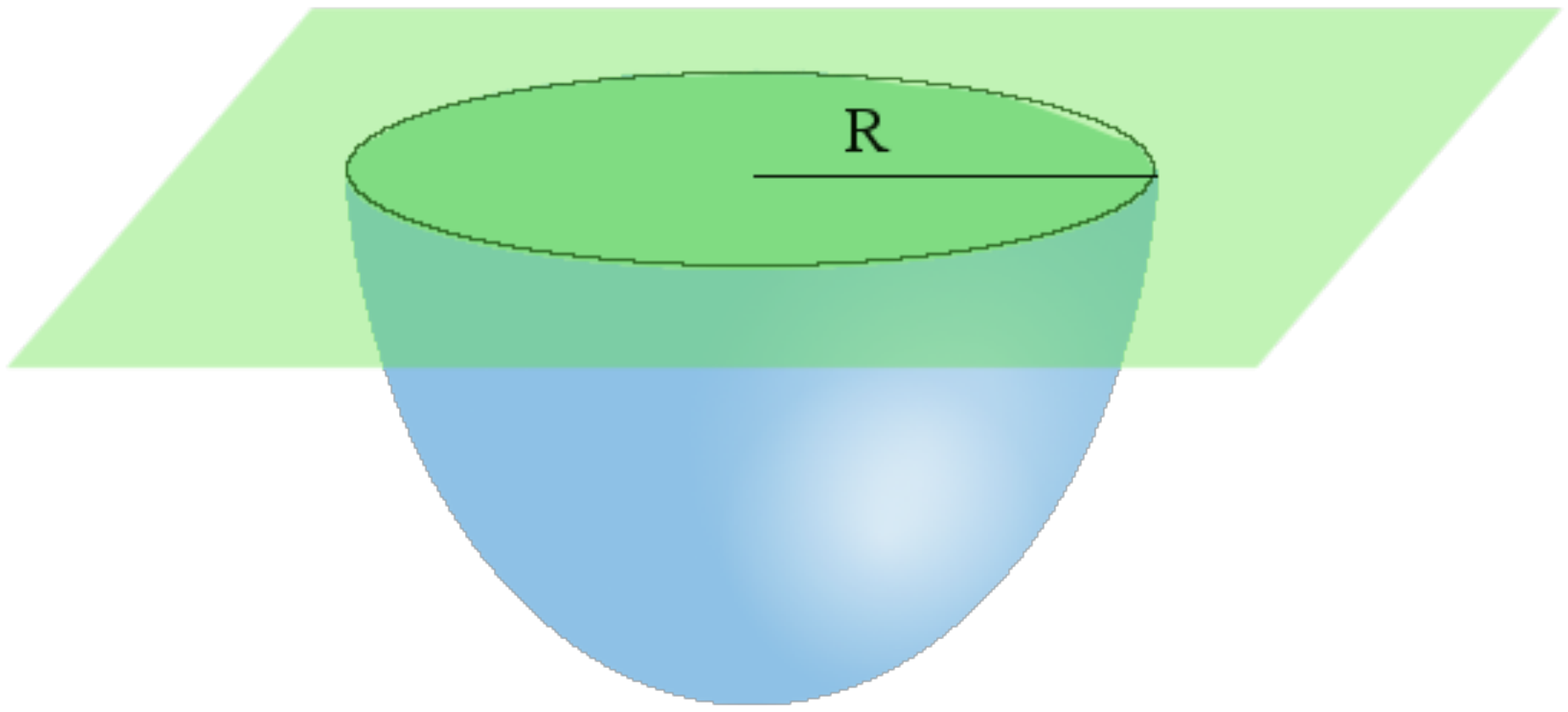} }
\caption{\small A schematic diagram of the Wilson loops of different shapes and the corresponding minimal area surfaces: rectangular in (a) and circular in (b).}
\label{rec_cir_shape}
\end{center}
\end{figure}

In a gauge theory, the Wilson loop operator is defined as a path ordered contour integral over a closed loop $\cC$ of the gauge field
\begin{eqnarray}
W (\cC) = \frac{1}{N} {\rm Tr} \left( \cP e^{\oint_{\cC} A}\right) \ ,
\end{eqnarray}
where $A$ is the gauge field and $N$ is the rank of the gauge group and $\cP$ denotes path ordering. In the AdS/CFT correspondence, the expectation value of the Wilson loop is related to the string partition function 
\begin{eqnarray}
\langle W(\cC) \rangle   = \int \cD \Sigma \, e^{- \cA (\Sigma)} \ ,
\end{eqnarray}
where $\Sigma$ is the string world sheet which extends in the bulk with the boundary condition $\partial \Sigma = \cC$ and $\cA(\Sigma)$ corresponds to the Nambu-Goto action for the string. In the strongly coupled limit, we can simplify the computation by making a saddle point approximation and evaluating the minimal area surface of the classical string with the same boundary condition $\partial \Sigma_0 = \cC$
\begin{eqnarray}
\langle W (\cC) \rangle = e^{- \cA (\Sigma_0)} \ ,
\end{eqnarray}
where $\Sigma_0$ represents the minimal area surface. In the AdS/CFT correspondence, such computation of Wilson loop operators were first done in \cite{Maldacena:1998im}. Here we will consider both rectangular and circular Wilson loops, which are schematically shown in fig.~\ref{rec_cir_shape}.
%

\subsubsection{Rectangular Wilson loop}

A rectangular strip Wilson loop can be parametrized by the boundary coordinates $\{x_1, x_2\}$ with the assumption that this infinite rectangular strip is invariant under the $x_2$-direction. Thus the corresponding minimal surface is parametrized by: $z(x)$ and $v(x)$, where $x \equiv x_1$. As in the case of geodesics, the boundary conditions are
\begin{eqnarray}
z (- \ell/2) = z_0 = z(\ell/2) \ , \quad v(- \ell/2) = t_0 = v(\ell/2) \ ,
\end{eqnarray}
where $\ell$ is the length of the rectangular Wilson loop along the $x^1$-direction. The Nambu-Goto action is given by
\begin{eqnarray} \label{areafunc}
\cA_{\rm NG} = \frac{RL^2}{2\pi\alpha'} \int_{-\ell/2}^{\ell/2} \frac{dx}{z^2} \left(1 - f v'^2 - 2 v' z' \right)^{1/2} \ ,
\end{eqnarray}
where $R$ is the length along the $x_2$-direction. Since there is no explicit $x$-dependence in the Lagrangian, the corresponding conservation equation is given by
\begin{eqnarray}
1 - f v'^2 - 2 v' z' = \left(\frac{z_*}{z}\right)^4 \ ,
\end{eqnarray}
where $z_*$ is the midpoint of $z(x)$. Using the definition of $v$ from (\ref{vdef}) we can obtain
\begin{eqnarray} \label{eomwlrec}
1 + \frac{z'^2}{f} = \left(\frac{z_*}{z}\right)^4 \quad \implies \quad \frac{dx}{dz} = \pm \frac{1}{\sqrt{f}} \left[ \left(\frac{z_*}{z}\right)^4 - 1\right]^{-1/2} \ ,
\end{eqnarray}
where the positive sign is taken for $x >0$ and the negative sign is taken for $x<0$. The boundary separation length is obtained by integrating the above equation from $z_0$ to $z_*$. Once again this length can be analytically computed for the pure AdS-case to give
\begin{eqnarray} \label{ellads2}
\ell _{\rm AdS} = z_* \sqrt{\pi} \frac{\Gamma\left(\frac{3}{4}\right)}{\Gamma\left(\frac{1}{4}\right)} \ .
\end{eqnarray}

\begin{figure}[!ht]
\begin{center}
\subfigure[] {\includegraphics[angle=0,
width=0.45\textwidth]{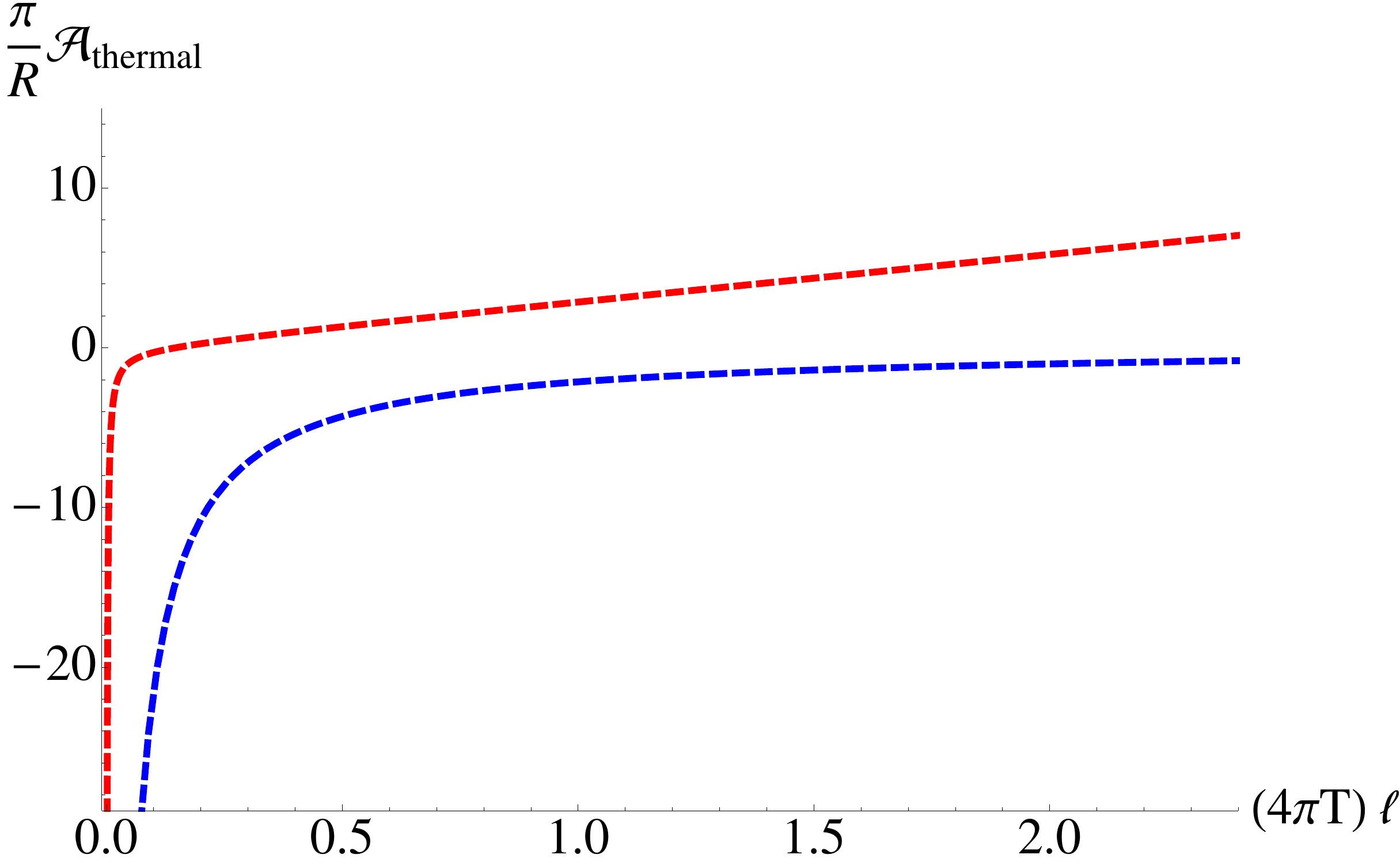} } 
\subfigure[] {\includegraphics[angle=0,
width=0.45\textwidth]{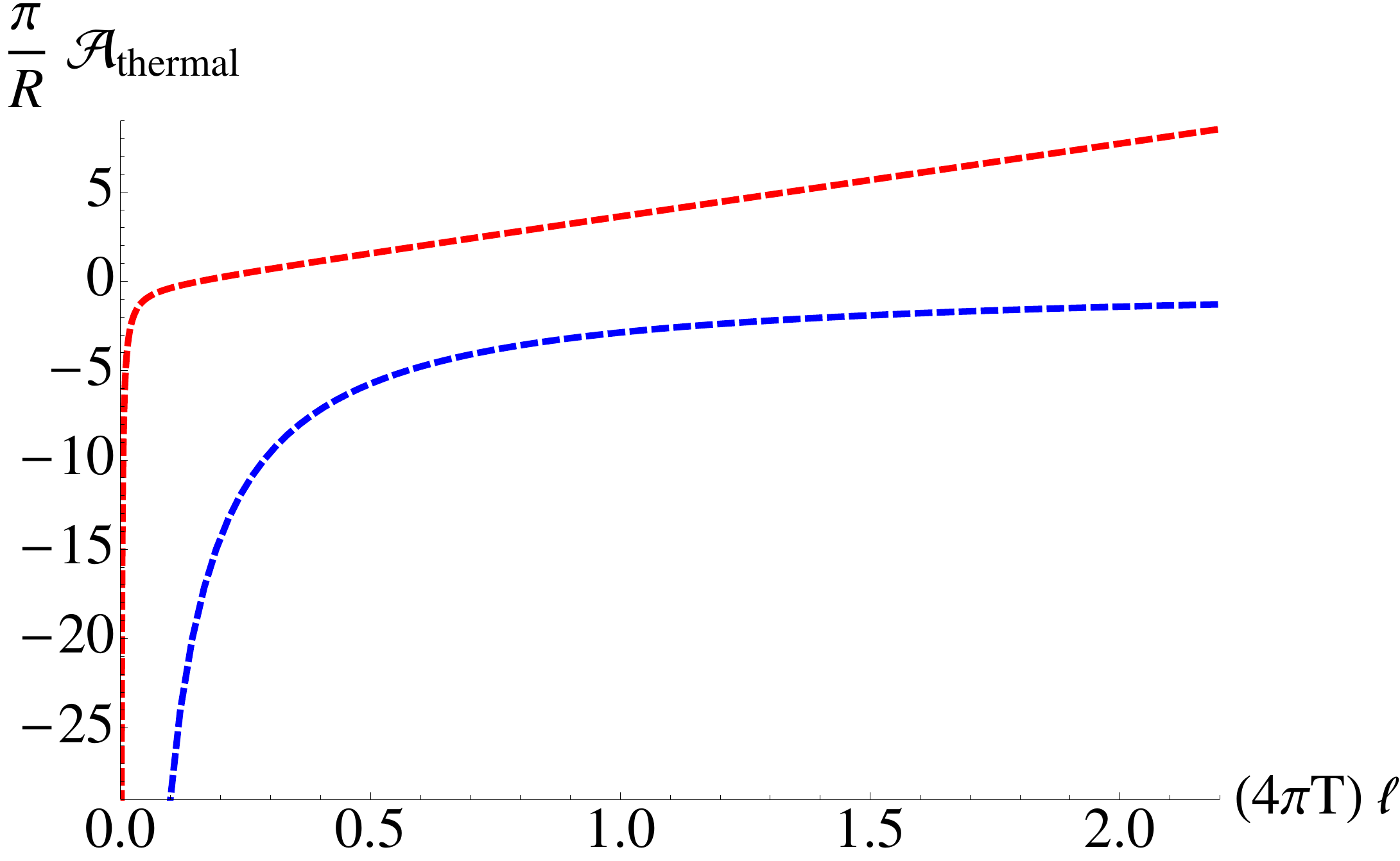} } 
\subfigure[] {\includegraphics[angle=0,
width=0.55\textwidth]{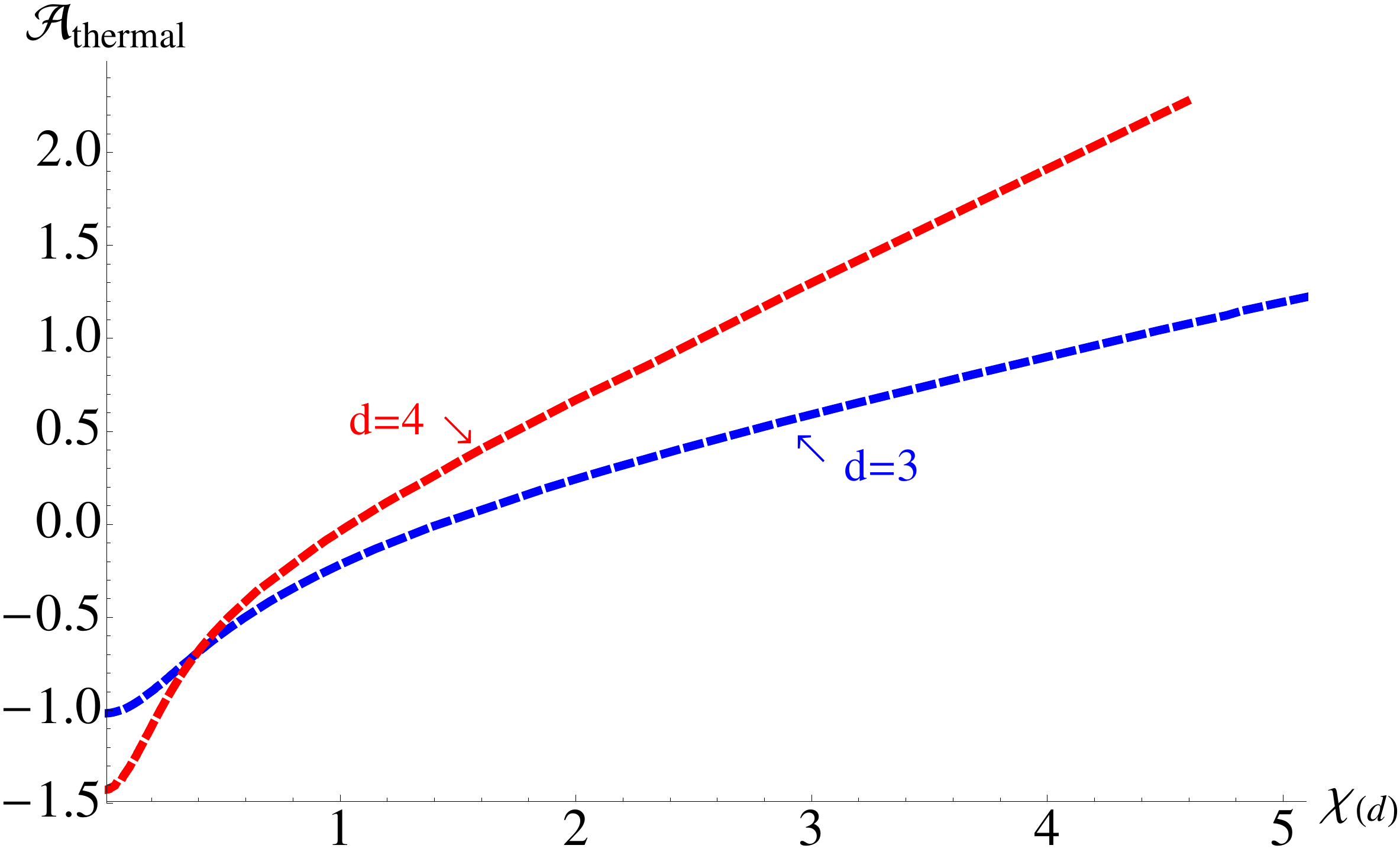} } 
\caption{\small Left panel (a): The case $d=3$. The blue curve corresponds to $\chi_{(3)} \approx 0.003$ and the red curve corresponds to $\chi_{(3)} \approx 22.5$. Right panel (b): The case $d=4$. The blue curve corresponds to $\chi_{(4)} \approx 0.002$ and the red curve corresponds to $\chi_{(4)} \approx 18.3$. In (c), we have shown how $\cA_{\rm thermal}$ scales with $\chi_{(d)}$ for both $d=3$ and $d=4$ for fixed $(4\pi T)\ell = 2$. $\cA_{\rm thermal}$ has been measured in units of the AdS-radius.}
\label{figd34RNwl}
\end{center}
\end{figure}

Using the solution in (\ref{eomwlrec}) in (\ref{areafunc}), the area of the minimal surface in purely AdS-background is given by
\begin{eqnarray}
\cA_{\rm AdS} = \alpha' \cA_{\rm NG} = \frac{RL^2}{\pi z_0} - \frac{RL^2}{\ell_{\rm AdS}} \left(\frac{\Gamma(3/4)}{\Gamma(1/4)}\right)^2 \ ,
\end{eqnarray}
where the first term in the above expression is divergent. Subtracting this diverging piece from the AdS-RN case, we get the following area of the minimal surface 
\begin{eqnarray} \label{Athermeq}
\cA_{\rm thermal} = \alpha' \cA_{\rm NG} = \frac{RL^2}{\pi} \int_{z_0}^{z_*} \frac{dz}{z^2} \frac{1}{\sqrt{f \left(1 - (z/z_*)^4\right)}} - \frac{1}{z_0} \frac{RL^2}{\pi} \ .
\end{eqnarray}
Note also that the formula in (\ref{Athermeq}) is valid for any $d$ since the information of the dimensionality is encoded in the function $f$. Here also we use a similar technique as outlined before to generate a curve of $\cA_{\rm thermal}$  (measured in units of appropriate powers of $L$) vs $T \ell$. The dependences are shown in fig.~\ref{figd34RNwl}.
The general observation is once again for increasing $\chi_{(d)}$, $\cA_{\rm thermal}$ increases monotonically: For large values of $T\ell$, we find $\cA_{\rm thermal} \sim (T\ell)$ with a slope that depends on $\chi_{(d)}$. In \ref{figd34RNwl}(c) we have shown how the area functional scales with the dimensionless ratio $\chi_{(d)}$ for fixed values of $(T\ell)$. Here also we find that the generic behavior is non-linear, but it becomes dominantly linear for large values of $\chi_{(d)}$ with a dimension-dependent slope. Non-linearities only show up for small values of $\chi_{(d)}$.

\subsubsection{Circular Wilson loop}

Let us now discuss the circular Wilson loop. At the boundary we choose a $2$-dimensional plane $\{x_1, x_2\}$ and rewrite it in the polar coordinate $\{\rho, \phi\}$
\begin{eqnarray}
dx_1^2 + dx_2^2 = d\rho^2 + \rho^2 d \phi^2 \ .
\end{eqnarray}
Using the azimuthal symmetry in the $\phi$-direction, the minimal area surface can be represented by $z(\rho)$ and $v(\rho)$. The Nambu-Goto action for this Wilson loop is given by
\begin{eqnarray}
\cA_{\rm NG} = \frac{L^2}{\alpha'} \int_0^R d\rho \frac{\rho}{z^2} \left(1 - f(z) v'^2 - 2 v' z' \right)^{1/2} \ ,
\end{eqnarray}
where $' \equiv d/d\rho$ and we have integrated over the azimuthal angular direction that yields a factor of $(2 \pi)$. Using the definition of $v$ once again we can write this as
\begin{eqnarray} \label{wleq}
\cA_{\rm NG} = \frac{L^2}{\alpha'} \int_0^R d\rho \frac{\rho}{z^2} \sqrt{1 + \frac{z'^2}{f(z)}} \ .
\end{eqnarray}
Note that in this case we do not have any conservation equation. The equation of motion for $z(\rho)$ obtained from (\ref{wleq}) is given by
\begin{eqnarray} \label{wleqn}
z'' + z'^3 \frac{1}{\rho f} + z'^2 \left( \frac{2}{z} - \frac{1}{2f} \frac{df}{dz} \right) + \frac{z'}{\rho} + \frac{2f}{z} = 0 \ .
\end{eqnarray}
The boundary conditions we should use are given by
\begin{eqnarray} \label{bcwlcir1}
z(0) = z_* \ , \quad z'(0) = 0 \ .
\end{eqnarray}
By expanding near the $\rho=0$ point, we get
\begin{eqnarray} \label{bcwlcir2}
z_{\rm middle}(\rho) = z_* - \left(\frac{1 - M z_*^3}{2 z_*} + \frac{Q^2}{4} z_*^3 \right) \rho^2 + \ldots \ .
\end{eqnarray}
We use the above expansion at $\rho=\epsilon$ to impose the boundary conditions and solve the equation (\ref{wleqn}) using Mathematica's NDSolve, where $\epsilon$ is some small number.\footnote{Typically we have used $\epsilon \sim 10^{-3}$.} To generate the desired curve we now have to carry out the following steps: First we fix $M$ (set to unity) and $Q$. Then we choose some $z_*$ which is numerically close to the event-horizon. For these choices of the input parameters and the above boundary conditions we first solve the differential equation numerically. To read off $R$, we now impose the (radial) IR boundary condition: $z(R) = z_0$, where $z_0$ is the IR cut-off. Using this numerical solution we can generate an unique value for $\cA_{\rm thermal}$ for a given $R$. Now we can vary $z_*$ and keep repeating the same process to generate the curve we want.

\begin{figure}[!ht]
\begin{center}
\subfigure[] {\includegraphics[angle=0,
width=0.45\textwidth]{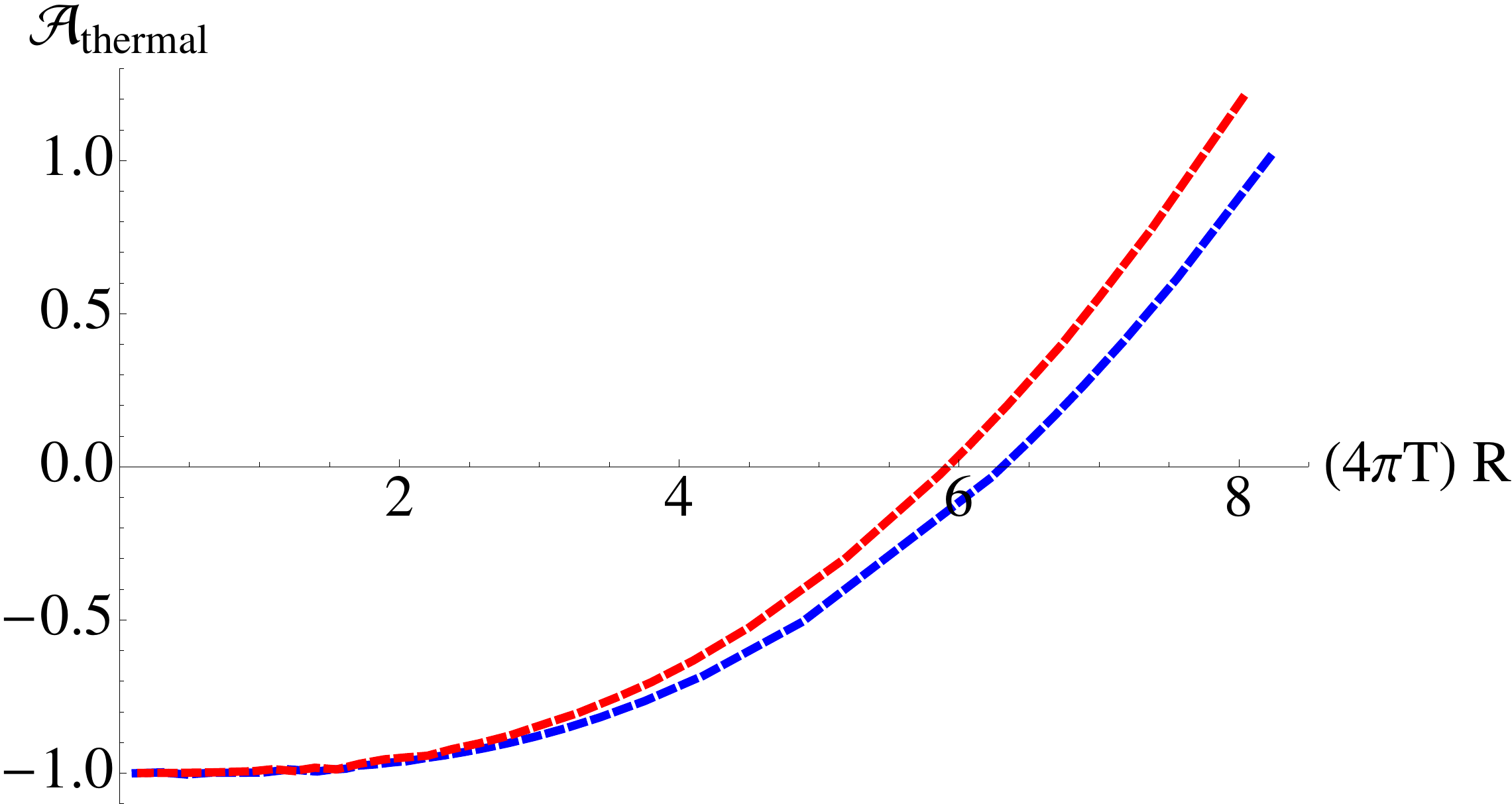} } 
\subfigure[] {\includegraphics[angle=0,
width=0.45\textwidth]{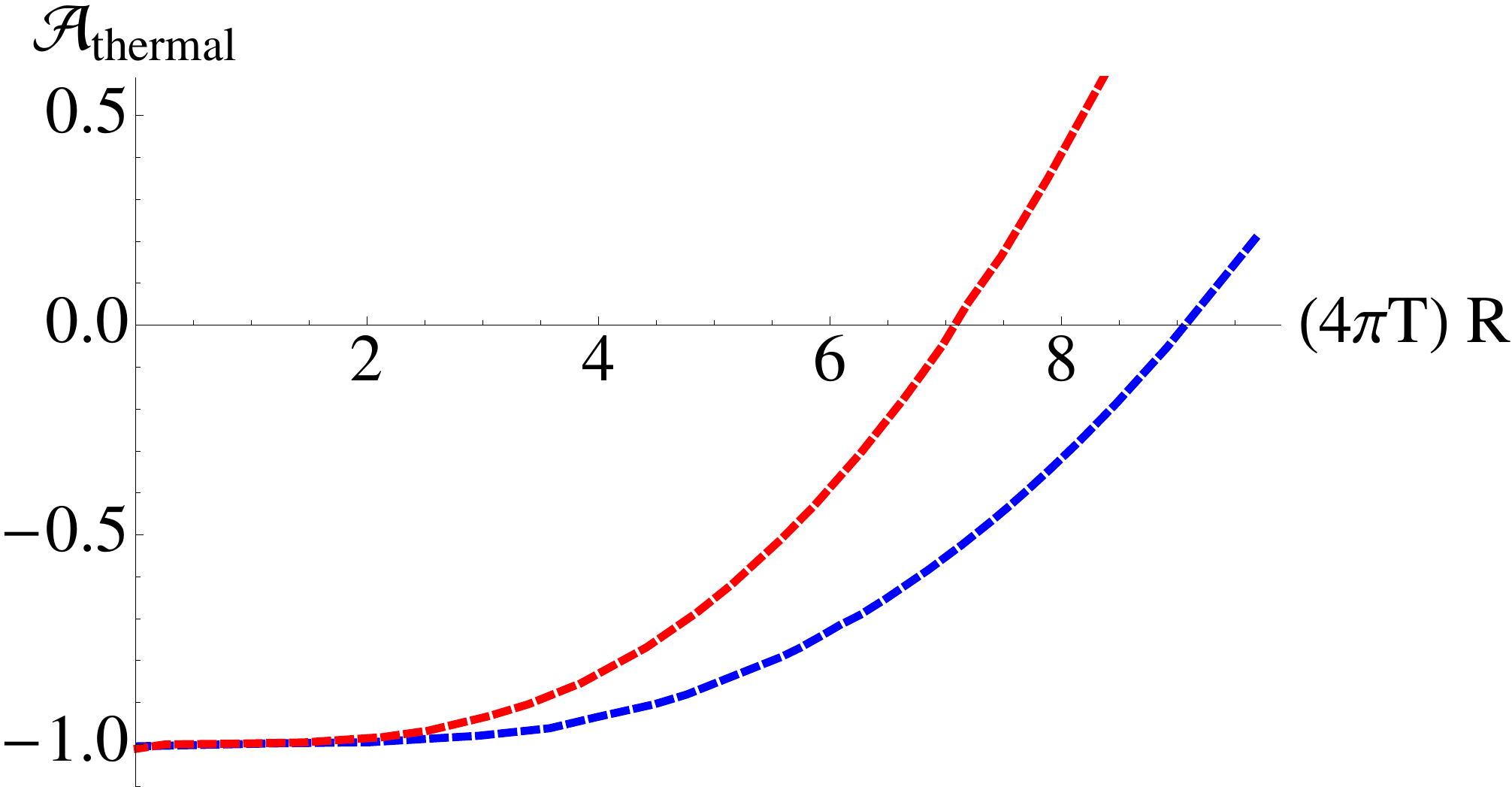} } 
\caption{\small Left panel: The case $d=3$. The blue curve corresponds to $\chi_{(3)} \approx 0.003$ and the red curve corresponds to $\chi_{(3)} \approx 0.19$. Right panel: The case $d=4$. The blue curve corresponds to $\chi_{(4)} \approx 0.002$ and the red curve corresponds to $\chi_{(4)} \approx 0.24$. $\cA_{\rm thermal}$ is measured in units of the AdS-radius.}
\label{figd34RNwlcir}
\end{center}
\end{figure}

Before presenting the results, let us recall again that this area is formally a divergent quantity. To extract the divergent piece, we can again just focus on the pure AdS-case. In this case (which corresponds to setting $M=0$ and $Q=0$), a simple solution of (\ref{wleqn}) is given by\footnote{This solution was obtained in \cite{Berenstein:1998ij} by taking an infinite straight Wilson line and then applying conformal transformation to map the line to a circle.}
\begin{eqnarray}
z(\rho) = \sqrt{R^2 - \rho^2} \ ,
\end{eqnarray}
which ultimately gives
\begin{eqnarray}
\cA_{\rm AdS} = \frac{L^2}{\alpha'} \frac{R}{z_0} - \frac{L^2}{\alpha'} \ ,
\end{eqnarray}
where $z_0$ is the radial IR cut-off. Thus the finite contribution to the thermal Wilson loop is obtained to be
\begin{eqnarray}
\cA_{\rm thermal} = \alpha' \cA_{\rm NG} - \frac{RL^2}{z_0} \ .
\end{eqnarray}
The numerical results are shown in fig.~\ref {figd34RNwlcir}. It is clear that the expectation value of the Wilson loop depends on the dimensionless parameter $\chi_{(d)}$. Our general observation again seems to hold here, {\it i.e.} increasing $\chi_{(d)}$ increases the area of the minimal surface for a given value of $(TR)$. Note that in this case, for large values of $(TR)$, $\cA_{\rm thermal} \sim (TR)^2$.
%

\subsection{Entanglement Entropy}

Consider a quantum field theory with many degrees of freedom and assume that the zero temperature ground state of the system is described by the pure ground state denoted by $\left. | \Psi \right >$, which does not have any degeneracy. The von Neuman entropy of this system, defined as $S_{\rm total} = - {\rm tr} \rho \log \rho = 0$, where $\rho = \left | \Psi \right > \left < \Psi \right |$.

Now we can imagine dividing the system into two subsystems $A$ and $B$. The total Hilbert space now factorizes as $\cH_{\rm total} = \cH_{A} \otimes \cH_{B}$. Now let us imagine an observer who has access only to the subsystem $A$, consequently the relevant density matrix for this observer is the reduced density matrix defined as
\begin{eqnarray}
\rho_A = {\rm tr}_B \, \rho \ .
\end{eqnarray}
The entanglement entropy is now defined as the von Neuman entropy using this reduced density matrix
\begin{equation}
S_A = - {\rm tr}_A \, \rho_A \log \rho_A \ .
\end{equation}
Entanglement entropy measures the quantum entanglement between the two subsystems $A$ and $B$ and it is non-zero even at zero temperature. Entanglement entropy is also an useful order parameter in quantum phase transitions. Here we will eventually use this observable to probe the physics of thermalization in the presence of a chemical potential and we will explicitly demonstrate that this is the observable that sets the time-scale of equilibration since of all the non-local operators considered, it thermalizes the latest.

A prescription for computing entanglement entropy using the AdS/CFT correspondence was suggested by \cite{Ryu:2006bv} and has been analyzed in details in the literature since then. For time-dependent background, the covariant proposal for entanglement entropy was suggested in \cite{Hubeny:2007xt}. According to this proposal one needs to consider extremal surfaces instead of minimal area surfaces For a recent review, see {\it e.g.} \cite{Nishioka:2009un}. Let us consider bulk AdS$_{d+1}$ space-time. Then the formula for the entanglement entropy is given by
\begin{eqnarray}
S_A = \frac{{\rm Area} \left(\gamma_A\right)}{4 G_N^{(d+1)}} \ ,
\end{eqnarray}
where $G_N^{(d+1)}$ is the $(d+1)$-dimensional Newton's constant; $\gamma_A$ denotes the $(d-1)$-dimensional minimal surface whose boundary coincide with the boundary of the region $A$: $\partial \gamma_A = \partial A$.

In $d=3$, the computation of the entanglement entropy is identical to the Wilson loop computation. Thus here we will only consider the case $d=4$. As before, we will consider two geometric shapes: infinite rectangular strip or the straight belt and the spherical region. Fig.~\ref{rec_cir_shape} again serves as a pictorial representation of the particular shape considered and the corresponding minimal surface. Thermalization of the entanglement entropy in $d=2$ was first studied in \cite{AbajoArrastia:2010yt} for vanishing chemical potential.

\subsubsection{The straight belt}

The straight belt can be parametrized by the boundary coordinates $\{x_1, x_2, x_3\}$ with the assumption that this infinite strip is invariant under both $x_2$ and $x_3$-directions. The corresponding minimal surface can be represented by $z(x)$ and $v(x)$ where $x\equiv x_1$. As before, the boundary conditions imposed are
\begin{eqnarray}
z(\ell/2) = z_0 = z(- \ell/2) \ , \quad v(\ell/2) = t_0 = v(- \ell/2)  \ . 
\end{eqnarray}
The volume functional is given by
\begin{eqnarray}
\cV = A \int_{-\ell/2}^{\ell/2} \frac{L^3dx}{z^3} \left( 1 - f v'^2 - 2 v' z' \right)^{1/2} \ , \quad {\rm where} \quad  \cV \equiv {\rm Area} (\gamma_A) \ .
\end{eqnarray}
Here $A$ is the area that results from integrating over the $x^2$ and the $x^3$-directions. The equation of motion resulting from the conservation equation is given by
\begin{eqnarray}
1 - fv'^2 - 2 v' z' = \left(\frac{z_*}{z}\right)^6 \ .
\end{eqnarray}
Using the definition of $v$ from (\ref{vdef}) we can obtain
\begin{eqnarray}
1 + \frac{z'^2}{f} = \left(\frac{z_*}{z}\right)^6 \quad \implies \frac{dz}{dx} = \pm \sqrt{f(z)} \left[ \left(\frac{z_*}{z} \right)^6 - 1 \right]^{1/2} \ ,
\end{eqnarray}
where the positive sign is taken for $x>0$ and the negative sign is taken for $x<0$. We can analytically obtain the length for pure AdS-background to give
\begin{eqnarray}
\ell_{\rm AdS} = z_* \sqrt{\pi} \frac{\Gamma(2/3)}{\Gamma(1/6)} \ .
\end{eqnarray}
Finally the volume functional is given by
\begin{eqnarray}
\cV = 2 A L^3 \int_{z_0}^{z_*} \frac{dz}{z^3} \frac{1}{\sqrt{f}} \frac{1}{\sqrt{1- (z/z_*)^6}} \ ,
\end{eqnarray}
which is again a formally divergent quantity. In pure AdS, this volume functional is obtained to be
\begin{eqnarray}
\cV_{\rm AdS} = 2 A L^3 \int_{z_0}^{z_*} \frac{dz}{z^3} \frac{1}{\sqrt{1- (z/z_*)^6}} & = & \frac{A L^3}{z_0^2} + A L^3 \frac{\sqrt{\pi}}{3 z_*^2} \frac{\Gamma(-1/3)}{\Gamma(1/6)} \nonumber\\
& = & \frac{A L^3}{z_0^2} + \frac{A L^3}{\ell_{\rm AdS}^2} \frac{\pi\sqrt{\pi}}{3} \frac{\Gamma(-1/3) \left(\Gamma(2/3)\right)^2}{\left( \Gamma(1/6) \right)^3} \ .
\end{eqnarray}
Thus the finite part of the volume functional can be obtained to be
\begin{eqnarray}
\cV_{\rm thermal}  = 2 A L^3 \int_{z_0}^{z_*} \frac{dz}{z^3} \frac{1}{\sqrt{f}} \frac{1}{\sqrt{1- (z/z_*)^6}}  - \frac{A L^3}{z_0^2} \ .
\end{eqnarray}
\begin{figure}[!ht]
\begin{center}
\subfigure[] {\includegraphics[angle=0,
width=0.5\textwidth]{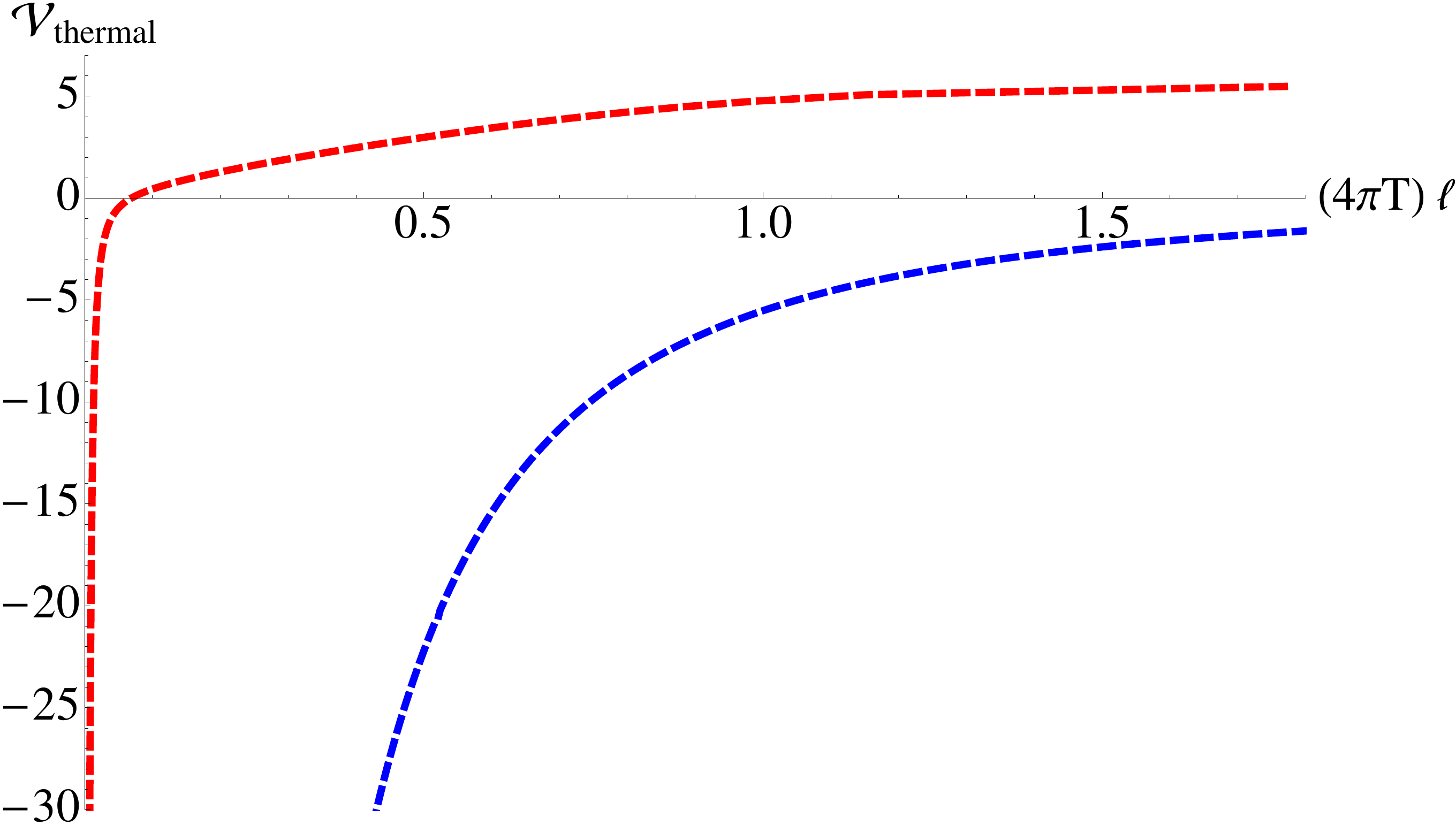} } 
\subfigure[] {\includegraphics[angle=0,
width=0.45\textwidth]{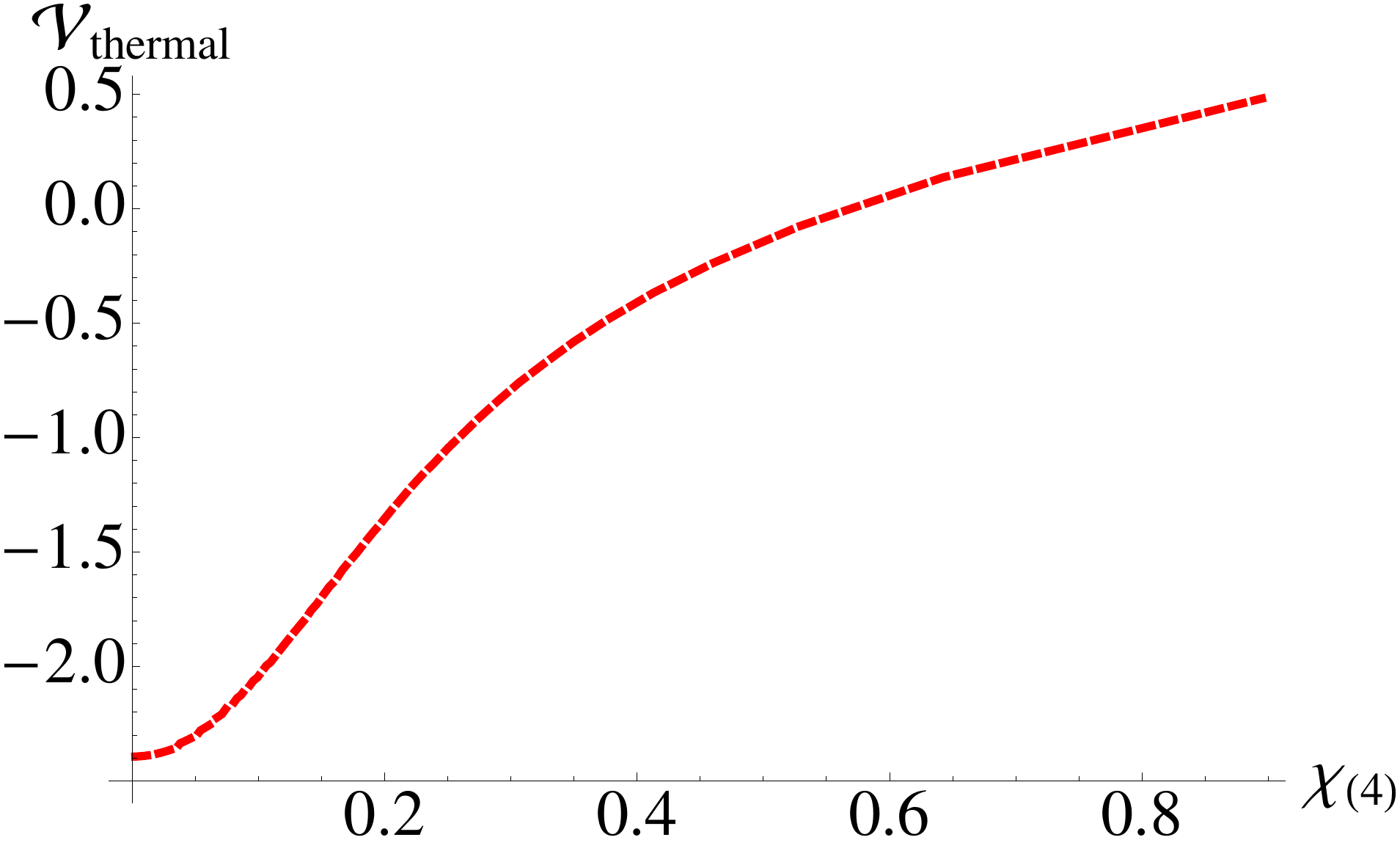} } 
\caption{The case $d=4$. Left panel: The blue curve corresponds to $\chi_{(4)} \approx 0.002$ and the red curve corresponds to $\chi_{(4)} \approx 18.3$. Right panel: the scaling of $\cV_{\rm thermal}$ with $\chi_{(4)}$ for fixed $(4\pi T) \ell = 1.5$. We measure $\cV_{\rm thermal}$ in units of the AdS-radius and also set $A=1$.}
\label{figd4RNeerec}
\end{center}
\end{figure}
The behaviour of the entanglement entropy is demonstrated in fig.~\ref{figd4RNeerec}. Here we also notice that for large values of $(T\ell)$, $\cV_{\rm thermal} \sim (T\ell)$ with a slope that depends on $\chi_{(4)}$. On the other hand, the scaling of $\cV_{\rm thermal}$ with $\chi_{(4)}$ displays the non-linear and monotonically increasing behavior where the non-linearities are washed out for large values of $\chi_{(4)}$. Although we have not displayed it in the figure, the slope of the linear behavior of $\cV_{\rm thermal}$ with $\chi_{(4)}$ for large $\chi_{(4)}$ depends on the value of $(T\ell)$.

\subsubsection{The spherical region}

Now we consider the case where one of the subsystem is a circular disc. To parametrize this circular disc in polar coordinate, we rewrite
\begin{eqnarray} \label{polarcir}
\sum_{i=1}^{d-1} dx_i^2 = d\rho^2 + \rho^2 d\Omega_{d-2}^2 \ .
\end{eqnarray}
The minimal area surface can now be parametrized by $z(\rho)$ and $v(\rho)$. The volume element in this case is given by
\begin{eqnarray} \label{ee5}
\cV & = & 4 \pi  \int_0^R d\rho \frac{L^3 \rho^2}{z^3} \left( 1 - f v'^2 - 2 z' v' \right)^{1/2} \nonumber\\
   & = & 4 \pi \int_0^R d\rho \frac{L^3 \rho^2}{z^3}  \left( 1 + \frac{z'^2}{f} \right)^{1/2} \ .
\end{eqnarray}
We do not have the integral of motion anymore and the full equations of motions are obtained by varying the functional in (\ref{ee5}). To avoid clutter, we do not present the specific form of the equation of motion here. The boundary conditions are once again
\begin{eqnarray}
z(0) = z_* \ , \quad z'(0) = 0 \ .
\end{eqnarray}
As before, solving the equation near the $\rho=0$ region we get
\begin{eqnarray}
z(\rho) = z_* - \frac{3 - 3 M z_*^4 + 2 Q^2 z_*^6}{6 z_*} \rho^2 + \ldots \ .
\end{eqnarray}
In practice we use the above expansion at $\rho = \epsilon$ to impose the boundary conditions. Here $\epsilon$ is a small number typically of the order of $10^{-3}$.

\begin{figure}[!ht]
\begin{center}
\includegraphics[width=8.5cm]{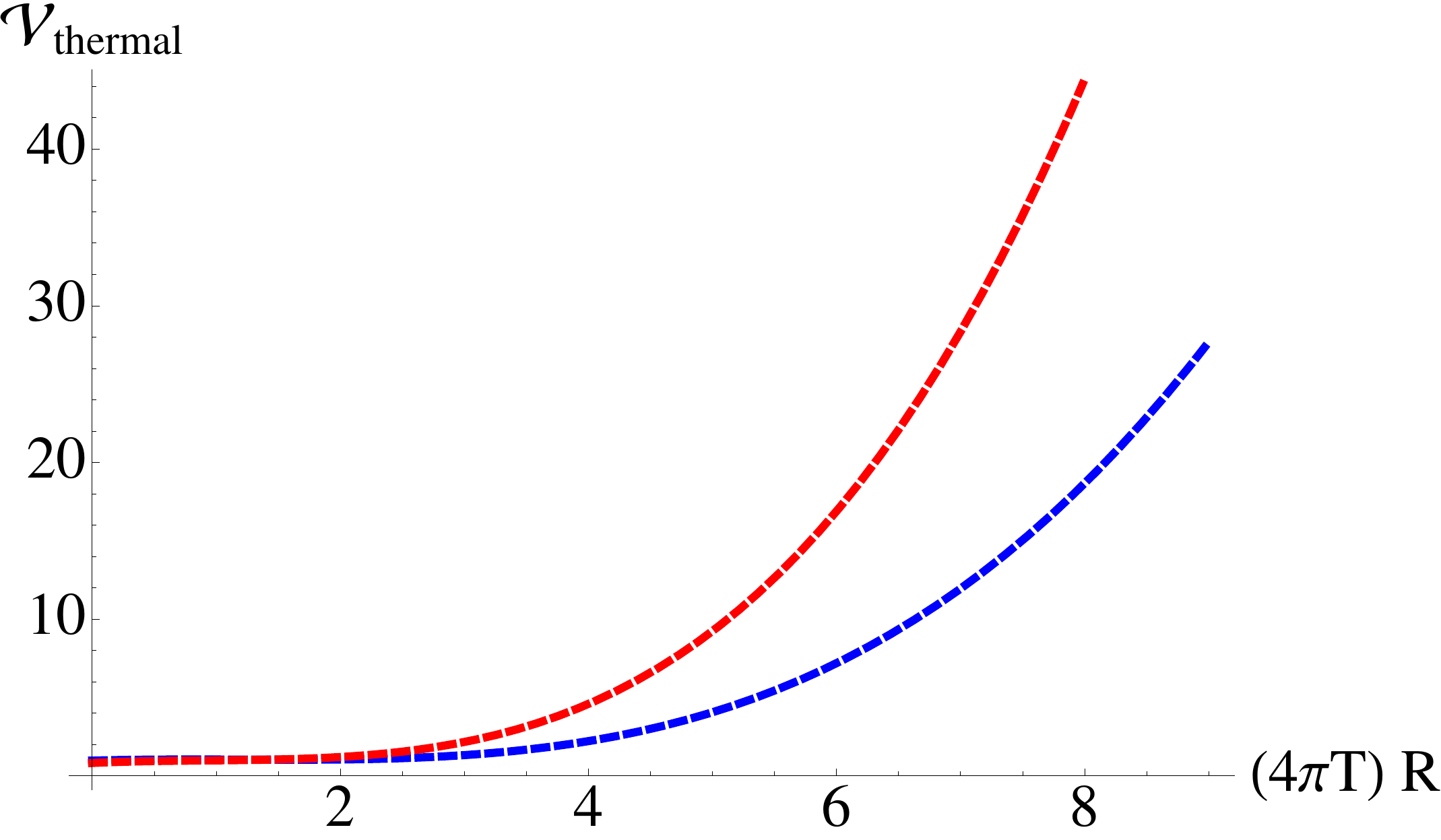}
\caption{The case $d=4$. The blue curve corresponds to $\chi_{(4)} \approx 0.002$ and the red curve corresponds to $\chi_{(4)} \approx 0.24$. The dimensionful quantity $\cV_{\rm thermal}$ is measured in units of the AdS-radius.}
\label{figd4RNeecir}
\end{center}
\end{figure}

To determine the divergent piece in the volume, let us evaluate it in the pure AdS-case. The solution of the minimal surface is simple: $z^2 = R^2 - \rho^2$. So we get
\begin{eqnarray}
\cV_{\rm AdS} = 4 \pi  R L^3 \int_0^{\rho(z_0)} d \rho \frac{\rho^2}{\left(R^2 - \rho^2 \right)^2} = 2\pi L^3 \left(\frac{R^2}{z_0^2} + \log \frac{z_0}{\sqrt{2} R}\right) + {\rm finite} \ .
\end{eqnarray}
Hence the finite part we are interested in is given by
\begin{eqnarray}
\cV_{\rm thermal} = 4 \pi L^3 \int_0^R d\rho \frac{\rho^2}{z^3}  \left( 1 + \frac{z'^2}{f} \right)^{1/2} - 2\pi L^3 \left(\frac{R^2}{z_0^2} + \log \frac{z_0}{\sqrt{2} R}\right) \ .
\end{eqnarray}
The dependence is shown in fig.~\ref{figd4RNeecir}. 
Once again we observe the behavior that increasing $\chi_{(4)}$ increases the volume functional for a spherical subsystem of a given radius. For large values of $(TR)$, we recover a quadratic behavior of the volume functional, which is expected from the general area-law scaling of entanglement entropy.

\section{Non-equilibrium physics}

\subsection{The bulk action and the background}

Our goal here is to study the same set of non-local observables in a time-dependent background. This time-dependent background should capture the physics of the formation of a black hole --- in the present context --- a black hole with a definite mass and charge. To that end, we will use a generalized version of the AdS-Vaidya background including a charge for the black hole. For obvious reasons, we will call this the AdS-RN-Vaidya background.

To find the corresponding Vaidya background, we have to couple the above action in (\ref{action1}) with an external source
\begin{eqnarray}
S = S_0 + \kappa S_{\rm ext} \ ,
\end{eqnarray}
where $\kappa$ is a constant and we do not specify the form of $S_{\rm ext}$. The equations of motion in this case will take the following form
\begin{eqnarray}
&& R_{\mu\nu} - \frac{1}{2} \left(R- 2 \Lambda \right) g_{\mu\nu} - g^{\alpha\rho} F_{\rho\mu} F_{\alpha\nu} + \frac{1}{4} g_{\mu\nu} \left(F^{\alpha\beta} F_{\alpha \beta}\right) = 2 \left( 8\pi G_N^{(d+1)} \kappa \right) T_{\mu\nu}^{\rm ext} \ , \\
&& \partial_\rho \left[ \sqrt{-g} g^{\mu\rho} g^{\nu\sigma} F_{\mu\nu} \right] = \left( 8\pi G_N^{(d+1)}\kappa\right) J_{\rm ext}^{\sigma} \ .
\end{eqnarray}
Let us start with the case where the matter field (corresponding to $S_{\rm ext}$) is neutral and subsequently the black hole formed will be the AdS-Schwarzschild. The metric is the $(d+1)$-dimensional infalling shell geometry described in the Poincar\`{e} patch by
\begin{eqnarray} \label{vaid1}
ds^2 = \frac{L^2}{z^2} \left( - f(v,z) dv^2 - 2 dv dz + d\vec{x}^2 \right) \ , \quad f(z,v) = 1 - m(v) z^d 
\end{eqnarray}
and is known as the AdS-Vaidya background. Here $m(v)$ is a function that captures the information of the black hole formation. On physical ground, $m(v)$ should interpolate between zero (in the limit $v\to -\infty$ corresponding to pure AdS) and a constant value (in the limit $v\to \infty$ corresponding to AdS-Sch). A {\it choice} of such a function is
\begin{eqnarray} \label{mchange}
m(v) = \frac{M}{2} \left(1 + \tanh \frac{v}{v_0} \right) \ .
\end{eqnarray}
Here $v_0$ is a parameter that denotes the thickness of the shell. With this choice, the external source must yield the following energy-momentum tensor
\begin{eqnarray}
2  \left( 8\pi G_N^{(d+1)} \kappa \right) T_{\mu\nu}^{\rm ext}  = \frac{d-1}{2} z^{d-1} \frac{dm}{dv} \delta_{\mu v} \delta_{\nu v} \ .
\end{eqnarray}
If we identify $k_\mu = \delta_{\mu v}$, then we get\cite{AbajoArrastia:2010yt}
\begin{eqnarray}
T_{\mu\nu}^{\rm ext} \sim k_\mu k_\nu \ , \quad {\rm with} \quad k^2 = 0 \ ,
\end{eqnarray}
which is characteristic of null dust. Thus the formation of the black hole is realized by a shell of infalling null dust. In \cite{Balasubramanian:2010ce, Balasubramanian:2011ur}, this metric has been used to study aspects of thermalization.

Our first goal here is to generalize the Vaidya metric for the AdS-RN background in $(d+1)$-dimensions. It is clear that now we need a charged null dust for the formation of the black hole. As before, we will present the corresponding Vaidya metric for specific values of $d$.\footnote{Note that here we are not careful about the regularity of the one-form $A_v$ at any particular point since the horizon is created only in the $v\to \infty$ limit. Perhaps a better way to write the background is to write the field strength instead of the gauge field itself. For all our purposes though, this gauge field does not play any role.}

In $d=2$ we get
\begin{eqnarray} \label{RNv2}
&& ds^2 = \frac{L^2}{z^2} \left( - f(z,v) dv^2 - 2 dv dz + dx^2 \right) \ , \quad A_v = q(v) \log z \ , \\
&& f(z,v) = 1 - m(v) z^2 + \frac{q(v)^2}{L^2} z^2 \log z \ , \quad \Lambda = - \frac{1}{L^2} \ , \\
&& 2 \kappa T_{\mu\nu}^{\rm ext} = \frac{1}{2} z \left( \frac{dm}{dv} - \frac{2}{L^2} \log (z) q(v) \frac{dq}{dv} \right) \delta_{\mu v } \delta_{\nu v} \ , \quad \kappa J_{\rm ext}^\mu = \frac{1}{L} \frac{dq}{dv} \delta^{\mu z} \ .
\end{eqnarray}

In $d=3$ we get
\begin{eqnarray} \label{RNv3}
&& ds^2 = \frac{L^2}{z^2} \left( - f(z,v) dv^2 - 2 dv dz + d\vec{x}^2 \right) \ , \quad A_v = q(v) z \ , \\
&& f(z,v) = 1 - m(v) z^3 + \frac{q(v)^2}{2 L^2} z^4 \ , \quad \Lambda = - \frac{3}{L^2} \ , \\
&& \kappa J_{\rm ext}^\mu = \frac{dq}{dv} \delta^{\mu z} \ , \quad 2 \kappa T_{\mu\nu}^{\rm ext} = z^2 \left[ \frac{dm}{dv} - \frac{z}{L^2} q(v) \frac{dq}{dv} \right] \delta_{\mu v} \delta_{\nu v} \ .
\end{eqnarray}

And, finally in $d=4$ we get
\begin{eqnarray} \label{RNv4}
&& ds^2 = \frac{L^2}{z^2} \left( - f(z,v) dv^2 - 2 dv dz + d\vec{x}^2 \right) \ , \quad A_v = q(v) z^2 \ , \\
&& f(z,v) = 1 - m(v) z^4 + \frac{2 q(v)^2}{3 L^2} z^6 \ , \quad \Lambda = - \frac{6}{L^2} \ , \\
&& \kappa J_{\rm ext}^\mu = 2 L \frac{dq}{dv} \delta^{\mu z} \ , \quad 2 \kappa T_{\mu\nu}^{\rm ext} = \frac{3}{2} z^3 \frac{dm}{dv} - \frac{2 z^5}{L^2} q(v) \frac{dq}{dv}\delta_{\mu v} \delta_{\mu v} \ .
\end{eqnarray}
As before, we will only consider the cases $d=3, 4$.\footnote{A general form of this background in $(d+1)$-dimensions in given in equation (\ref{genvaidya}).} For all these metrics, we can choose the profile for $m(v)$ as given in (\ref{mchange}) and we are also free to pick an analogous profile for $q(v)$. For future purposes, we will set the radius of the AdS-space, $L=1$, and thus all dimensionful quantities (the length, area or the volume functional as the case may be) will be measured in units of this scale.

\begin{figure}[!ht]
\begin{center}
\subfigure[] {\includegraphics[angle=0,
width=0.45\textwidth]{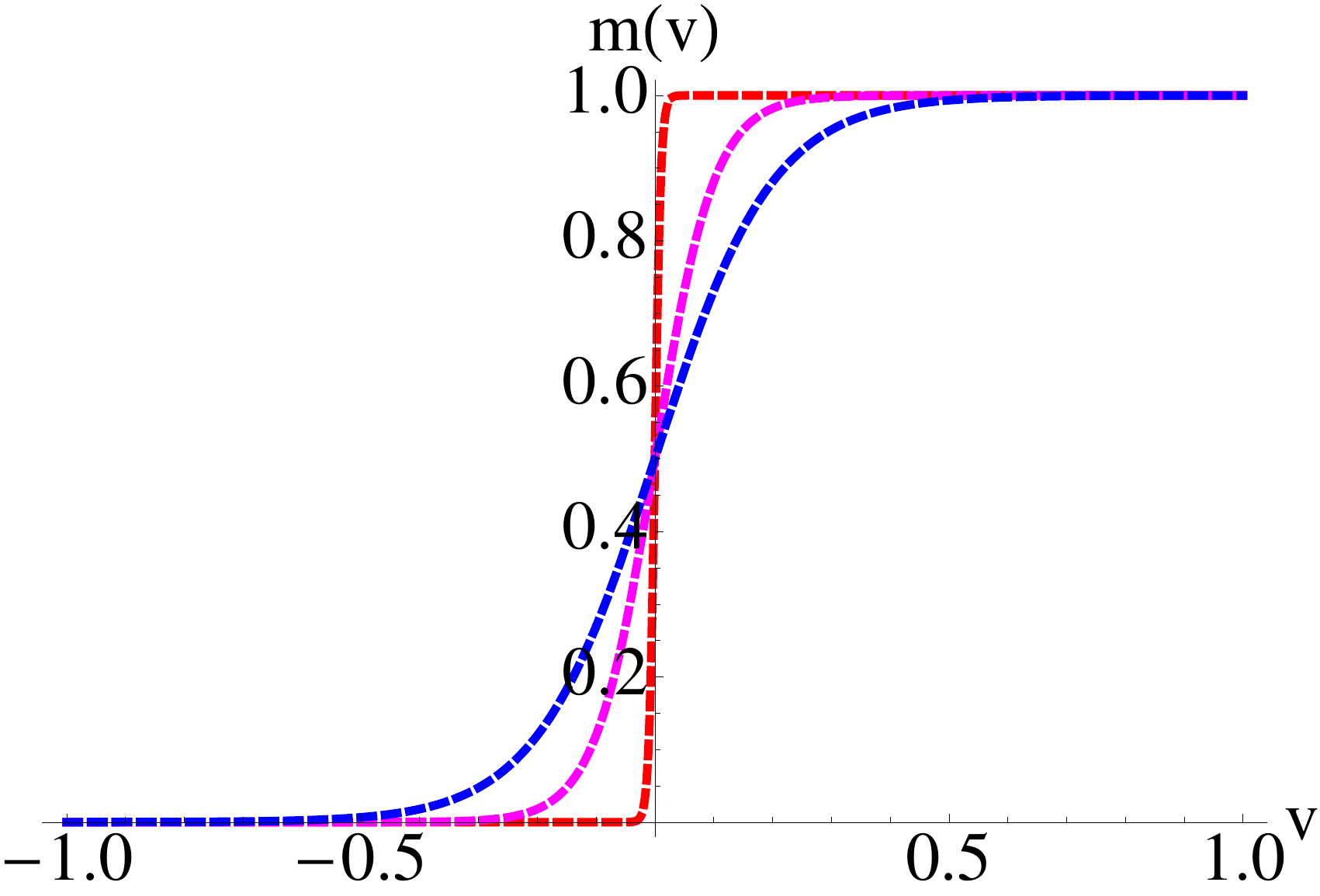} } 
\subfigure[] {\includegraphics[angle=0,
width=0.45\textwidth]{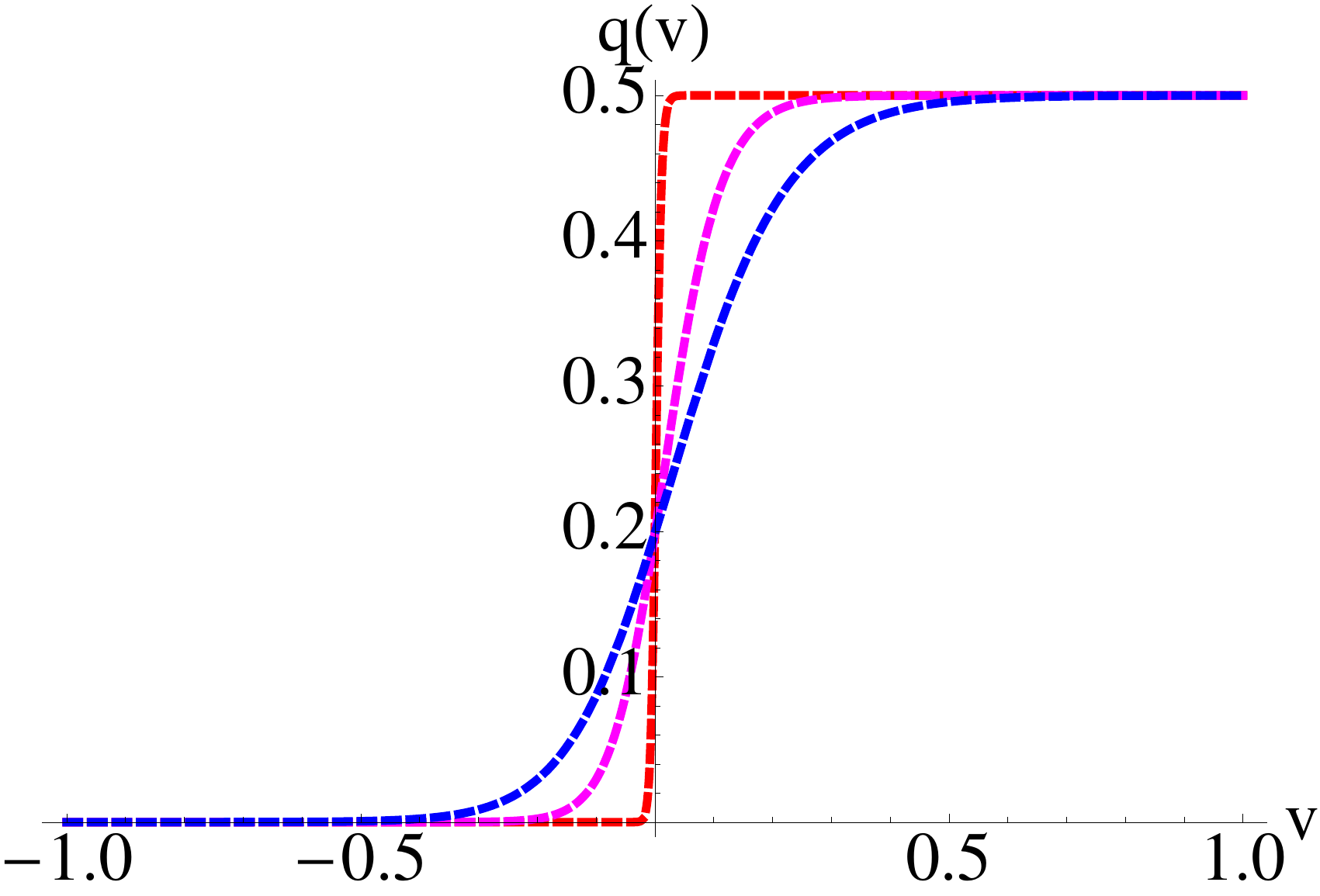} } 
\caption{\small The interpolating functions $m(v)$ and $q(v)$ in $d=3$ for various values of $v_0=0.01, 0.1, 0.2$ (red, pink and blue). We have set $Q=0.5$.}
\label{d3mqv}
\end{center}
\end{figure}

\begin{figure}[!ht]
\begin{center}
\subfigure[] {\includegraphics[angle=0,
width=0.45\textwidth]{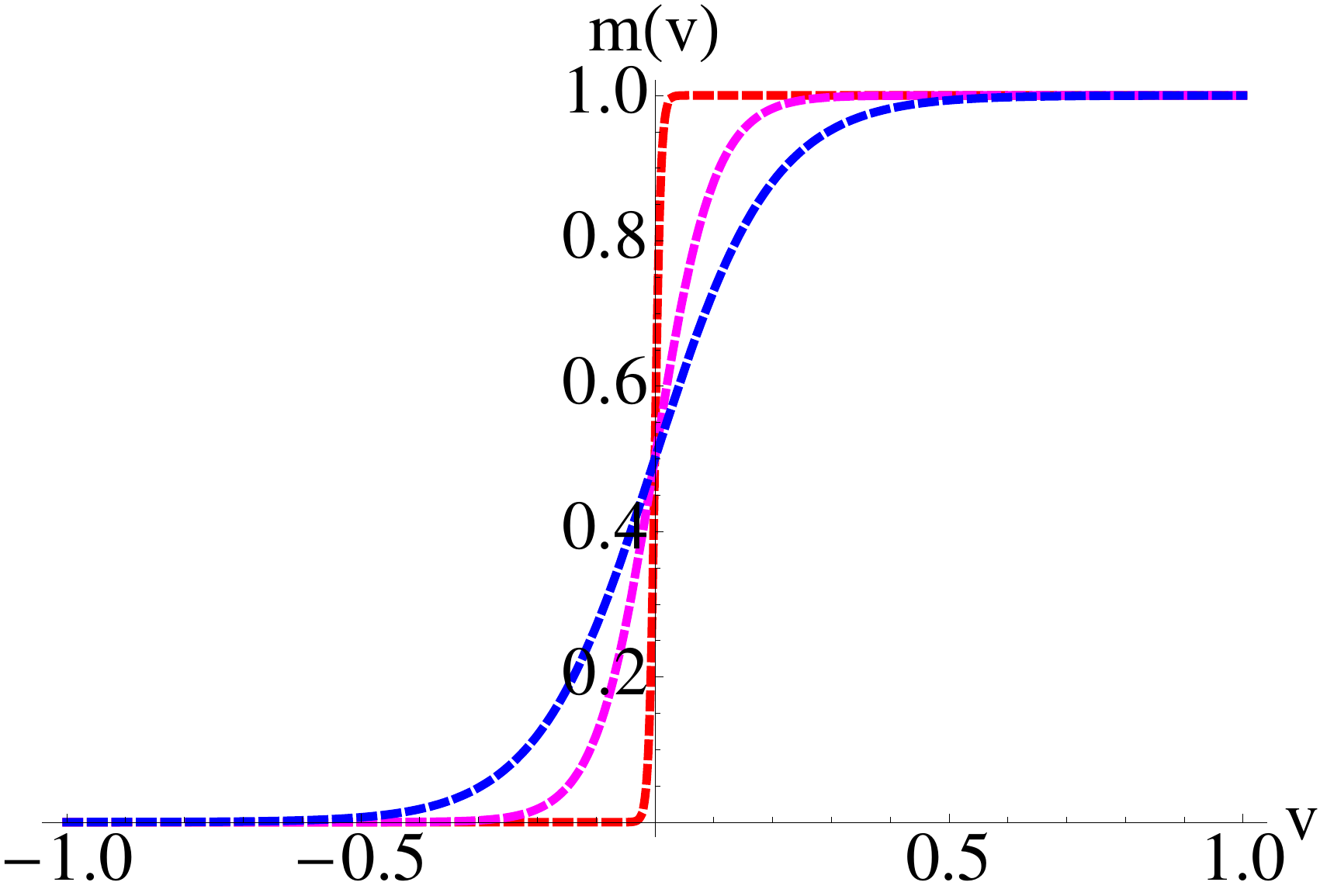} } 
\subfigure[] {\includegraphics[angle=0,
width=0.45\textwidth]{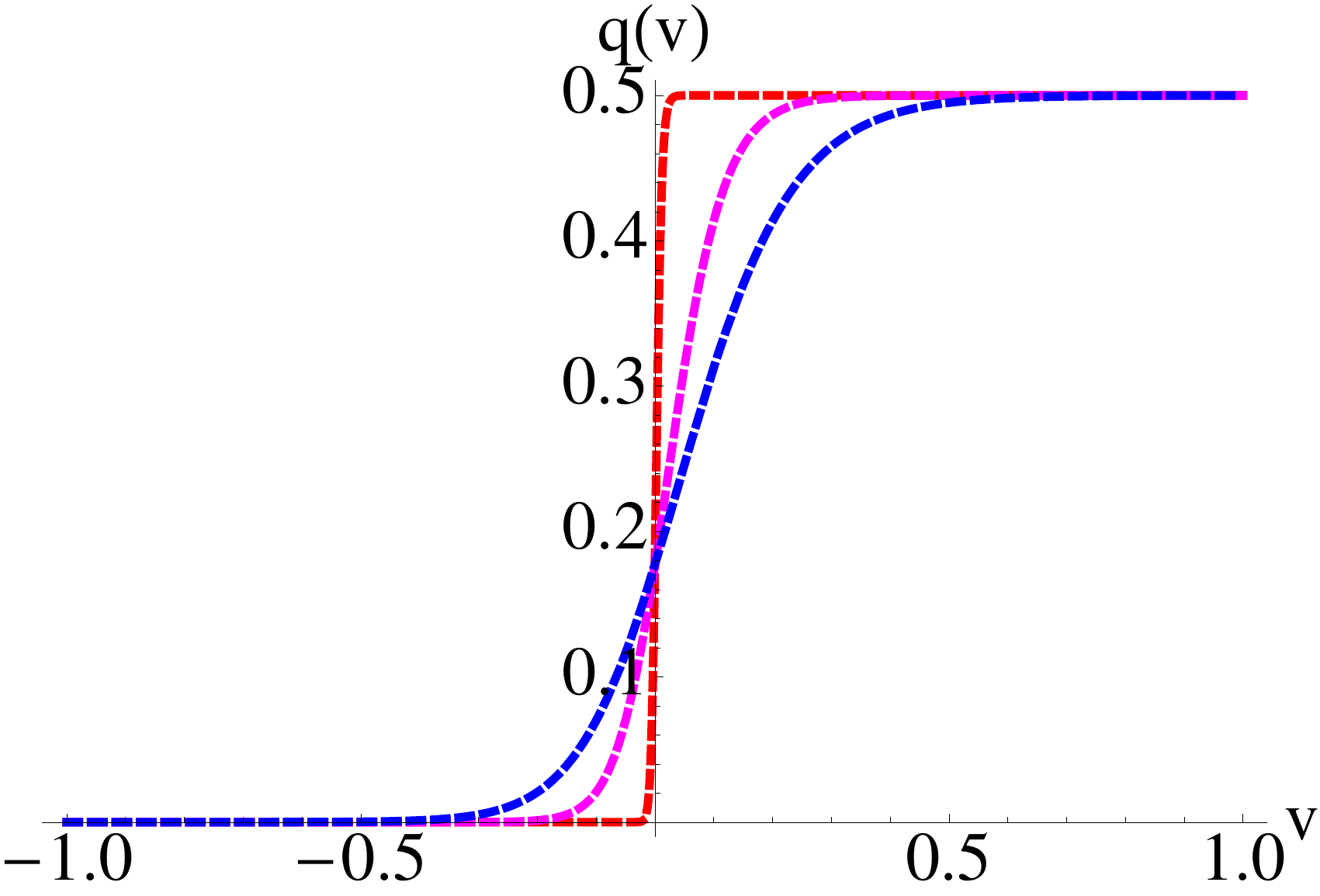} } 
\caption{\small The interpolating functions $m(v)$ and $q(v)$ in $d=4$ for various values of $v_0=0.01, 0.1, 0.2$ (red, pink and blue). We have set $Q=0.5$.}
\label{d4mqv}
\end{center}
\end{figure}

Now we will work with the backgrounds given in (\ref{RNv3}) and (\ref{RNv4}). As previously stated we will work with the following mass function
\begin{eqnarray} \label{mv}
m(v) = \frac{M}{2} \left(1 + \tanh\frac{v}{v_0}\right) \ .
\end{eqnarray}
To pick a charge function, we note the following\cite{Albash:2010mv}: For $d=3$ we can rewrite the function $f(z,v)$ as follows
\begin{eqnarray} 
f = 1 - \left(\frac{z}{z_H(v)}\right)^3 + \frac{Q^2}{2} \left(\frac{z}{z_H(v)}\right)^4 \ ,
\end{eqnarray}
which gives
\begin{eqnarray} \label{qv3}
z_H^3 = \frac{1}{m(v)} \ , \quad q(v)^2 = Q^2 m(v)^{4/3} \ .
\end{eqnarray}
For $d=4$ we can rewrite the function $f(z,v)$ as follows
\begin{eqnarray}
f = 1 - \left(\frac{z}{z_H(v)}\right)^4 + \frac{2 Q^2}{3} \left(\frac{z}{z_H(v)}\right)^6 \ ,
\end{eqnarray}
which gives
\begin{eqnarray} \label{qv4}
z_H^4 = \frac{1}{m(v)} \ , \quad q(v)^2 = Q^2 m(v)^{3/2} \ .
\end{eqnarray}
From now on, we also set $M=1$. To appeal to the visual cortex, let us plot how the interpolating function $m(v)$ and $q(v)$ behave in various dimensions in fig.~\ref{d3mqv} and fig.~\ref{d4mqv}.
It is clear that in the thin shell limit, characterized by $v_0 \to 0$, the variation of the mass and the charge functions are sharply changing around $v=0$ and approximates a step function behavior. In the remainder of this paper, we will analyze the observables in these dynamic (time-dependent) backgrounds in the thin-shell approximation.

Before proceeding to the analysis of the probes of thermalization, a notational comment is in order: In the previous sections, we have denoted the regularized length, area or the volume functional with a subscript ``thermal". From here on, we will keep the subscript ``thermal" in direct comparison with the equilibrium cases but the quantities are not be interpreted as thermal ones, rather they have explicit time-evolution.

\subsection{Spacelike geodesics}

We start by analyzing the two-point function. We consider geodesics with a boundary separation along $x_1=x$-direction (all other spatial directions at both the end-points are the same). The profile is described by two functions $z(x)$ and $v(x)$. The length element is
\begin{eqnarray}
\cL = \int_{-\ell/2}^{\ell/2} \frac{dx}{z} \left( 1 - f v'^2 - 2 v' z' \right)^{1/2} \ .
\end{eqnarray}
The conservation equation is given by
\begin{eqnarray}
1- f v'^2 - 2 v' z' = \left(\frac{z_*}{z}\right)^{2} \ .
\end{eqnarray}
Using this conservation equation, the two equations of motion corresponding to the variation of $z(x)$ and $v(x)$ are obtained to be
\begin{eqnarray}
&& z v'' + 2 z' v' -1 + v'^2 \left(f - \frac{1}{2}z \frac{\partial f}{\partial z}\right) = 0 \ , \\
&& z'' + f v'' + \frac{\partial f}{\partial z} z' v' + \frac{1}{2} \frac{\partial f}{\partial v} v'^2 = 0 \ .
\end{eqnarray}
We solve the above two differential equations subject to the following boundary conditions
\begin{eqnarray} \label{bc}
z(\epsilon) = z_* \ , \quad z'(\epsilon) = 0 + {\rm corrections} \ , \quad v(\epsilon) = v_* \ , \quad v'(\epsilon) = 0 + {\rm corrections}  \ ,
\end{eqnarray}
where $\epsilon$ is a small number. In practice, we fix the slopes $z'$ and $v'$ at $x = \epsilon$ from the equations of motion themselves, which gives the ``corrections" terms above. So far $z_*$ and $v_*$ are two free parameters that generate the numerical solutions for $z(x)$ and $v(x)$. The boundary data can be obtained from this numerical solution
\begin{eqnarray}
z(\pm \ell/2) = z_0 \ , \quad v (\pm \ell/2) = t \ .
\end{eqnarray}
Here $z_0$ is the radial IR cut-off and $t$ is the boundary time. To generate the corresponding thermalization curves, we do the following: for a fixed value of $z_*$ we keep varying $v_*$ till the read-off value of $z_0$ is sufficiently low. This generates one desired profile for $z(x)$ and $v(x)$, which we can use to compute the length functional. Now we vary $z_*$ and repeat the process again.

\begin{figure}[!ht]
\begin{center}
\subfigure[] {\includegraphics[angle=0,
width=0.45\textwidth]{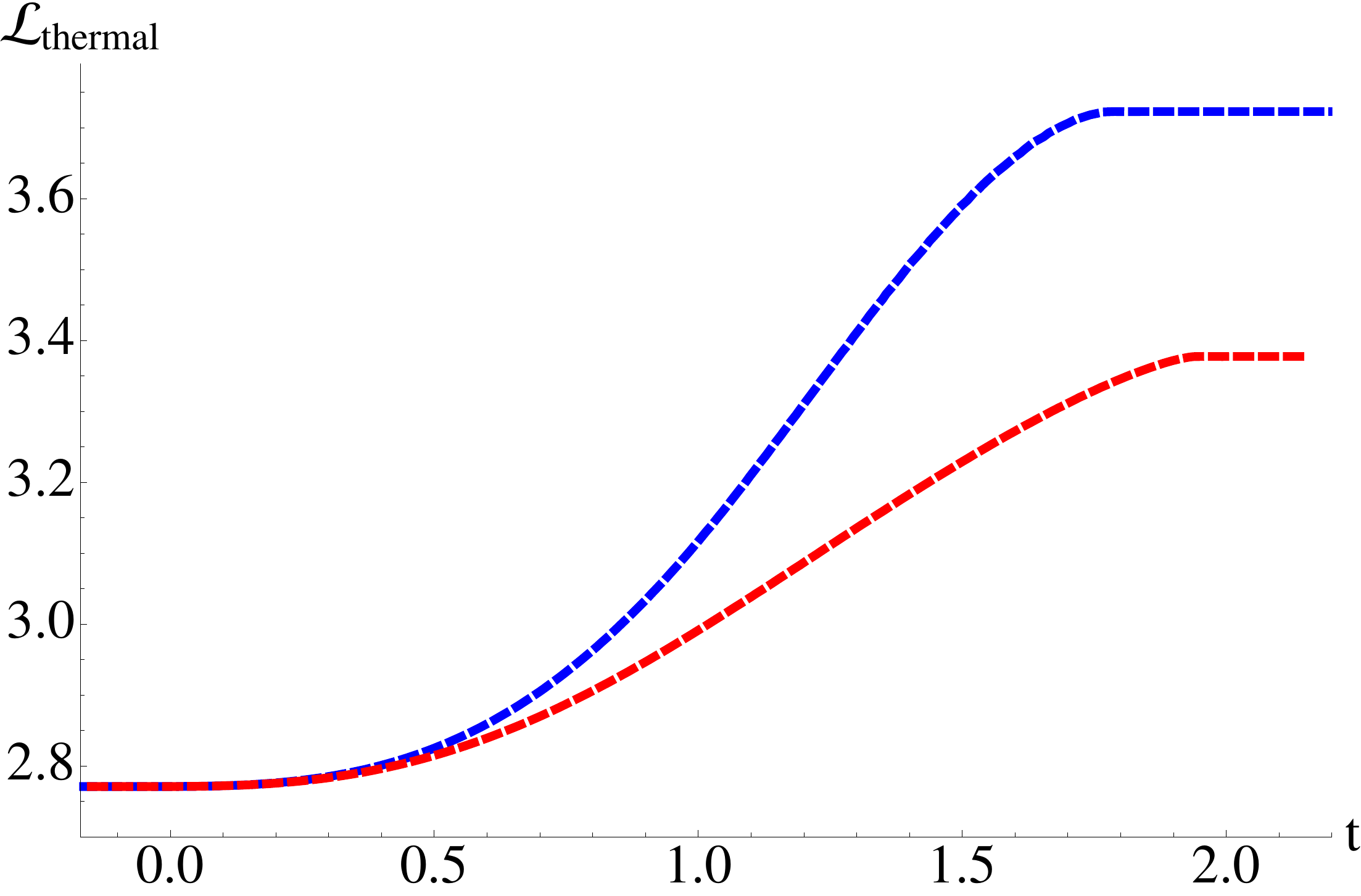} } 
\subfigure[] {\includegraphics[angle=0,
width=0.45\textwidth]{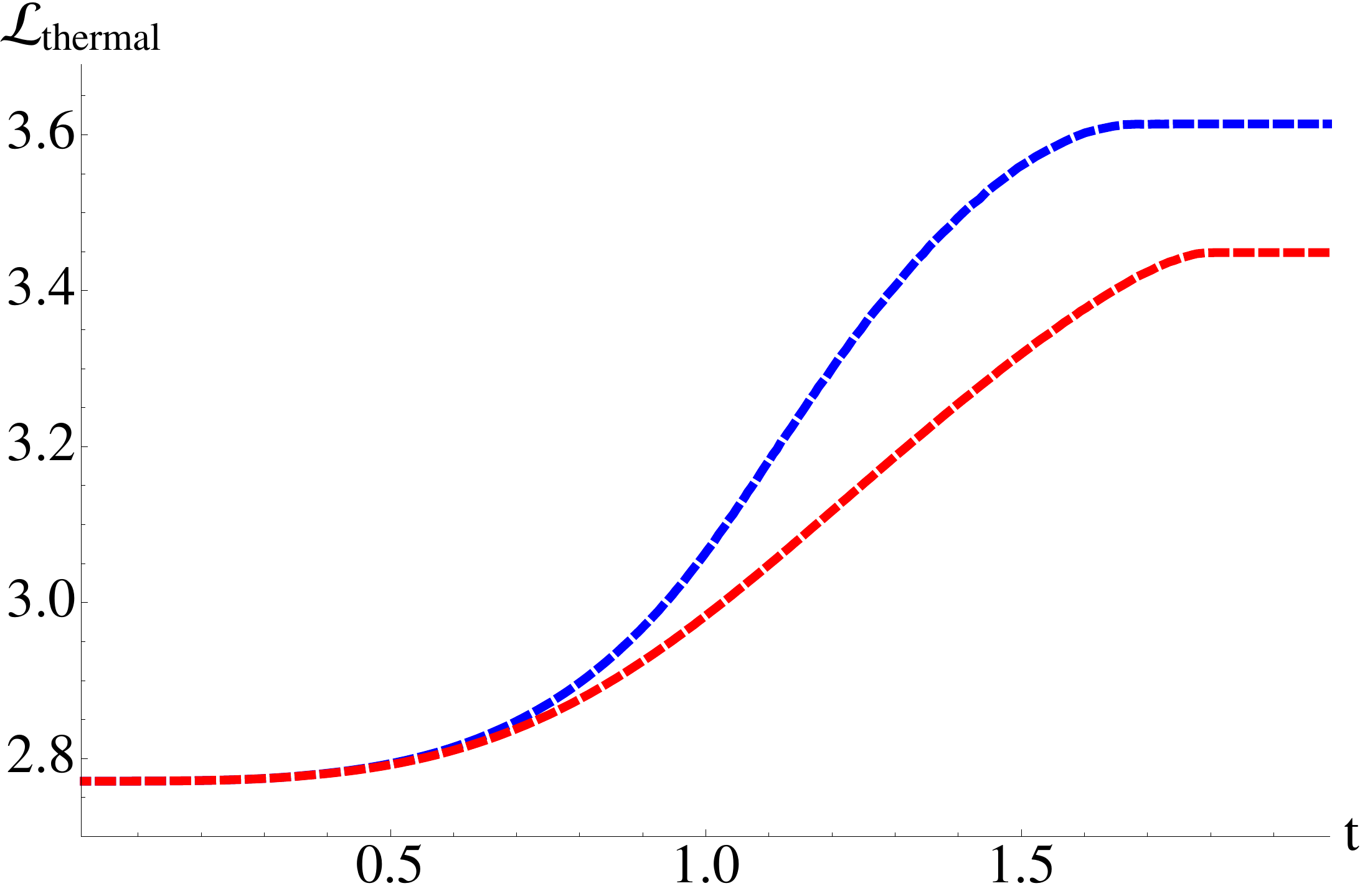} } 
\caption{\small Left panel: The case $d=3$. The blue and the red curve corresponds to $\chi_{(3)} \approx 0.003, \, 4.45$ respectively. Right panel: The case $d=4$. The blue and the red curve corresponds to $\chi_{(4)} \approx 0.002, \, 0.4$ respectively. $\cL_{\rm thermal}$ is measured in units of the AdS-radius and the boundary times measured in units of the black hole mass $M$. All curves here correspond to different values of $(T\ell)$, which we do not specify here.}
\label{rnv2pt}
\end{center}
\end{figure}

Using the data obtained as explained above we show how the thermalization occurs in fig.~\ref{rnv2pt}. It should be emphasized at this point that we are not being careful about the units in which we measure the boundary time in this figure. In what we have shown in fig.~\ref{rnv2pt}, the boundary time is measured in units of the black hole mass, $M$, which strictly speaking does not have a simple interpretation in terms of any physical quantity of the boundary theory. Moreover, the different curves corresponding to different values of $\chi_{(d)}$ are not obtained for the same equilibrium temperature. Thus fig.~\ref{rnv2pt} and all such subsequent ones in the later sections should be viewed as schematic representation of the physics that is going on.

We need to define a ``thermalization time" in order to investigate how it depends on the length scale for various values of $\chi_{(d)}$. Following \cite{Balasubramanian:2010ce}, we define two different time scales denoted by $\tau_{1/2}$ and $\tau_{\rm crit}$ respectively:

\noindent (i) $\tau_{1/2}$ is defined as the time required for the curves to reach half of their equilibrium value.

\noindent (ii) $\tau_{\rm crit}$ is defined as the critical time at which the geodesic grazes the middle of the shell at $v=0$. This is simply given by the following formula: 
\begin{eqnarray}
\tau_{\rm crit} = \int_{z_0}^{z_*} \frac{dz}{f(z)}  \ ,
\end{eqnarray}
where, as before, $z_0$ is the IR radial cut-off and $z_*$ determines the value of the boundary separation. Once again, we need to specify the scale in which we measure these thermalization times. From the point of view of the boundary theory the only relevant dimensionful quantity is the equilibrium temperature. Thus we will always study the behavior of the dimensionless quantities $(T\tau_{1/2})$ or $(T\tau_{\rm crit})$.

\begin{figure}[!ht]
\begin{center}
\subfigure[] {\includegraphics[angle=0,
width=0.48\textwidth]{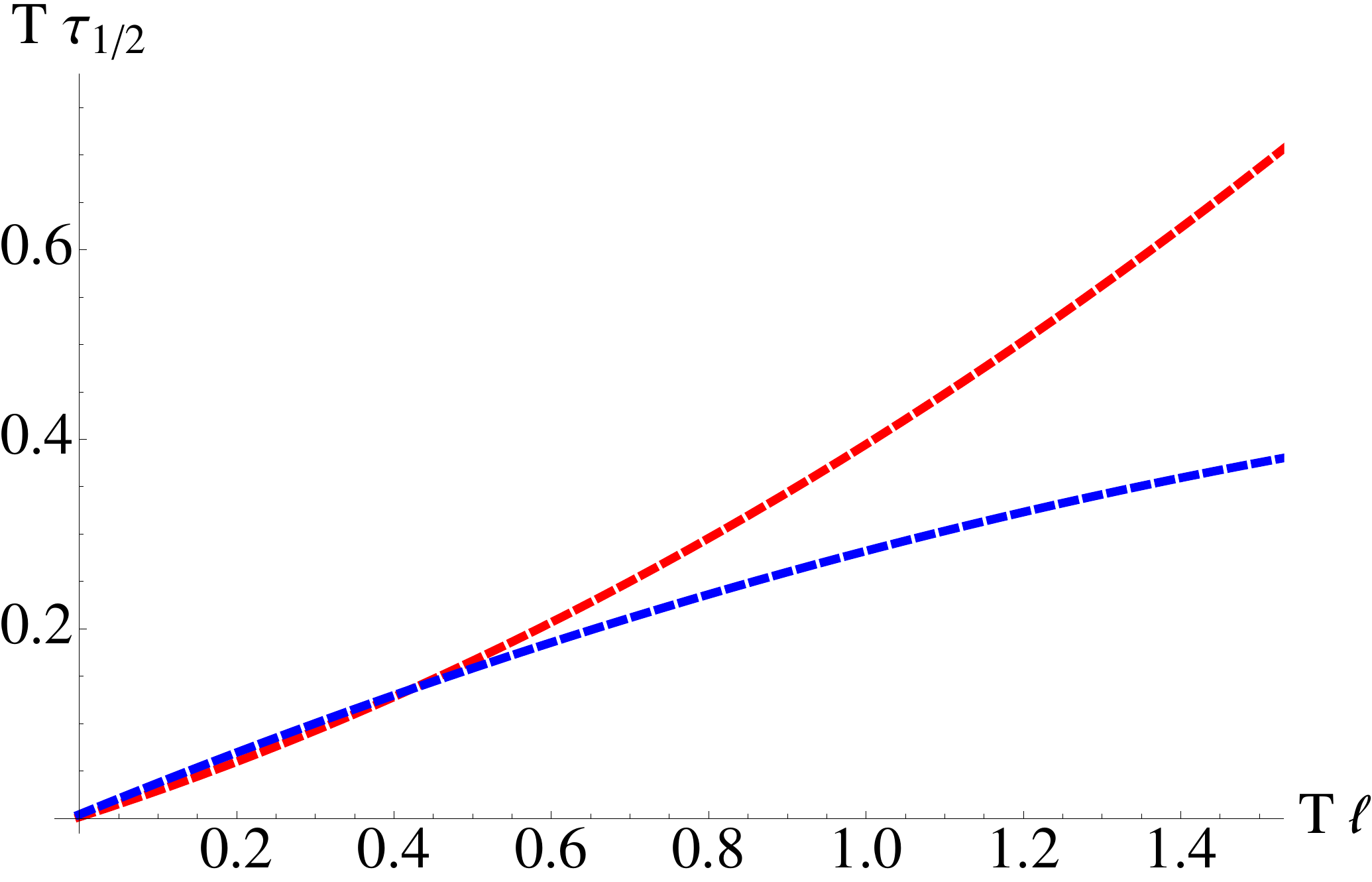}} 
\subfigure[] {\includegraphics[angle=0,
width=0.48\textwidth]{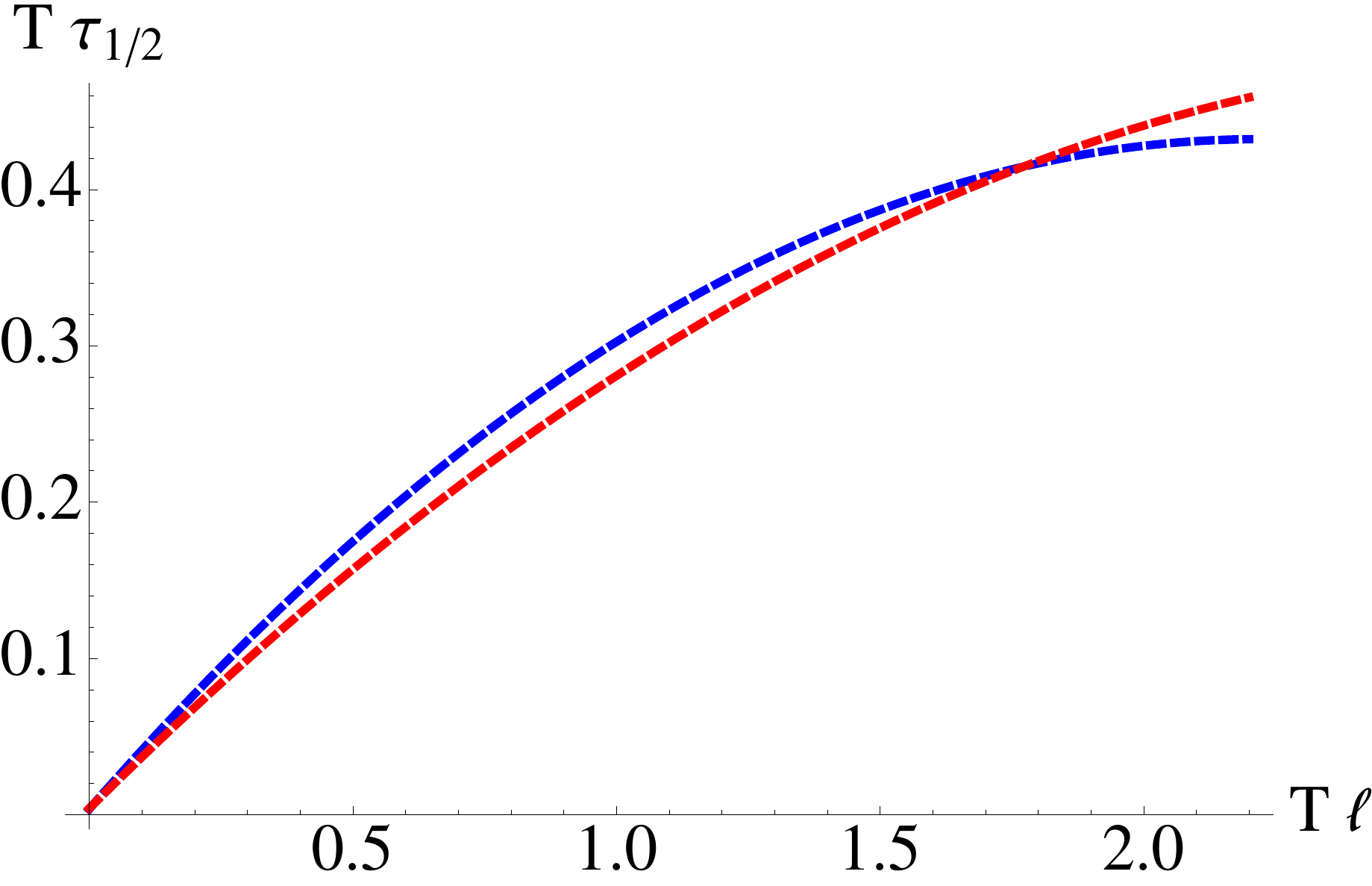}} 
\caption{\small Left panel: The case $d=3$. The blue and the red curve corresponds to $\chi_{(3)} \approx 0.003, \, 4.45$ respectively. Right panel: The case $d=4$. The blue and the red curve corresponds to $\chi_{(4)} \approx 0.001, \, 0.3$ respectively.}
\label{thalf_2pt}
\end{center}
\end{figure}

Now we can investigate how these thermalization times depend on the length-scale $\ell$ for various fixed values of $\chi_{(d)}$. The results for $\tau_{1/2}$ are shown in fig.~\ref{thalf_2pt}.
For small $\ell$, $\tau_{1/2}$ grows linearly with the property that $\tau_{1/2} < \ell/2$. The deviation from linearity for large values of $\ell$ is clear. For increasing $\chi_{(d)}$ this deviation occurs either in the opposite direction or occurs more slowly and for large value of $\ell$ increasing $\chi_{(d)}$ increases the value of $\tau_{1/2}$. For small values of $\ell$, $\tau_{1/2}$ can decrease with increasing $\chi_{(d)}$. Thus we already observe possibly interesting and different physics dominating two different regimes: $(T\ell) \ll 1$ and $(T\ell)\gg 1$ for increasing values of $\chi_{(d)}$. It should be noted however that for very small values of $\ell$, $\tau_{1/2}$ is not very sensitive to the value of $\chi_{(d)}$.

\begin{figure}[!ht]
\begin{center}
\subfigure[] {\includegraphics[angle=0,
width=0.48\textwidth]{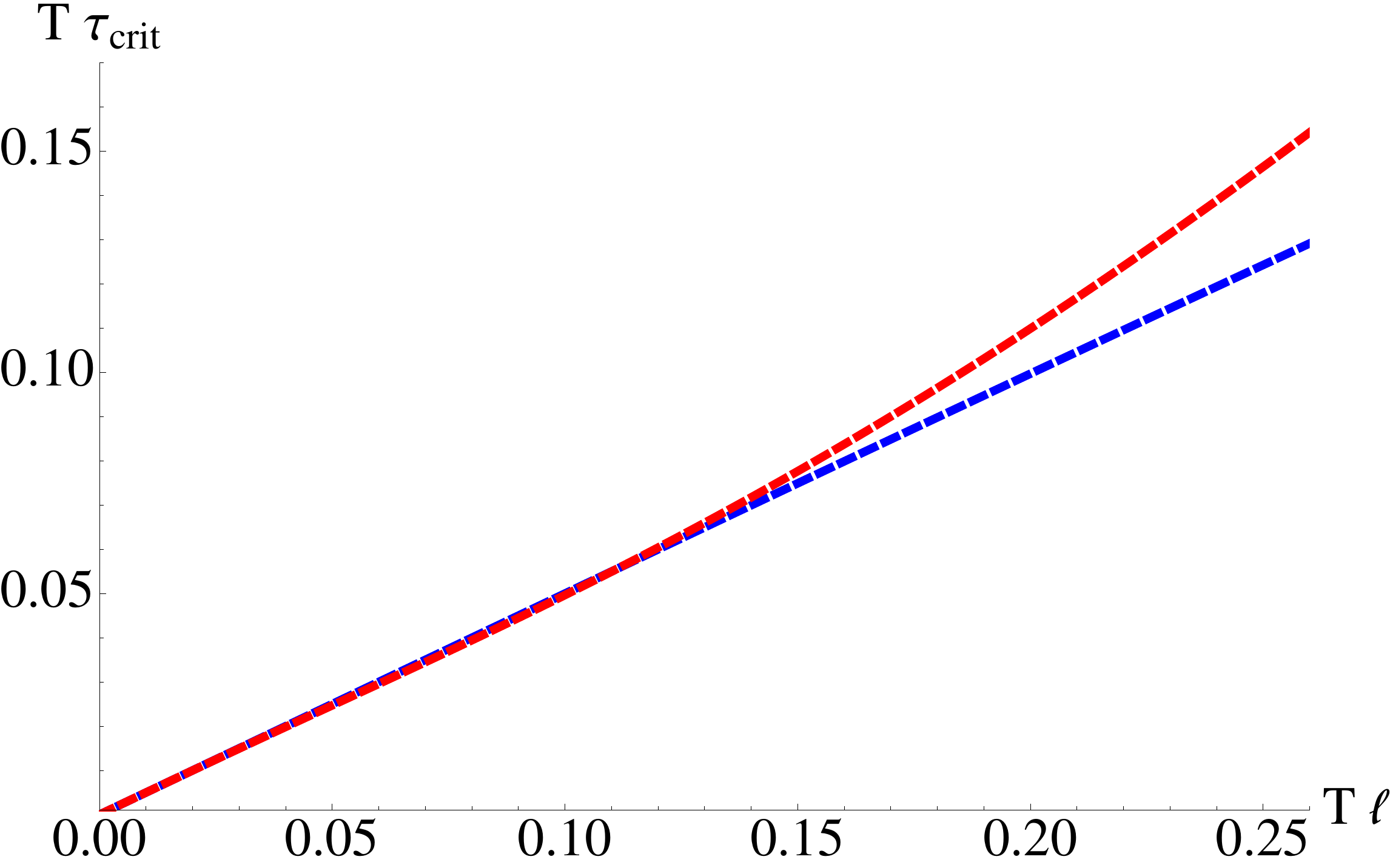}} 
\subfigure[] {\includegraphics[angle=0,
width=0.48\textwidth]{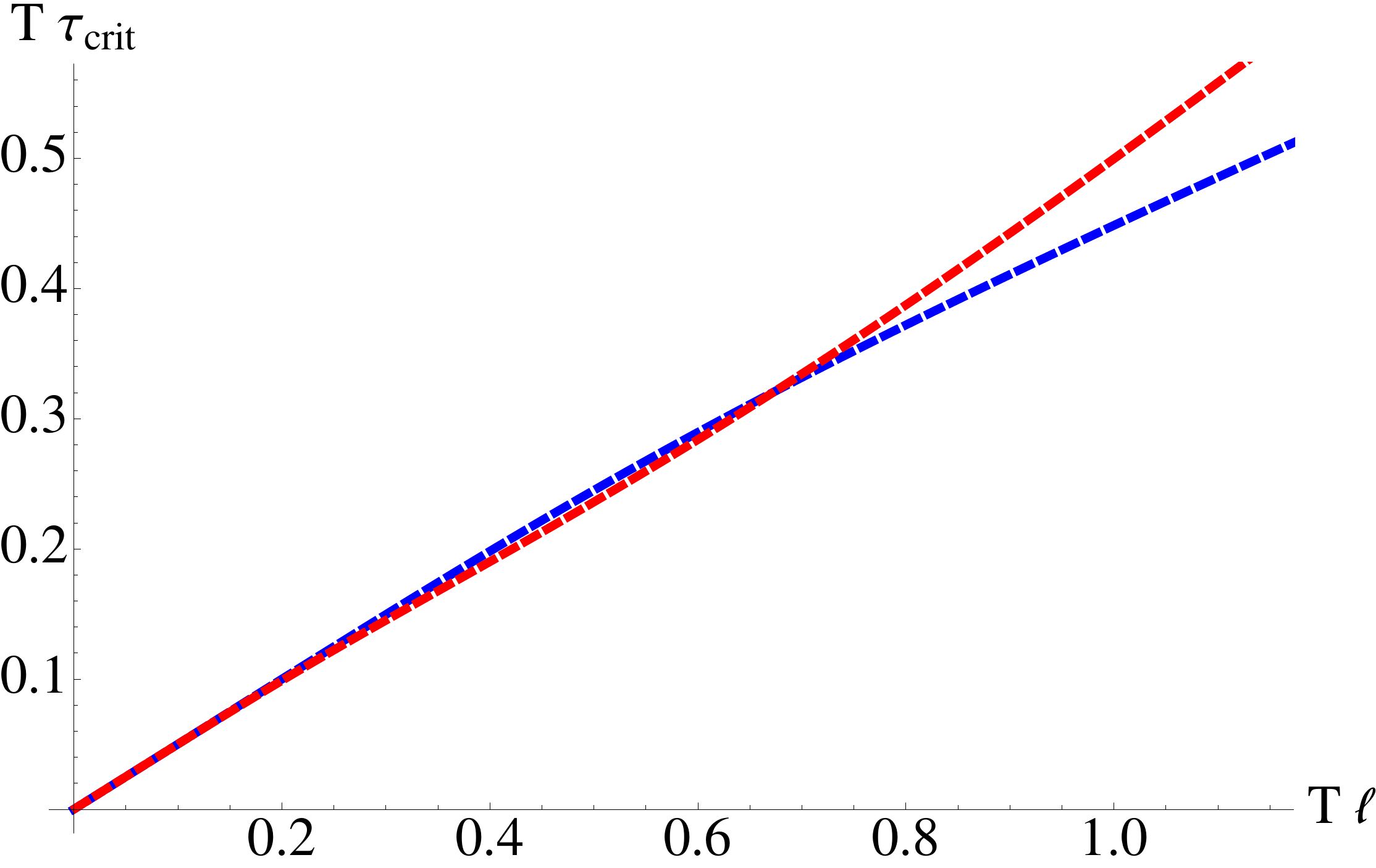}}
\caption{\small Left panel (a): The case $d=3$. The blue and the red curve corresponds to $\chi_{(3)} \approx 0.003, \, 4.45$ respectively. Right panel (b): The case $d=4$. The blue and the red curve corresponds to $\chi_{(4)} \approx 0.002, \, 1.1$ respectively.}
\label{tcrit_2pt}
\end{center}
\end{figure}

On the other hand, the dependence of $\tau_{\rm crit}$ with the boundary separation length is shown in fig.~\ref{tcrit_2pt}. The qualitative behavior of this thermalization time is much like the one observed for $\tau_{1/2}$: For small boundary separation, the thermalization time is not very sensitive to the parameter $\chi_{(d)}$ --- although there is a range within which increasing $\chi_{(d)}$ actually decreases $\tau_{\rm crit}$. However, for large values of $\ell$, $\tau_{\rm crit}$ clearly increases with increasing $\chi_{(d)}$. The deviation from linearity for large values of $\ell$ is also quite interesting. For small values of $\chi_{(d)}$ this deviation from linearity results in $\tau_{\rm crit} < \ell/2$; whereas for larger values of $\chi_{(d)}$ this deviation results in $\tau_{\rm crit} > \ell/2$, which is seen from fig.~\ref{tcrit_2pt} as the blue and the red curves bending away from each other as $\ell$ increases. This bending away effect is more pronounced for $d=3$ as compared to $d=4$.

We can also investigate the dependence of $\tau_{\rm crit}$ with $\chi_{(d)}$. To this end, let us define 
\begin{eqnarray}
\tau_{\rm crit}^0 = \lim_{\chi_{(d)} \to 0} \tau_{\rm crit} \left(\chi_{(d)}\right) 
\end{eqnarray}
and normalize the measured thermalization time in units of $\tau_{\rm crit}^0$. 
\begin{figure}[!ht]
\begin{center}
\subfigure[] {\includegraphics[angle=0,
width=0.45\textwidth]{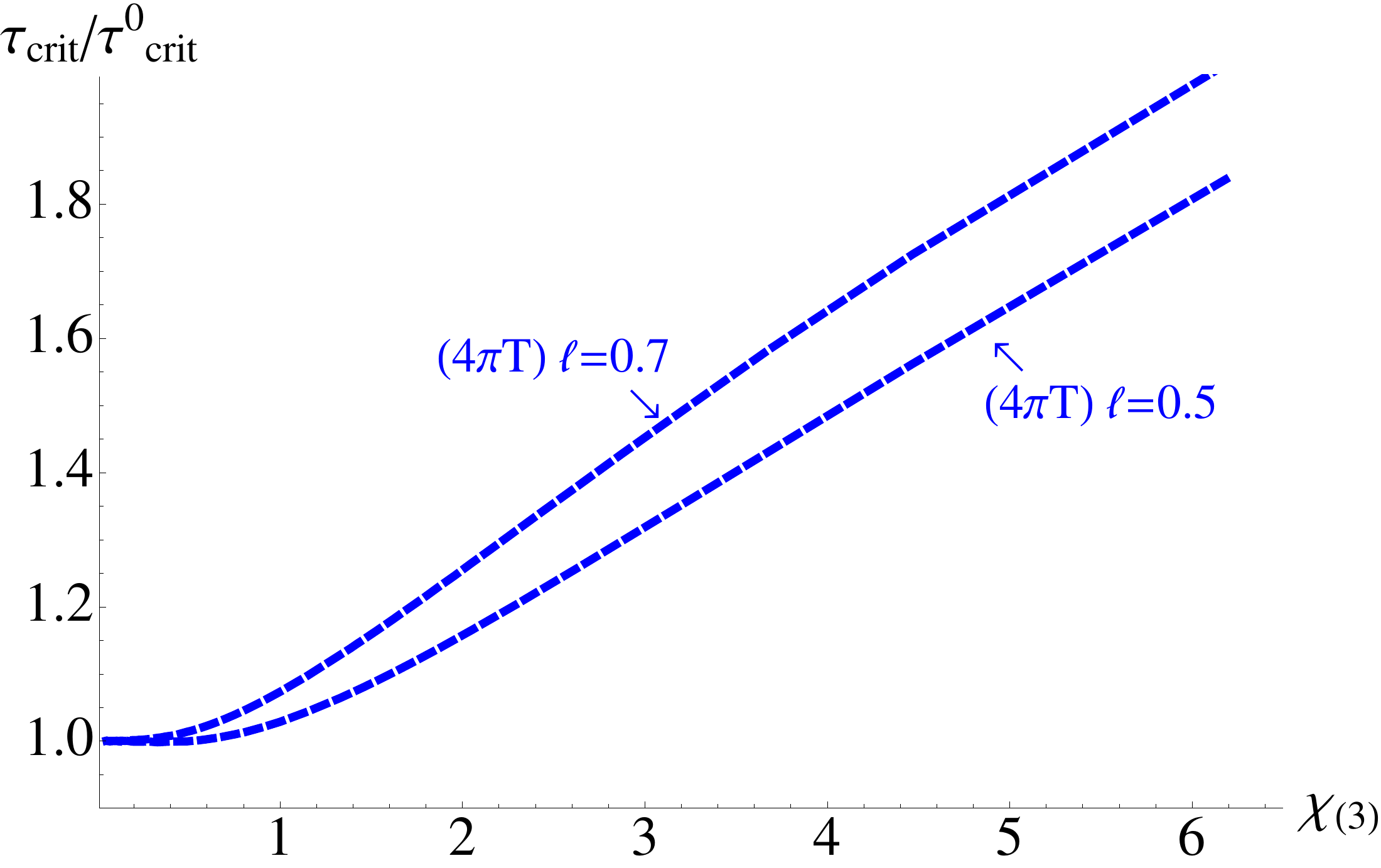} } 
\subfigure[] {\includegraphics[angle=0,
width=0.45\textwidth]{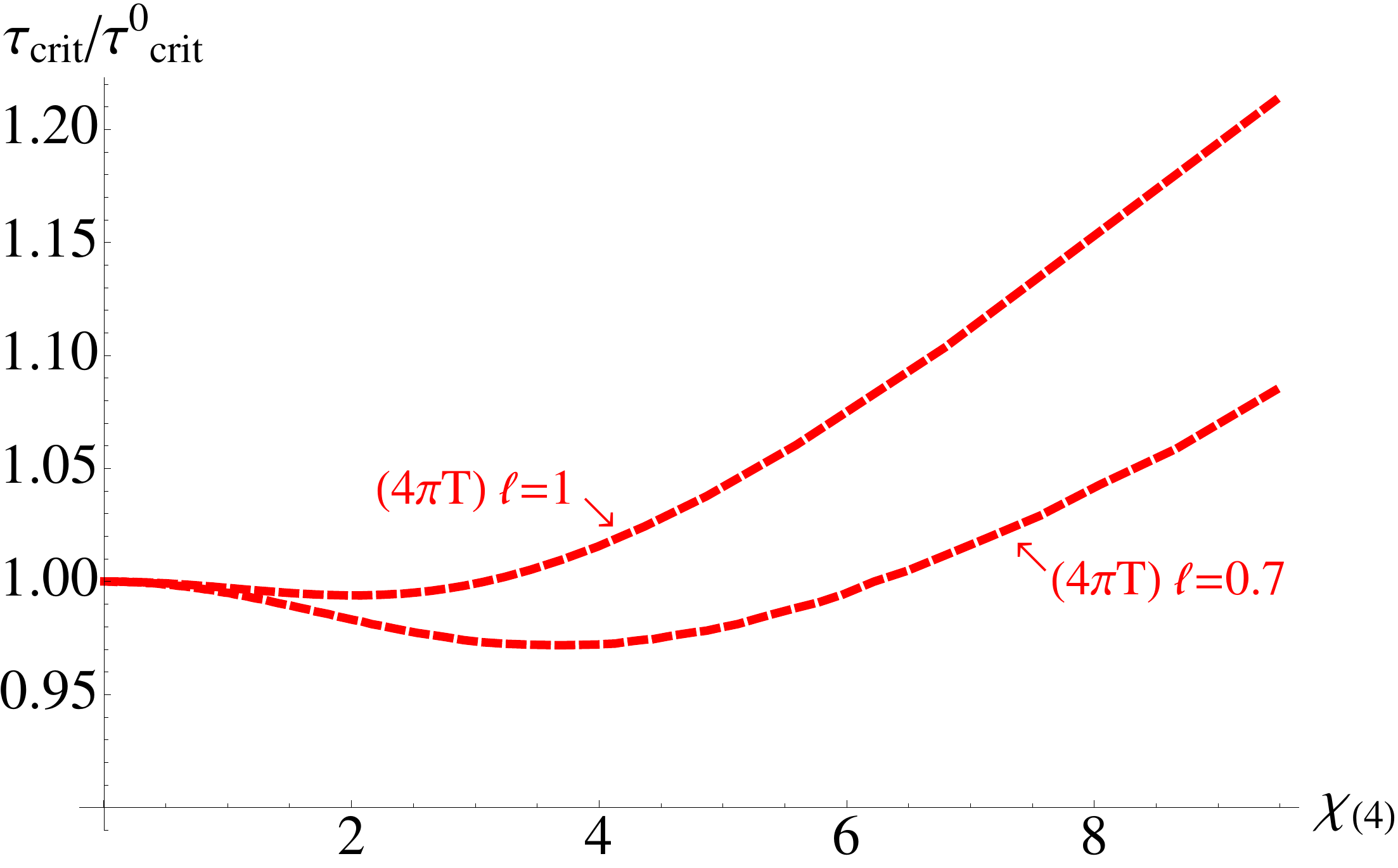} }
\caption{\small Left panel (a): The case of $d=3$. Right panel (b): The case of $d=4$. Here $\tau^0_{\rm crit}$ denotes the value of $\tau_{\rm crit}$ evaluated at $\chi_{(d)}=0$.}
\label{tcrit_chi_2pt}
\end{center}
\end{figure}
In fig.~\ref{tcrit_chi_2pt} we have shown the results. Once again the results display an interesting interplay of physics for small and large values of $\chi_{(d)}$. This effect is relatively milder in $d=3$ compared to $d=4$. From fig.~\ref{tcrit_chi_2pt}(b) we observe that for fixed values of $(T\ell)$, different physics dominates two different regimes: small $\chi_{(4)}$ and large $\chi_{(4)}$. Initially the curve decreases before turning back and increasing monotonically. This monotonically incasing behavior seems to be a linear one, with the slope depending on the fixed value of $(T\ell)$ and also the dimension we are in. This growth seems to be unbounded, implying that for infinitely strong chemical potential, it takes infinitely longer to reach thermalization. On the other hand, $\tau_{\rm crit}$ --- measured in units of $\tau_{\rm crit}^0$ --- has a minima and both the location and the minimum value depends on the fixed value of $(T\ell)$. From what we have shown in fig.~\ref{tcrit_chi_2pt}(b), we have about $4\%$-reduction in thermalization for $(T\ell) = 0.7/(4\pi)$. We will observe that all these features are very generic for all non-local probes that we consider in the subsequent sections.


\subsection{Wilson loops}

Once again we will consider two geometric shapes: the rectangular and the circular one and we appeal to fig.~\ref{rec_cir_shape} for a schematic representation.

\subsubsection{Rectangular strip}

\begin{figure}[!ht]
\begin{center}
\subfigure[] {\includegraphics[angle=0,
width=0.45\textwidth]{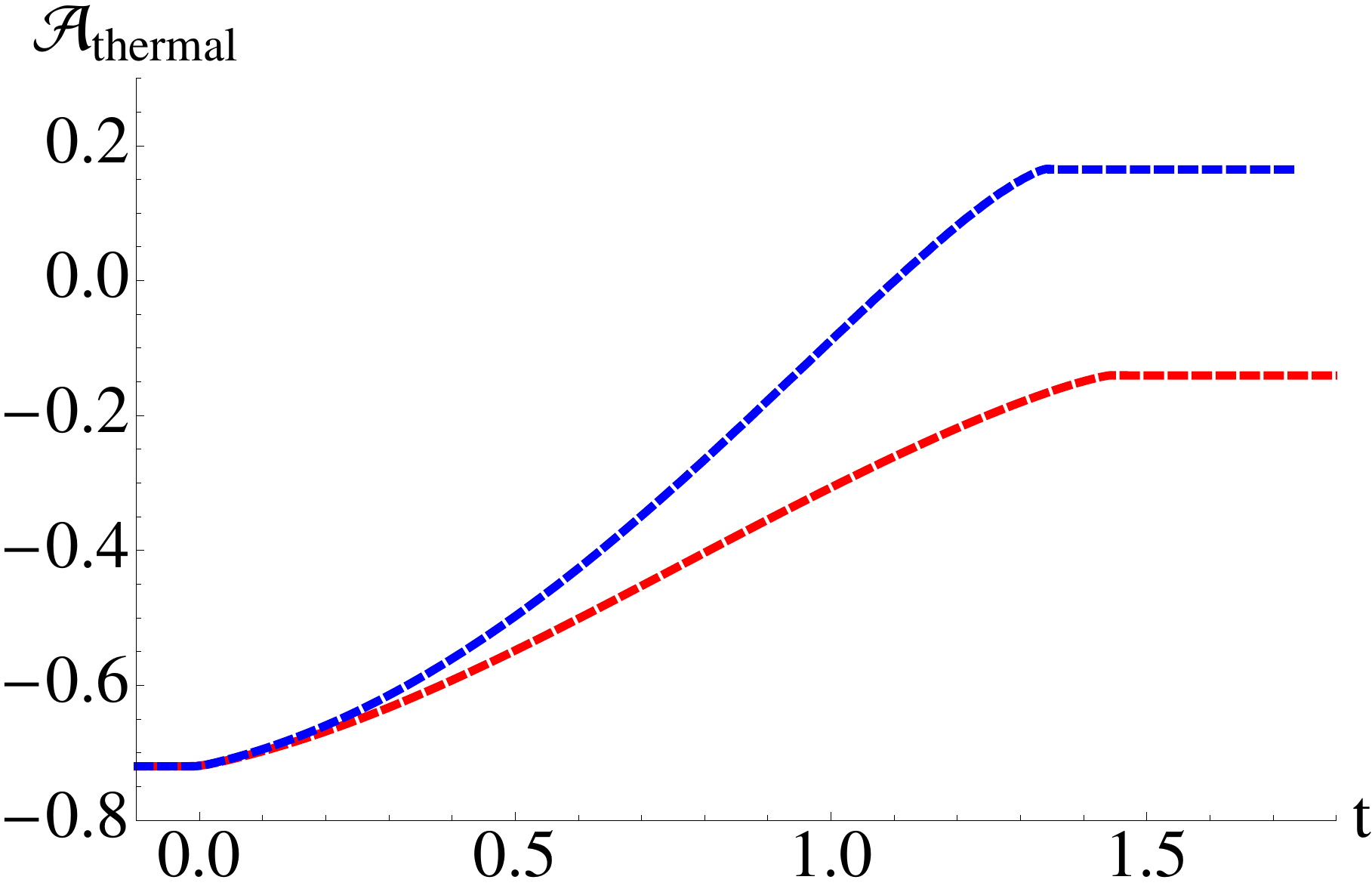} } 
\subfigure[] {\includegraphics[angle=0,
width=0.45\textwidth]{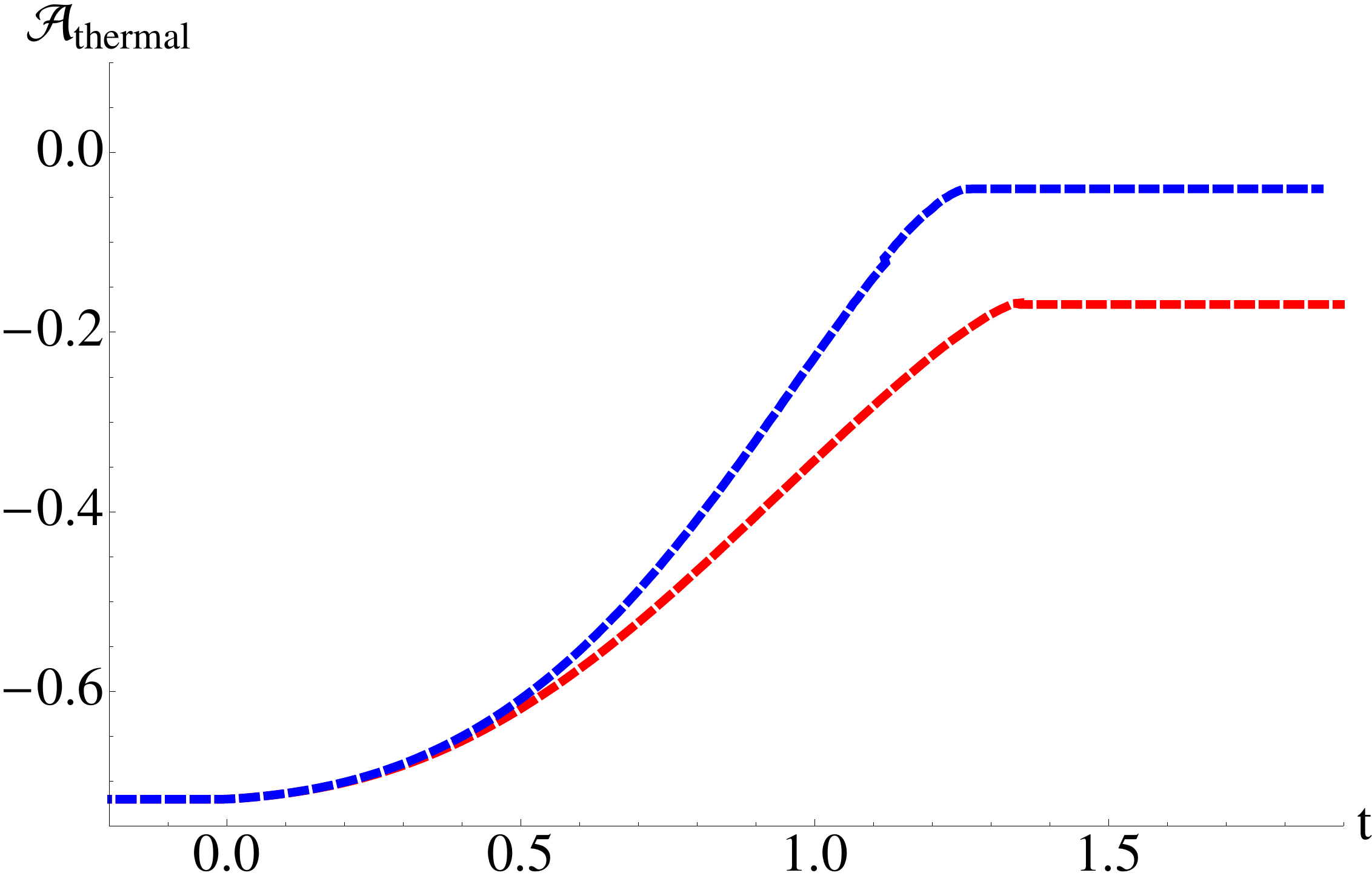} } 
\caption{\small Left panel: The case $d=3$. The blue and curves correspond to $\chi_{(3)} \approx 0.003, \, 4.47$ respectively. Right panel: The case $d=4$. The blue and red curves correspond to $\chi_{(3)} \approx 0.002, \, 0.4$ respectively. We measure $\cA_{\rm thermal}$ in units of the AdS-radius and the boundary time $t$ in units of the black hole mass.}
\label{rnvwlrec}
\end{center}
\end{figure}

\begin{figure}[!ht]
\begin{center}
\subfigure[] {\includegraphics[angle=0,
width=0.48\textwidth]{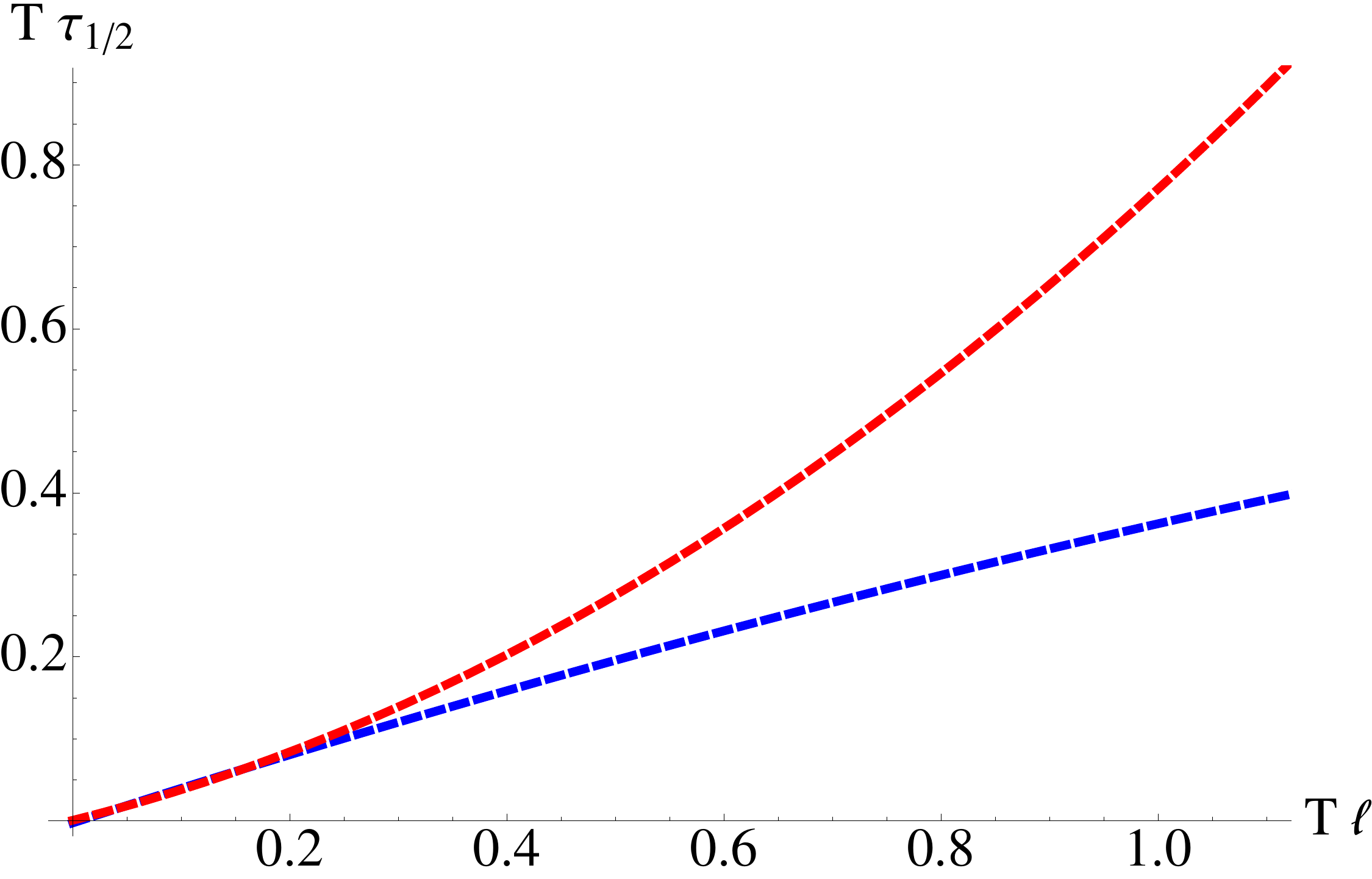} } 
\subfigure[] {\includegraphics[angle=0,
width=0.48\textwidth]{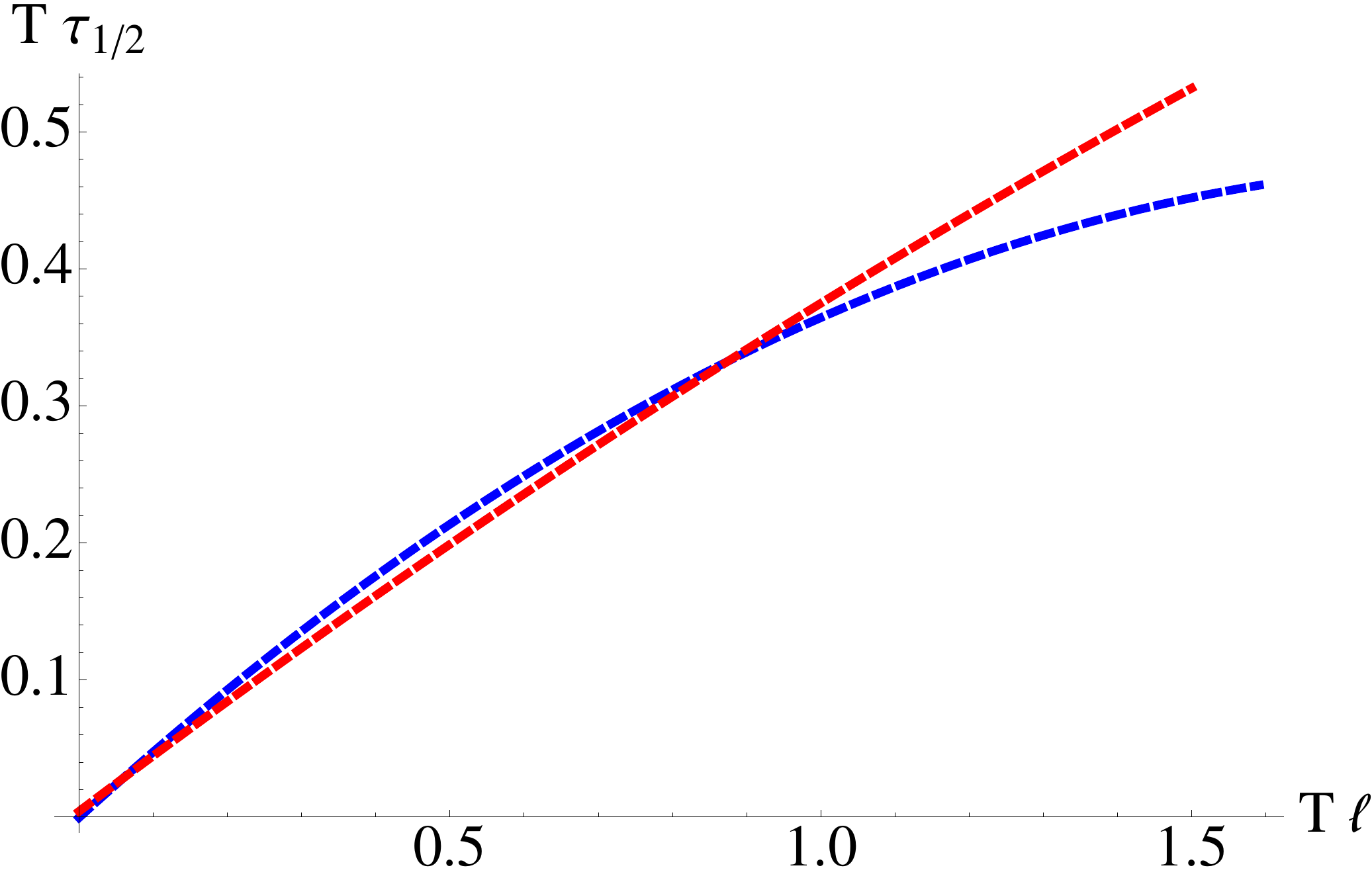} } 
\caption{\small Left panel: The case $d=3$. The blue and red curves correspond to $\chi_{(3)} \approx 0.003, \, 4.47$ respectively. Right panel: The case $d=4$. The blue and the red curve corresponds to $\chi_{(4)} \approx 0.002, \, 0.4$ respectively.}
\label{thalfWLrec}
\end{center}
\end{figure}

As before the area functional for the rectangular strip is given by
\begin{eqnarray}
\cA = \frac{R}{2\pi} \int_{-\ell/2}^{\ell/2} \frac{dx}{z^2} \left( 1 - f v'^2 - 2 v' z' \right)^{1/2} \ .
\end{eqnarray}
The conservation equation gives
\begin{eqnarray}
1 - f v'^2 - 2 v' z' = \left(\frac{z_*}{z}\right)^4 \ .
\end{eqnarray}
Using the conservation equation and extremizing the area functional we get the following two equations of motion
\begin{eqnarray}
&& z'' + v'' f + z' v' \frac{\partial f}{\partial z} + \frac{1}{2} v'^2 \frac{\partial f}{\partial z} = 0 \ , \\
&& z v'' + 4 z' v' -2 + v'^2 \left( 2 f - \frac{1}{2} z \frac{\partial f}{\partial z }\right) = 0 \ .
\end{eqnarray}
Note that compared to the computation of the geodesics, only the second equation changes. We again use the same boundary conditions as outlined in (\ref{bc}). Some of the resulting plots are shown in fig.~\ref{rnvwlrec}.
The behavior of the thermalization time with the length of the Wilson loop operator has been displayed in fig.~\ref{thalfWLrec}. The general behavior of $\tau_{1/2}$ is sub-linear. The deviation from linearity of $\tau_{1/2}$ decreases with increasing $\chi_{(d)}$ and for large enough $\ell$, the thermalization time increases with increasing value of $\chi_{(d)}$. From fig.~\ref{thalfWLrec}(b), we can clearly identify a regime of $(T\ell)$ where increasing $\chi_{(4)}$ actually decreases $\tau_{1/2}$.

\begin{figure}[!ht]
\begin{center}
\subfigure[] {\includegraphics[angle=0,
width=0.45\textwidth]{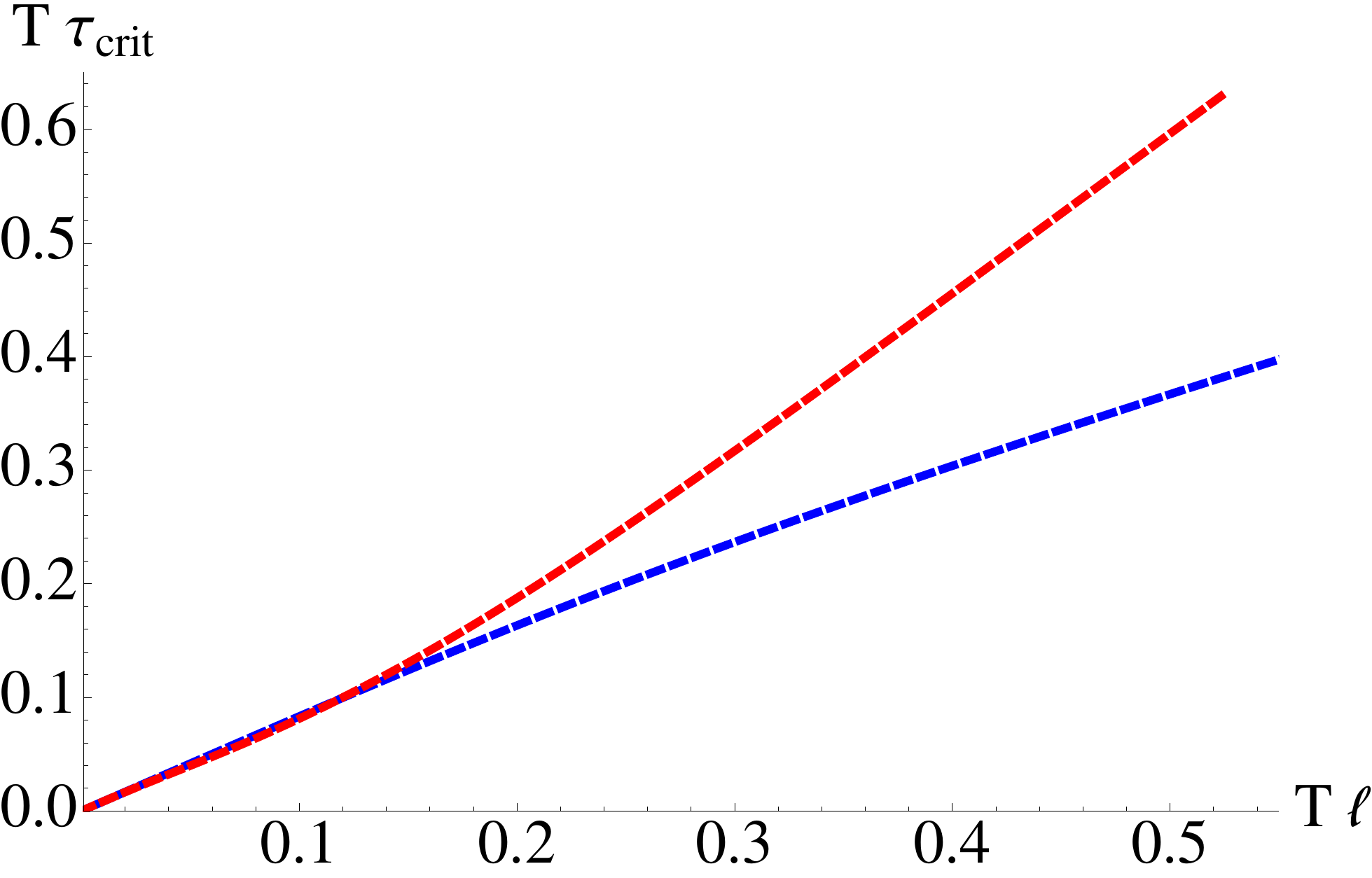} } 
\subfigure[] {\includegraphics[angle=0,
width=0.45\textwidth]{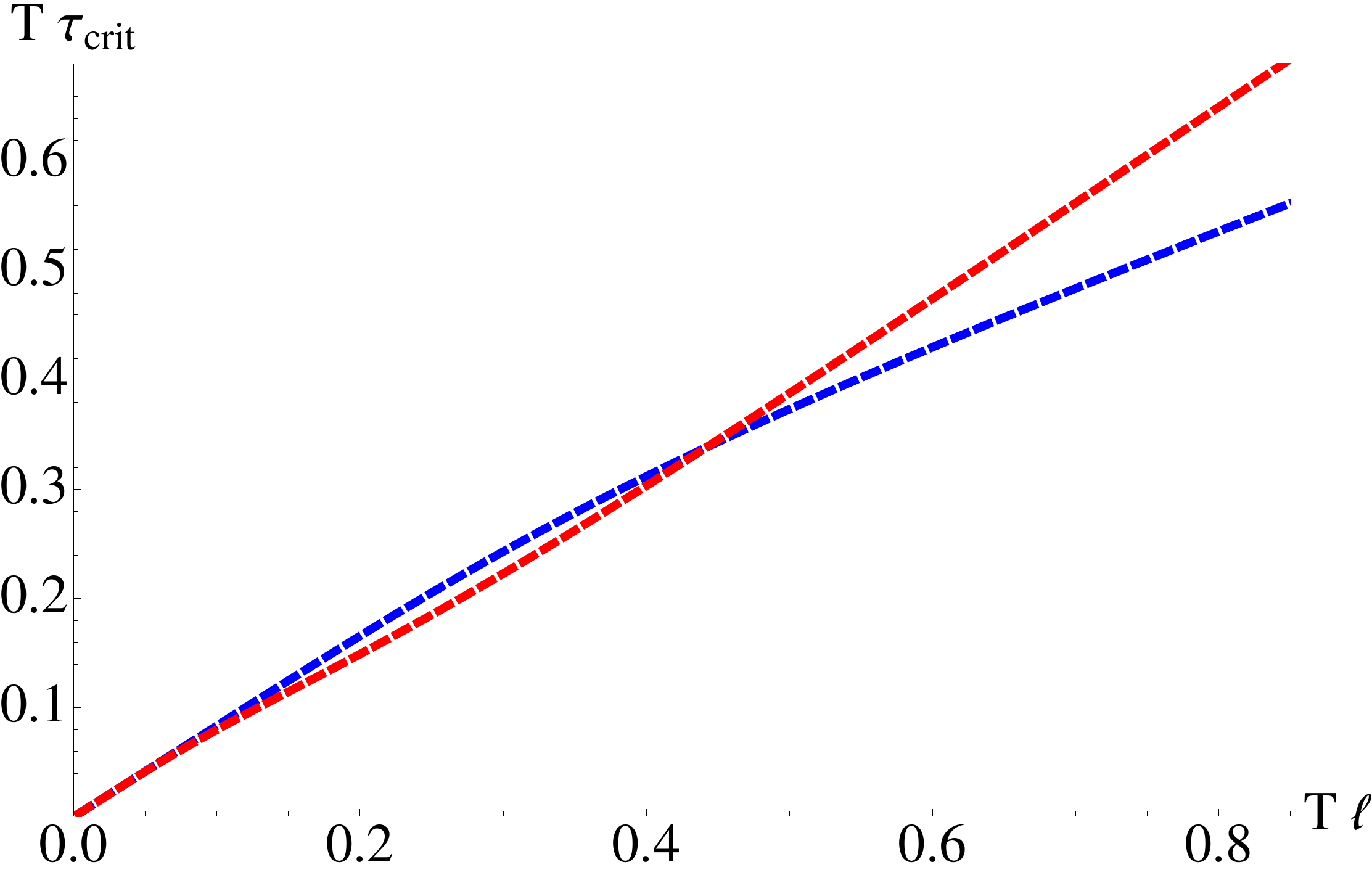} }
\caption{\small Left panel: The case $d=3$. The blue and red curves correspond to $\chi_{(3)} \approx 0.003, \, 4.47$ respectively. Right panel: The case $d=4$. The blue and the red curve corresponds to $\chi_{(4)} \approx 0.002, \, 1.1 $ respectively.}
\label{tcritWLrec}
\end{center}
\end{figure}

On the other hand, the behavior of $\tau_{\rm crit}$ with the length of the Wilson loop operator is shown in fig.~\ref{tcritWLrec}. It turns out that $\tau_{\rm crit}$ exceeds the linear relation in $\ell/2$ and for large enough $\ell$, the red and the blue curves bend away from each other. So increasing $\chi_{(d)}$ enhances the departure from the linear behavior for $\tau_{\rm crit}$. From fig.~\ref{tcritWLrec}(b) we can also identify a regime where $\tau_{\rm crit}$ decreases with increasing chemical potential and hence the thermalization happens faster. 
\begin{figure}[!ht]
\begin{center}
\subfigure[] {\includegraphics[angle=0,
width=0.45\textwidth]{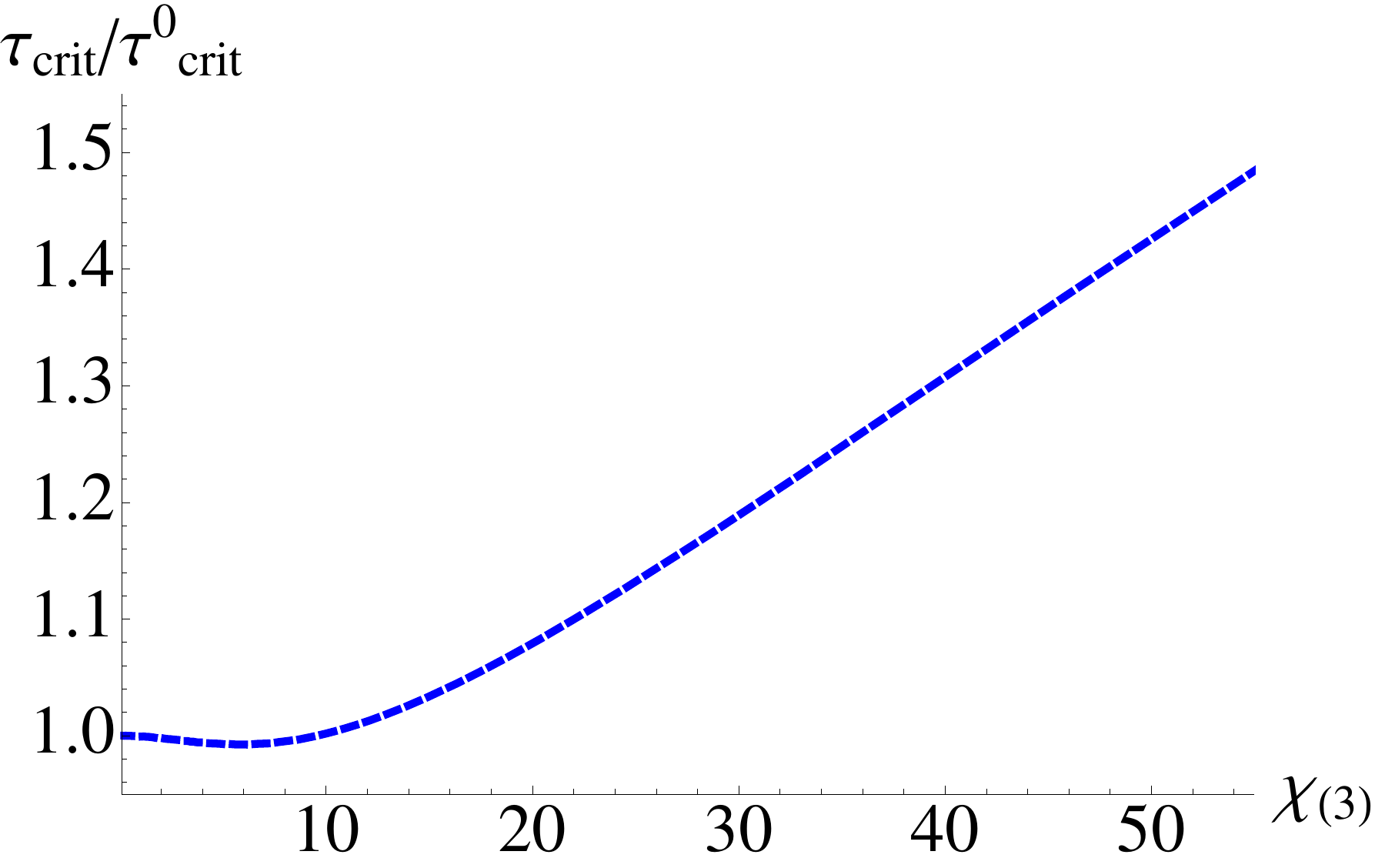} } 
\subfigure[] {\includegraphics[angle=0,
width=0.45\textwidth]{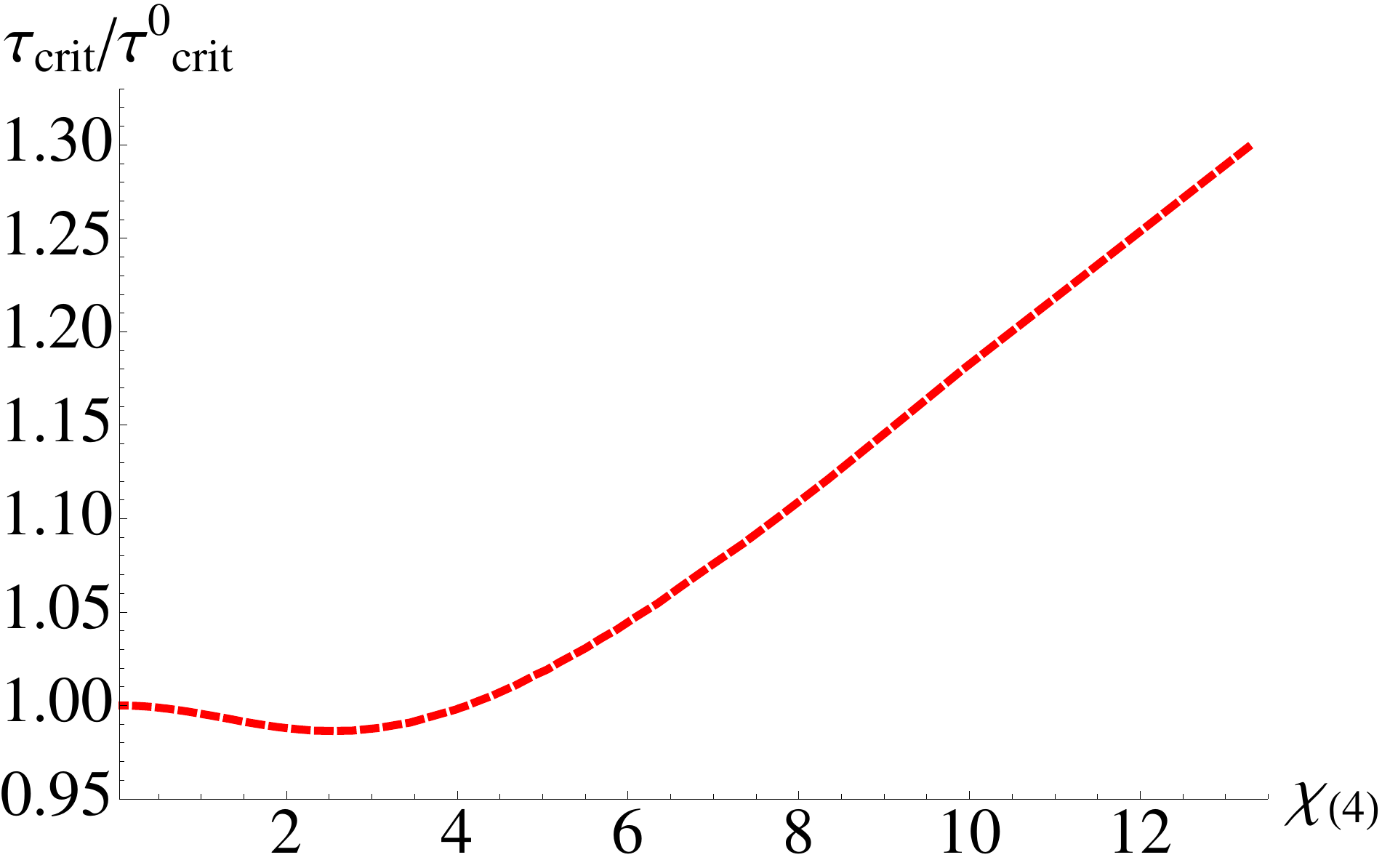} }
\caption{\small Left panel (a): The case $d=3$ for $(4\pi T) \ell = 0.4$. Right panel (b): The case $d=4$ for $(4\pi T) \ell = 0.7$.}
\label{tcrit_chi_WLrec}
\end{center}
\end{figure}

In fig.~\ref{tcrit_chi_WLrec}, we demonstrate how the thermalization time depends on $\chi_{(d)}$ for both $d=3$ and $d=4$. As in the case of 2-pt function, this response consists of two different regimes: for small values of $\chi_{(d)}$, thermalization seems faster and for larger values of $\chi_{(d)}$ thermalization becomes slower and approaches infinitely large values for infinitely strong chemical potential. The percentage reduction in thermalization time for small chemical potential is about $4\%$ for $d=4$ where this effect is more pronounced.

\subsubsection{Circular Wilson loop} \label{wlcir}

Let us now consider circular Wilson loops in the AdS-RN-Vaidya background. The corresponding minimal surface is parametrized as before and the area functional has the following form
\begin{eqnarray}
\cA (t, R) = \int_0^R d \rho \frac{\rho}{z^2} \sqrt{1- f v'^2 - 2 z' v'} \ ,
\end{eqnarray}
where $' \equiv d/d\rho$. Here the mass and the charge functions are time-dependent and are given by (\ref{mv}) and (\ref{qv3}) or (\ref{qv4}) for $d=3$ or $d=4$ respectively. As before, the equations motion resulting from the area functional are messy and therefore we do not present them explicitly. 

To impose the boundary conditions, we can use the rotational symmetry along $\phi$-direction and impose
\begin{eqnarray}
z(\epsilon) = z_* + {\rm corrections}\ , \quad v(\epsilon) = v_*  + {\rm corrections} \ ,
\end{eqnarray}
where $z_*$ and $v_*$ are the two free parameters, which we will describe how to fix. The ``corrections" in the above is obtained in the following manner: we expand the equations motion near $\rho=\epsilon$, where $\epsilon$ is a small number typically of the order of $10^{-3}$. Using these expansions we can fix the ``corrections" as we did in, {\it e.g.} (\ref{bcwlcir1}) and (\ref{bcwlcir2}).\footnote{The only difference is we now have to set the boundary conditions for two functions $z(\rho)$ and $v(\rho)$.} Note that this expansion also allows us to set $z'(\epsilon)$ and $v'(\epsilon)$.

\begin{figure}[!ht]
\begin{center}
\subfigure[] {\includegraphics[angle=0,
width=0.45\textwidth]{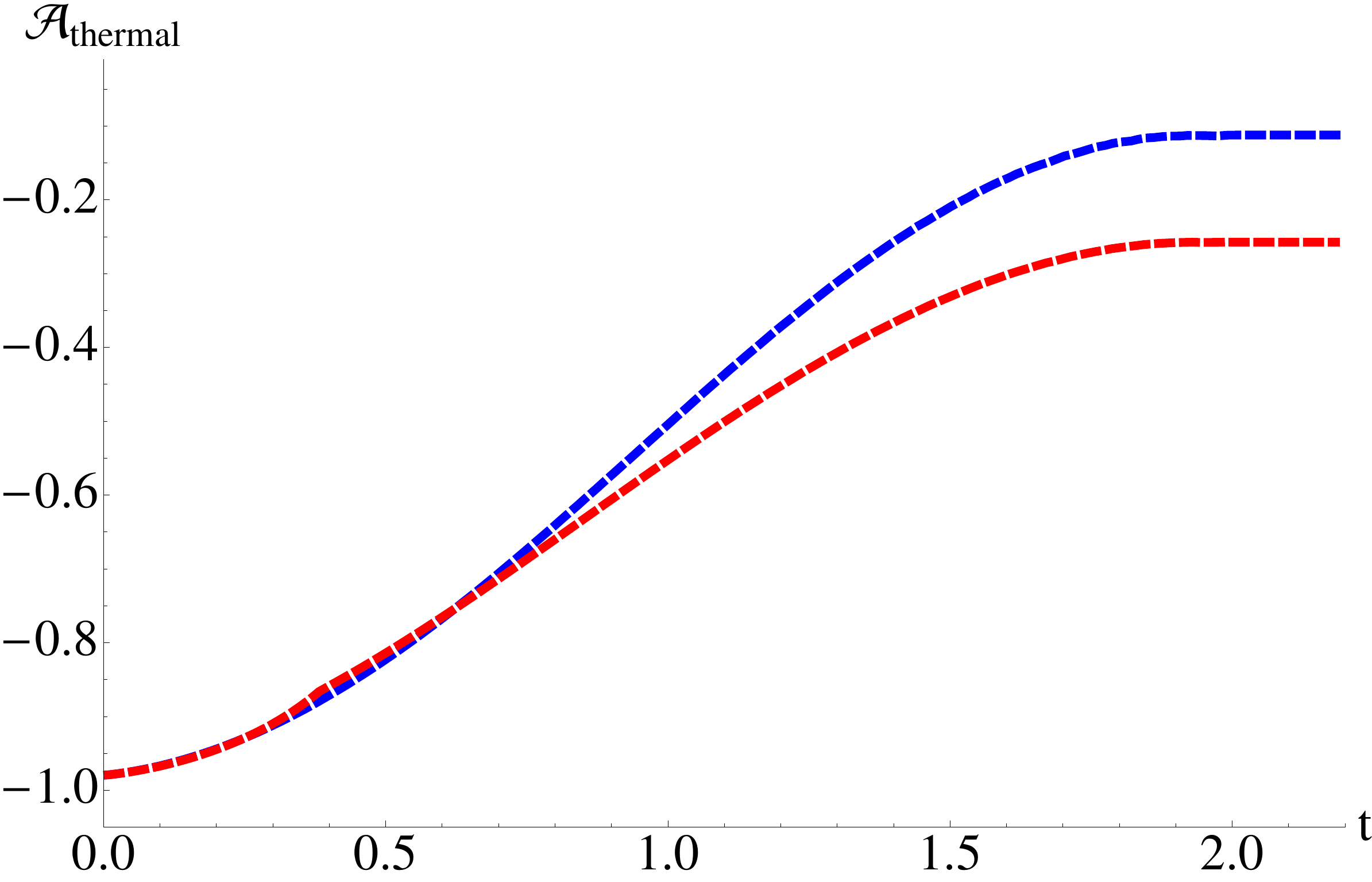} } 
\subfigure[] {\includegraphics[angle=0,
width=0.45\textwidth]{d=4RNV_WLrec.pdf} } 
\caption{\small Left panel: The case $d=3$. The blue and the red curves correspond to $\chi_{(3)} \approx 0.003, \, 0.33$ respectively. Right panel: The case $d=4$. The blue and red curves correspond to $\chi_{(3)} \approx 0.002, \, 1.1 $ respectively. We measure $\cA_{\rm thermal}$ in units of the AdS-radius and the boundary time in units of the black hole mass. The diameter, also measured in units of the black hole mass, is fixed to $D=4$.}
\label{rnvwlcir}
\end{center}
\end{figure}

Once these boundary conditions are fixed, we can obtain a numerical solution for $z(\rho)$ and $v(\rho)$ for every value of $z_*$ and $v_*$. From this solution we can read off the boundary data
\begin{eqnarray}
z(R) = z_0 \ , \quad v(R) = t \ ,
\end{eqnarray}
where $R$ is the radius of the circular Wilson loop, $z_0$ is the radial IR cut-off and $t$ is the boundary time. In practice we do the following: for a given $R$, we fix $z_*$ and keep varying $v_*$ till the obtained value of the radial IR cut-off, denoted by $z_0$, is small enough. Then we vary $z_*$ and repeat the process. Ultimately this generates the required data to produce the thermalization curves for the Wilson loop operator. Some of these curves have been shown in fig.~\ref{rnvwlcir}.

\begin{figure}[!ht]
\begin{center}
\subfigure[] {\includegraphics[angle=0,
width=0.45\textwidth]{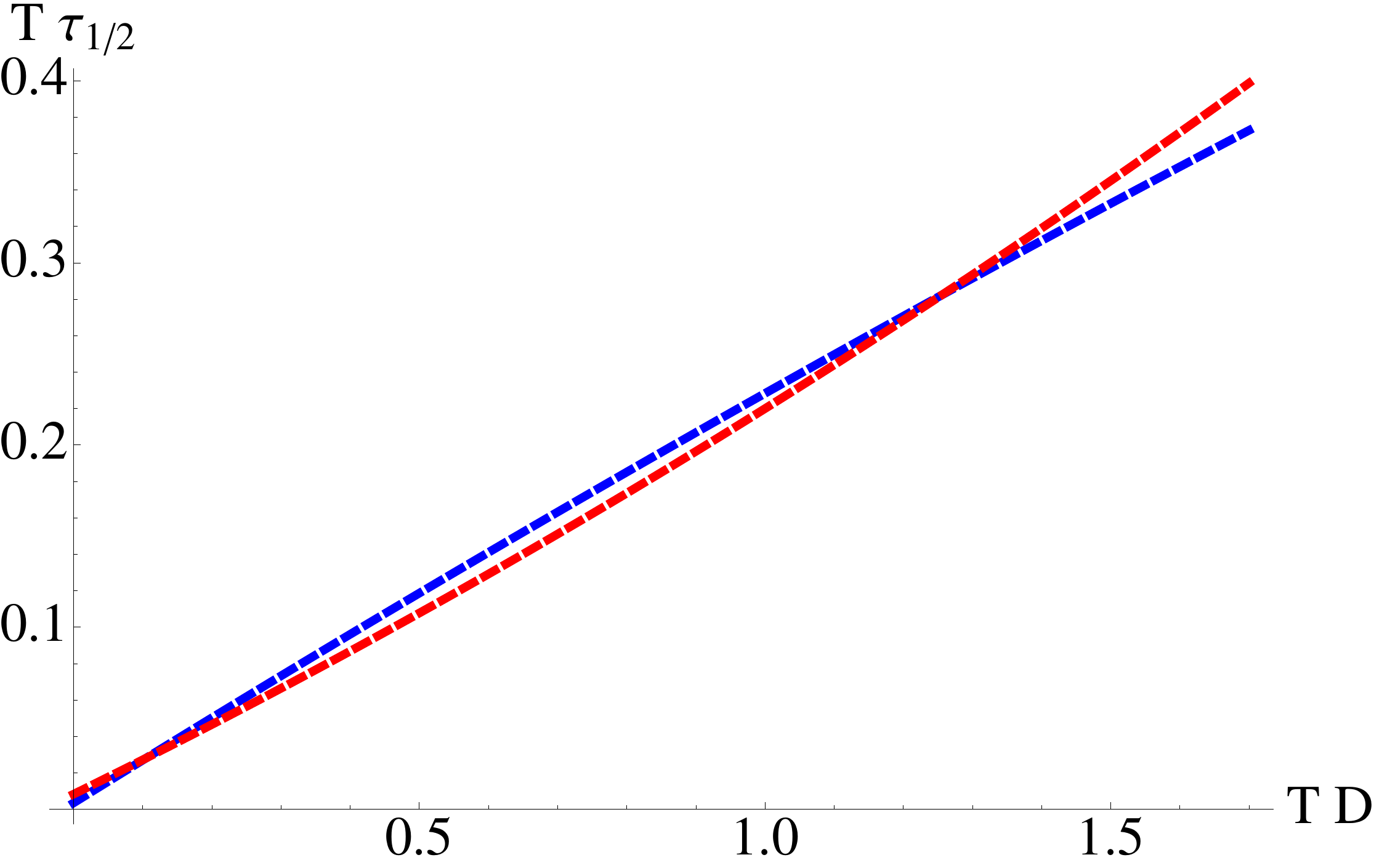} } 
\subfigure[] {\includegraphics[angle=0,
width=0.45\textwidth]{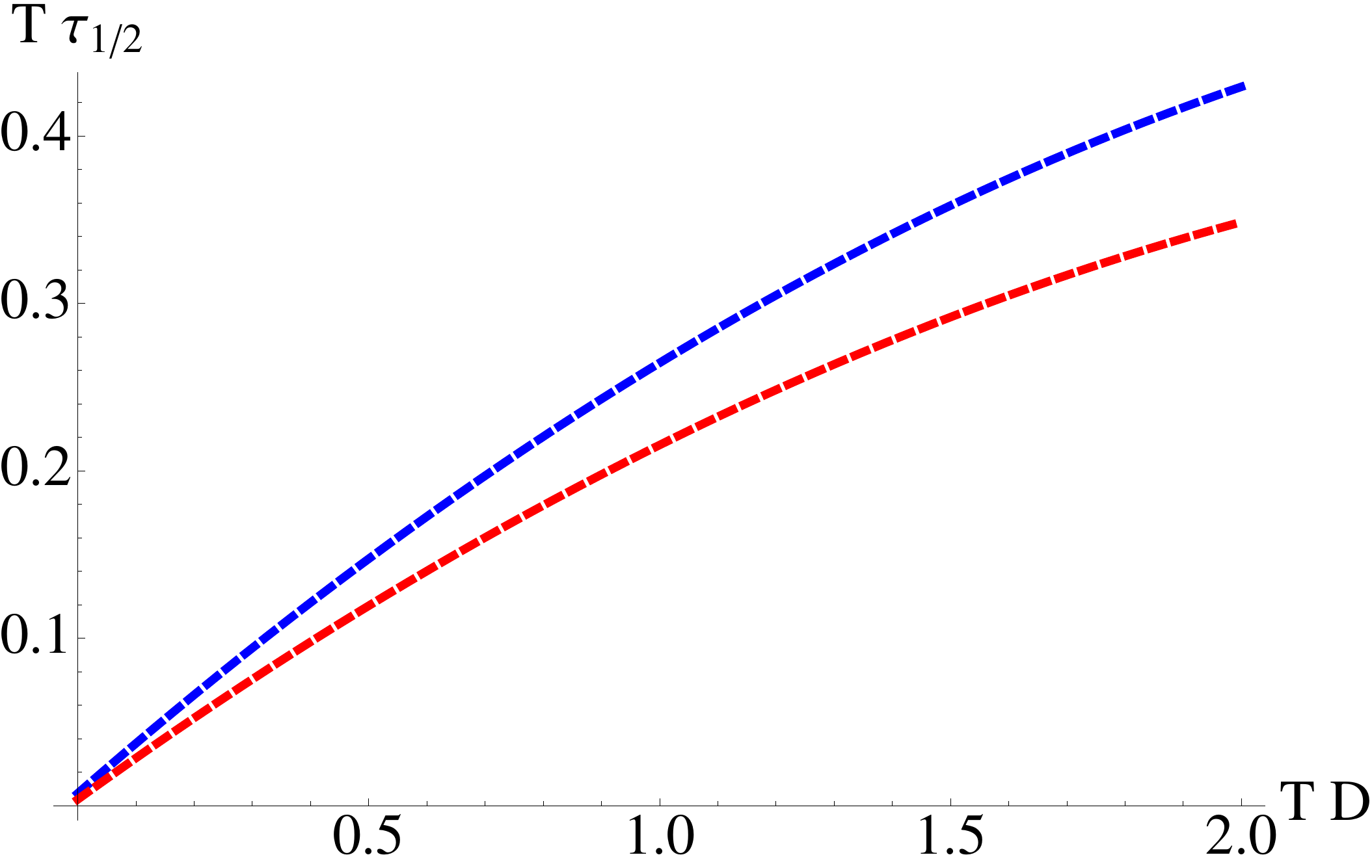} } 
\caption{\small Left panel: The case $d=3$. The blue and red curves correspond to $\chi_{(3)} \approx 0.003, \, 0.33$ respectively. Right panel: The case $d=4$. The blue and the red curve corresponds to $\chi_{(4)} \approx 0.002, \, 1.1 $ respectively.}
\label{thalfWLcir}
\end{center}
\end{figure}

To extract the behavior of the thermalization time $\tau_{1/2}$, we repeat this process for various values of the radius and finally obtain the fig.~\ref{thalfWLcir}. For small values of $\chi_{(d)}$, the behavior of $\tau_{1/2}$ is primarily sub-linear. The linear regime for small $D$ has a smaller slope in $d=4$ as compared to the one in $d=3$. Increasing $\chi_{(d)}$ suppresses the sub-linear behavior of $\tau_{1/2}$ and for large enough $D$, $\tau_{1/2}$ increases for increasing $\chi_{(d)}$. For small values of $D$, from fig.~\ref{thalfWLcir}(b) we can definitely identify the regime where increasing chemical potential leads to faster thermalization. Also, it is clear from the $d=3$ case that for moderate values of $\chi_{(3)}$, $\tau_{1/2}$ does not change much unless we go to very high values of $D$. 

\begin{figure}[!ht]
\begin{center}
\subfigure[] {\includegraphics[angle=0,
width=0.45\textwidth]{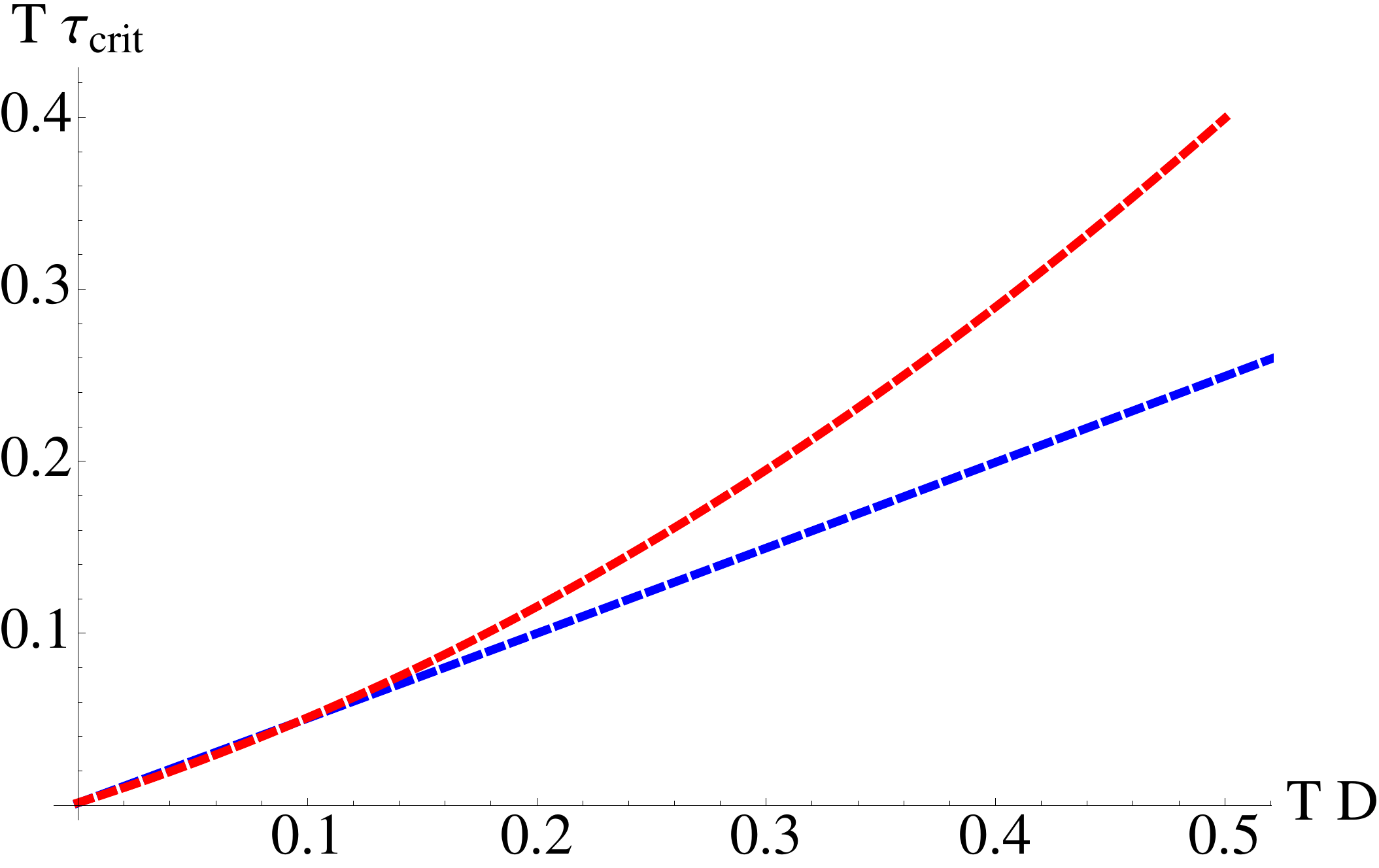} } 
\subfigure[] {\includegraphics[angle=0,
width=0.45\textwidth]{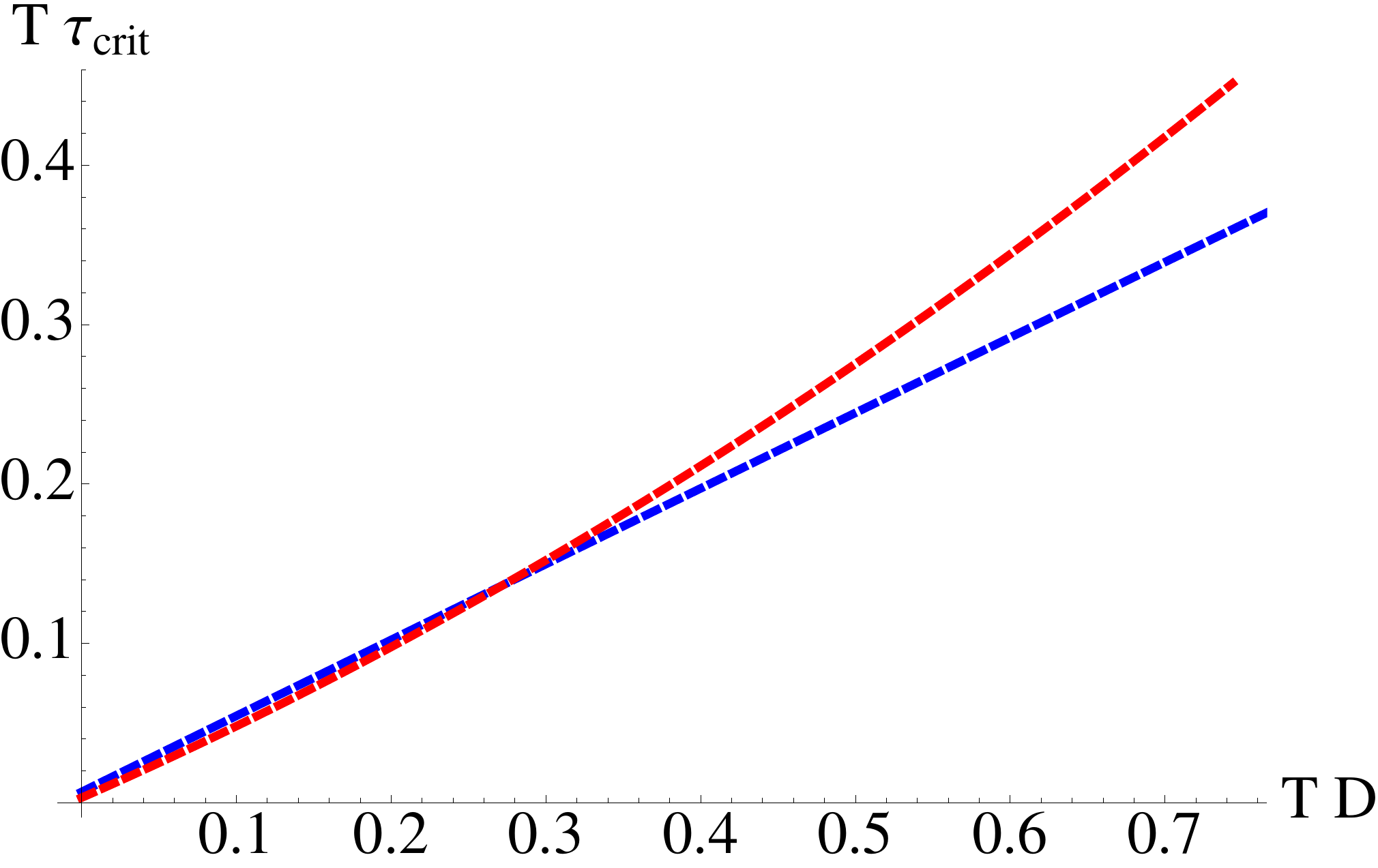} } 
\caption{\small Left panel: The case $d=3$. The blue and red curves correspond to $\chi_{(3)} \approx 0.003, \, 4.47$ respectively. Right panel: The case $d=4$. The blue and the red curve corresponds to $\chi_{(4)} \approx 0.002, \, 1.1 $ respectively.}
\label{tcritWLcir}
\end{center}
\end{figure}

On the other hand the thermalization time, denoted by $\tau_{\rm crit}$, is shown in fig.~\ref{tcritWLcir}. For small values of $\chi_{(d)}$, for a relatively large regime of values for $D$, $\tau_{\rm crit}$ behaves linearly with a slope very close to $1/2$. Increasing $\chi_{(d)}$, promotes a deviation from this linearity for larger values of $D$ and the corresponding curve bends away. This means for large enough $D$, increasing $\chi_{(d)}$ increases $\tau_{\rm crit}$.
%


\subsection{Entanglement entropy}

\subsubsection{The rectangular belt}

\begin{figure}[!ht]
\begin{center}
\includegraphics[angle=0,
width=0.65\textwidth]{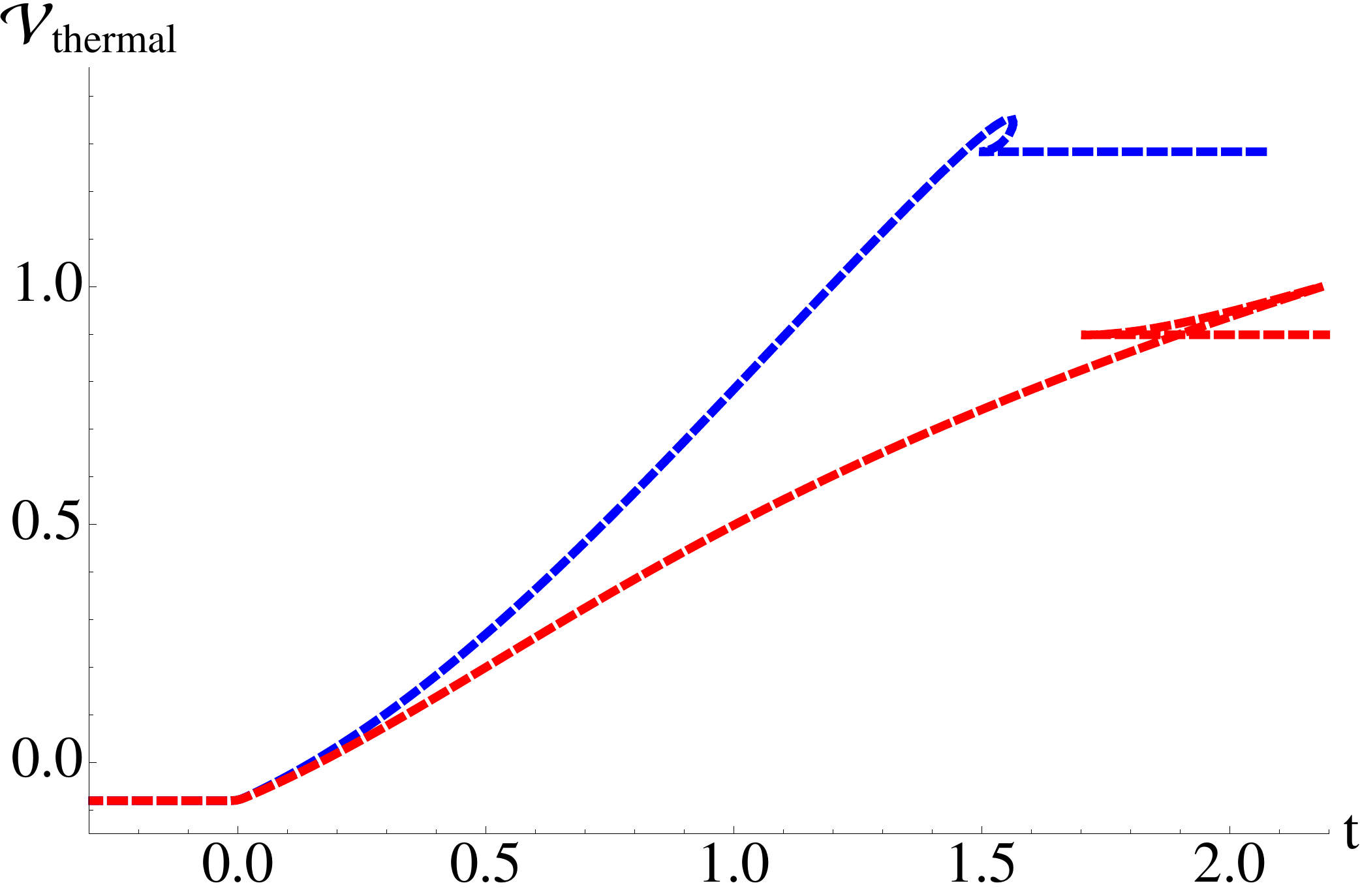}
\caption{The case of $d=4$. The blue and the red curve corresponds to $\chi_{(4)} \approx 0.002, \, 1.1$ respectively. $\cV_{\rm thermal}$ is measured in units of the AdS-radius. Here we have fixed $\ell=2$ in units of the black hole mass. We do see the appearance of the swallow-tail behavior in the thermalization curve for entanglement entropy.}
\label{rnveerec}
\end{center}
\end{figure}

Let us first begin with the rectangular strip geometry. The volume functional corresponding to the minimal area surface can be parametrized by $z(x)$ and $v(x)$, where $x\equiv x_1$, with the assumption that this minimal area surface is invariant under the other two planar directions. The corresponding volume functional is now given by
\begin{eqnarray}
\cV = \int_{-\ell/2}^{\ell/2} \frac{dx}{z^3} \left(1 - f v'^2 - 2 v' z' \right)^{1/2} \ ,
\end{eqnarray}
where $'\equiv d/dx$. This gives the conservation equation
\begin{eqnarray}
1 - f v'^2 - 2 v' z' = \left(\frac{z_*}{z}\right)^6 \ .
\end{eqnarray}
The two equations of motion resulting from the variation of the volume functional can be written as
\begin{eqnarray}
&& z'' + f v'' + \frac{\partial f}{\partial z} z' v' + \frac{1}{2} v'^2 \frac{\partial f}{\partial v} = 0 \ , \\
&& z v'' + 6 v' z' -3 + v'^2 \left( 3 f - \frac{1}{2} z \frac{\partial f}{\partial v} \right) = 0 \ .
\end{eqnarray}
Once again only the second equation above changes. We will again solve these two equations subject to the boundary conditions outlined in (\ref{bc}) and using similar method as outlined earlier. Some of the curves obtained in this way are shown in fig.~\ref{rnveerec}.

\begin{figure}[!ht]
\begin{center}
\subfigure[]{\includegraphics[angle=0,
width=0.45\textwidth]{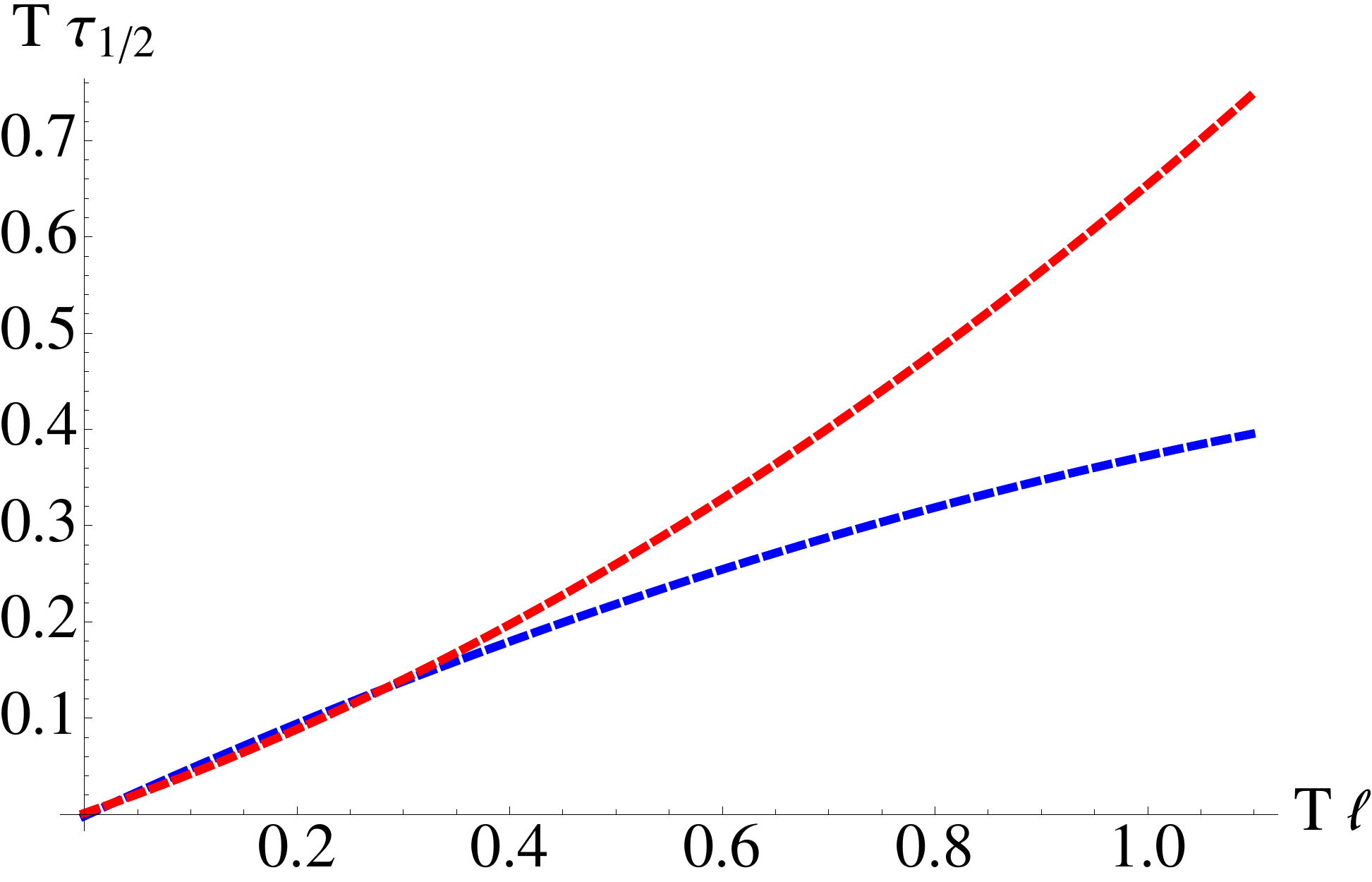}}
\subfigure[]{\includegraphics[angle=0,
width=0.45\textwidth]{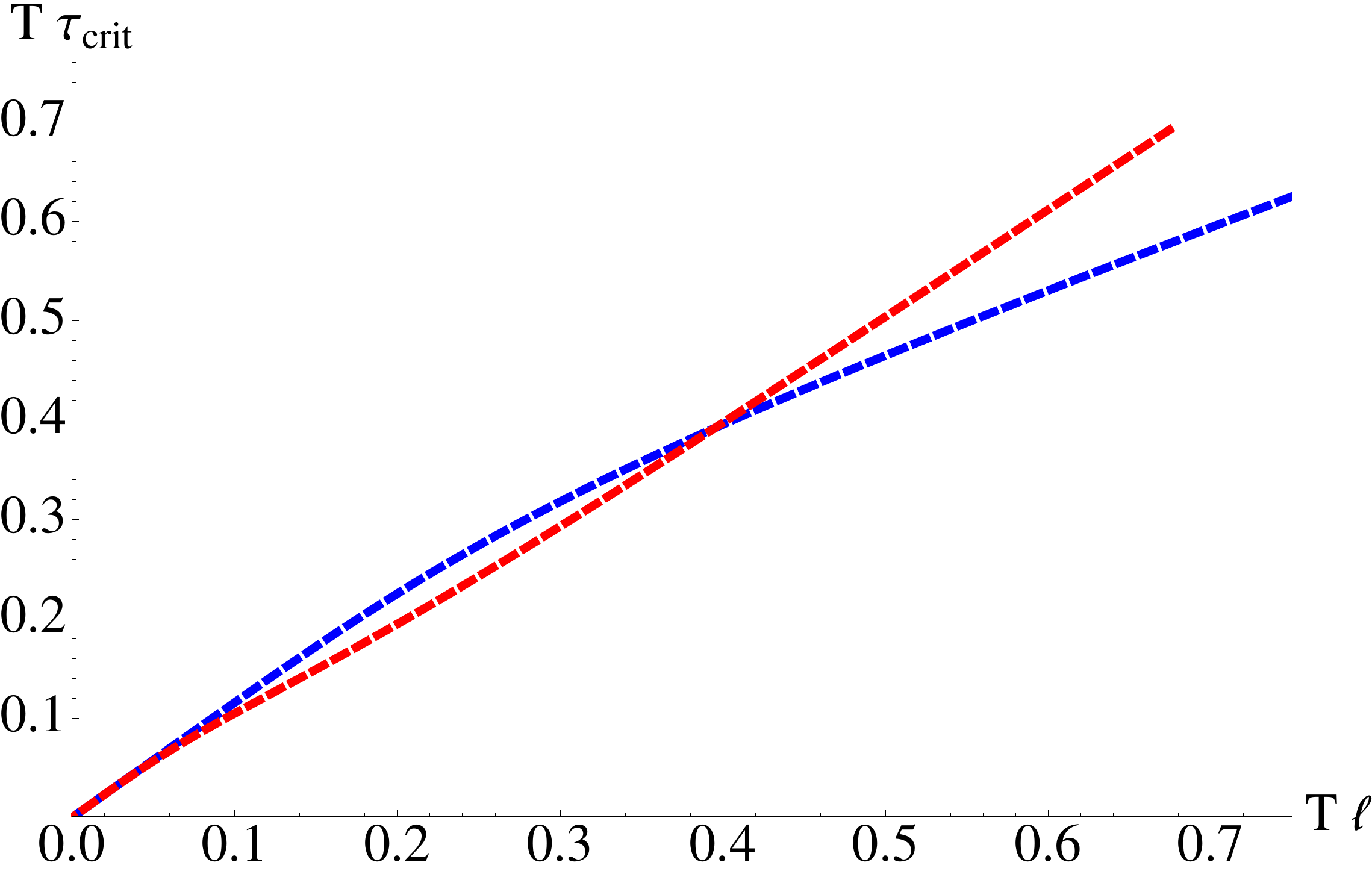}}
\subfigure[]{\includegraphics[angle=0,
width=0.5\textwidth]{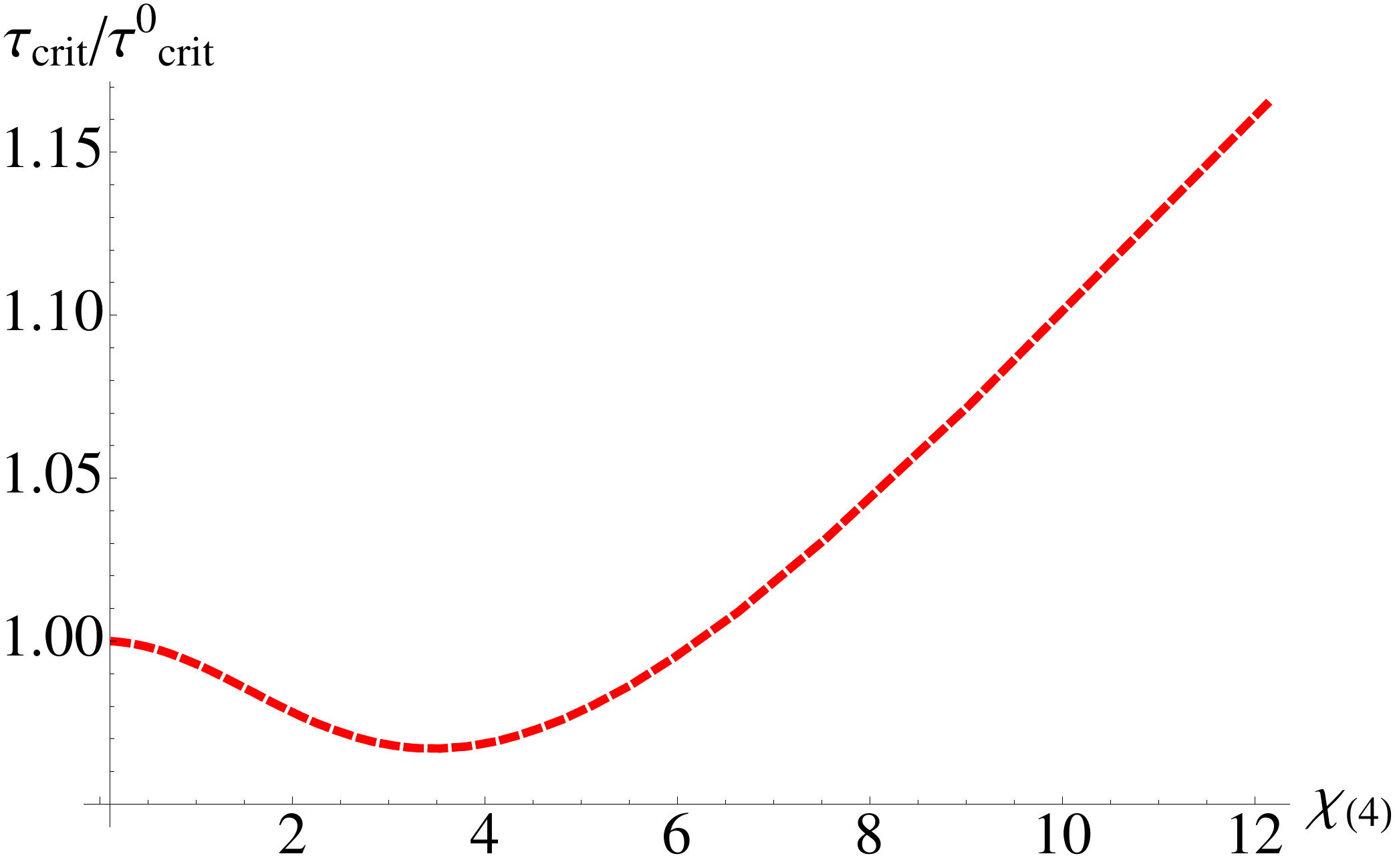}}
\caption{The case of $d=4$. Left panel (a): $\tau_{1/2}$ as a function of the boundary separation $\ell$. The blue and the red curve corresponds to $\chi_{(4)} \approx 0.002, \, 1.1$ respectively. Right panel (b): $\tau_{\rm crit}$ as a function of the boundary separation $\ell$. The blue and the red curve corresponds to $\chi_{(4)} \approx 0.002, \, 1.1$ respectively. In (c) we have shown how $\tau_{\rm crit}$ behaves with $\chi_{(4)}$ for fixed $(4\pi T)\ell = 0.5$.}
\label{td4eerec}
\end{center}
\end{figure}

In fig.~\ref{td4eerec} we have shown how the two thermalization times $\tau_{1/2}$ and $\tau_{\rm crit}$ depend on the length $\ell$ of the rectangular region. For small values of $\chi_{(4)}$, $\tau_{1/2}$ behaves linearly with $\ell/2$ for small values of $\ell$: the corresponding slope of this linear behavior is unity. For larger length, $\tau_{1/2}$ becomes sub-linear. On the other hand, for larger values of $\chi_{(4)}$ the deviation from linear behavior of $\tau_{1/2}$ is suppressed and for larger length, increasing $\chi_{(4)}$ increases $\tau_{1/2}$. 

The qualitative behavior of $\tau_{\rm crit}$ is similar to that of $\tau_{1/2}$. However, the slope of the linear regime is quite different. For $\tau_{\rm crit}$, in the linear regime, $\tau_{\rm crit} \sim \ell$ and therefore has almost twice the slope compared to the curves for $\tau_{1/2}$. Once again we observe that for large enough $\ell$, larger $\chi_{(4)}$ leads to larger $\tau_{\rm crit}$. Finally, in \ref{td4eerec}(c) we have shown the dependence of $\tau_{\rm crit}$ with the dimensionless ratio $\chi_{(4)}$, which gives a similar result as we have observed earlier, specifically we do seem to have two different regimes for $\chi_{(4)}$ distinguished by a faster or a slower thermalization corresponding to small and large values of $\chi_{(d)}$. A similar behavior is observed as $(T\ell)$ is varied for fixed values of $\chi_{(4)}$.

\subsubsection{Spherical region}

\begin{figure}[!ht]
\begin{center}
 {\includegraphics[angle=0,
width=0.65\textwidth]{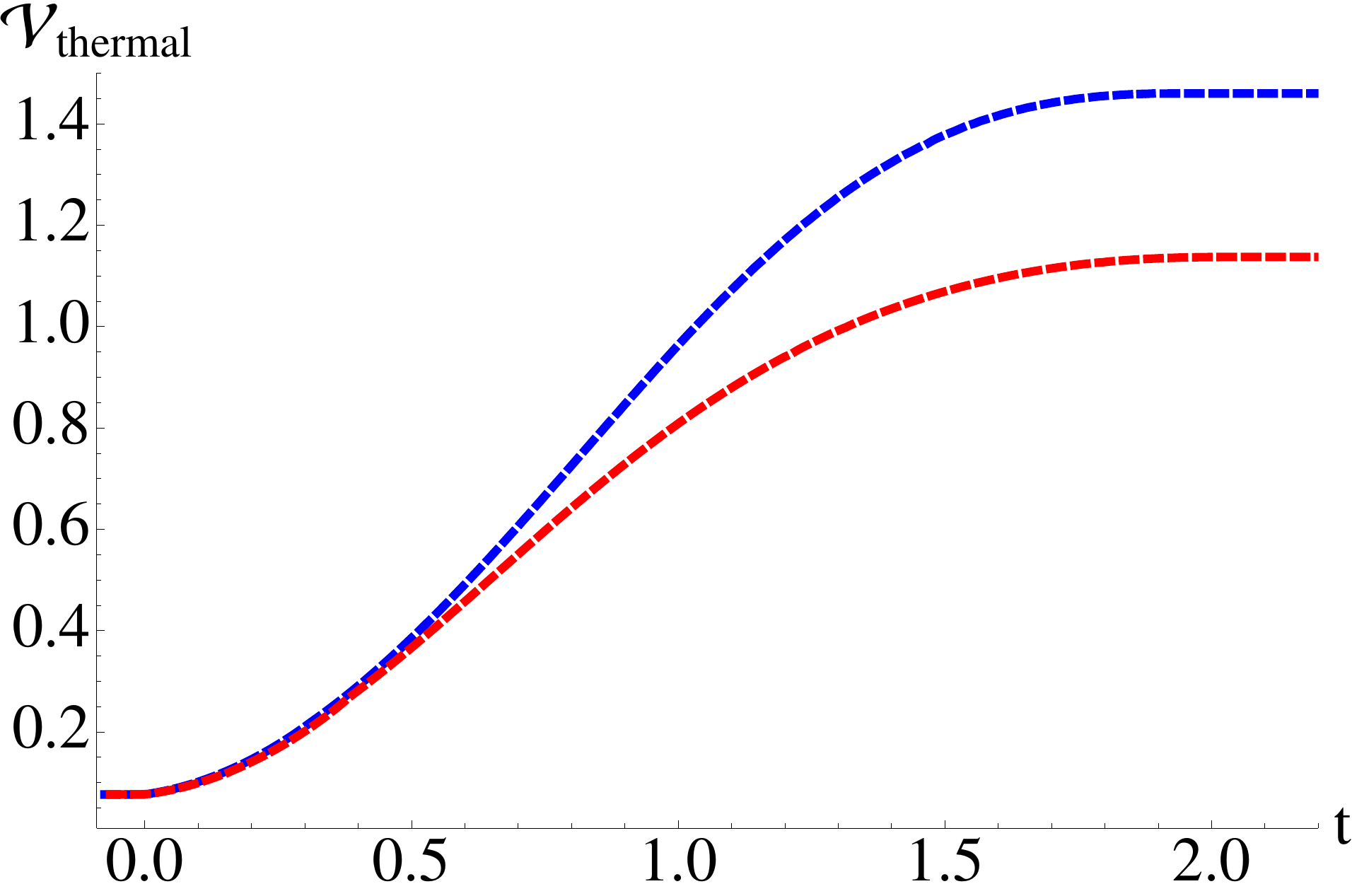} }
\caption{\small The case of $d=4$. The blue and the red curve corresponds to $\chi_{(4)} \approx 0.002, \, 1.1$ respectively for fixed diameter $D=4$ (in units of the black hole mass). $\cV_{\rm thermal}$ is measured in units of the AdS-radius.}
\label{eecir}
\end{center}
\end{figure}

Now we will investigate the case of a spherical region. To this end, we parametrize the spherical disc in polar coordinates as in (\ref{polarcir}) and represent the corresponding minimal are surface by $z(\rho)$ and $v(\rho)$. The corresponding volume functional is given by
\begin{eqnarray}
\cV = 4 \pi \int_0^R d\rho \frac{\rho^2}{z^3} \sqrt{1 - f v'^2 - 2 v' z'} \ ,
\end{eqnarray}
where $'\equiv d/d\rho$. We follow similar methods as outlined in section \ref{wlcir} in obtaining the thermalization curves and a few representative curves are shown in fig.~\ref{eecir}.

\begin{figure}[!ht]
\begin{center}
\subfigure[] {\includegraphics[angle=0,
width=0.48\textwidth]{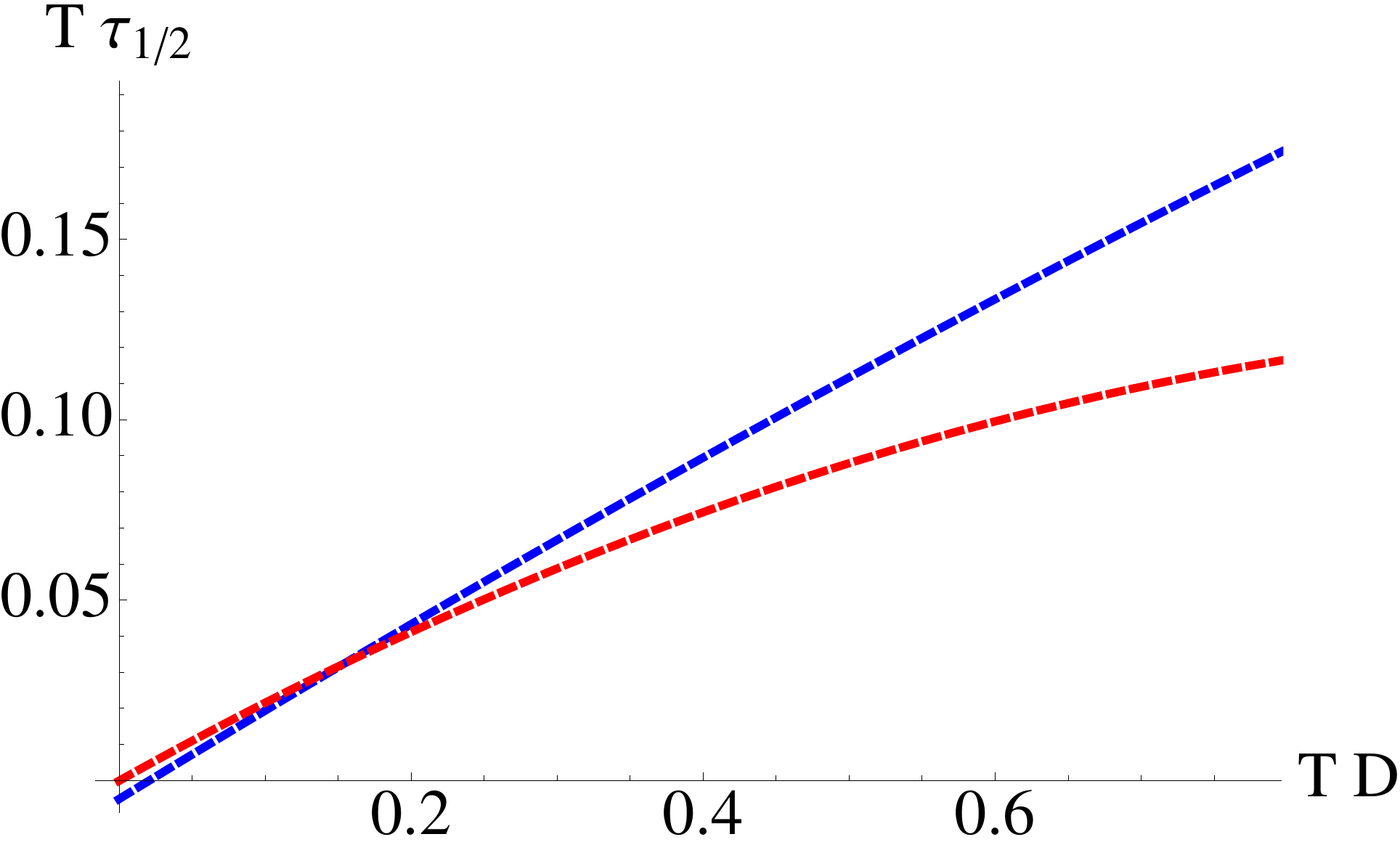} }
\subfigure[] {\includegraphics[angle=0,
width=0.48\textwidth]{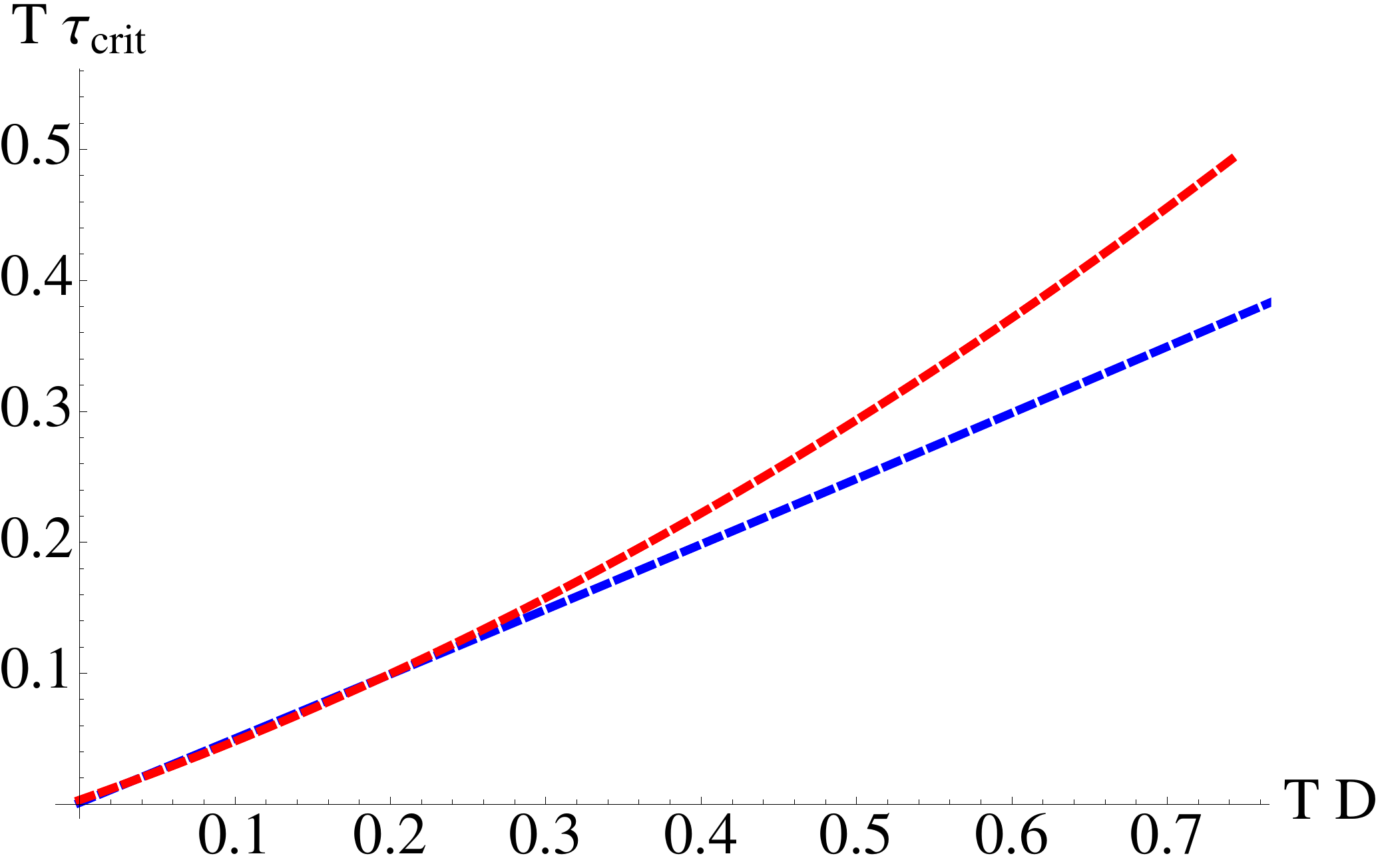} }
\caption{\small The case of $d=4$. The blue and the red curve corresponds to $\chi_{(4)} \approx 0.002, \, 1.1$ respectively.}
\label{td4eecir}
\end{center}
\end{figure}

We have demonstrated the behavior of the thermalization time with the diameter of the spherical region in fig.~\ref{td4eecir}. In this case, $\tau_{1/2}$ has a sub-linear behavior for the range of the diameter we have explored for both small and relatively large values of $\chi_{(4)}$. The linear regime grows slower than $D/2$. Also, for the range of the diameter that we have explored, increasing $\chi_{(4)}$ decreases $\tau_{\rm 1/2}$.

The behavior of $\tau_{\rm crit}$, on the other hand, is in agreement with previously obtained qualitative features. For small value of $\chi_{(4)}$, $\tau_{\rm crit} \sim D/2$ and this linear behavior persists till larger values of $D$. However, as $\chi_{(4)}$ is increased, $\tau_{\rm crit}$ exceeds this linear behavior and the corresponding curve bends away.

\section{Summary and Discussion}

We have explored, in details, the behavior of thermalization time as the chemical potential is varied by probing the following various non-local observables: two-point function, Wilson loop and entanglement entropy. In this section, we briefly summarize and review some crucial features of the observations we made. For simplicity, we will only comment on the behavior of $\tau_{\rm crit}$, but the behavior of $\tau_{1/2}$ is qualitatively similar.

At vanishing chemical potential, which in our notation would be denoted by $\chi_{(d)} = 0$, we reproduce the results obtained in \cite{Balasubramanian:2010ce, Balasubramanian:2011ur}. The generic features in the absence of a chemical potential are the following: for the two-point function, the thermalization time $\tau_{\rm crit} = \ell/2$ in AdS$_3$ but $\tau_{\rm crit} < \ell/2$ in AdS$_{4,5}$ and deviates from linearity. The $\tau_{\rm crit} = \ell/2$ behavior is understood from the dual $(1+1)$-dim CFT. On the other hand, the deviation from linearity in higher dimensions is interpreted as an indicator of a faster-than-causal thermalization possibly resulting from the homogeneity of the initial configuration\cite{Balasubramanian:2010ce}. In higher dimensions, the Wilson loop or the entanglement entropy also behaves in a similar fashion.

\begin{figure}[!ht]
\begin{center}
 {\includegraphics[angle=0,
width=0.65\textwidth]{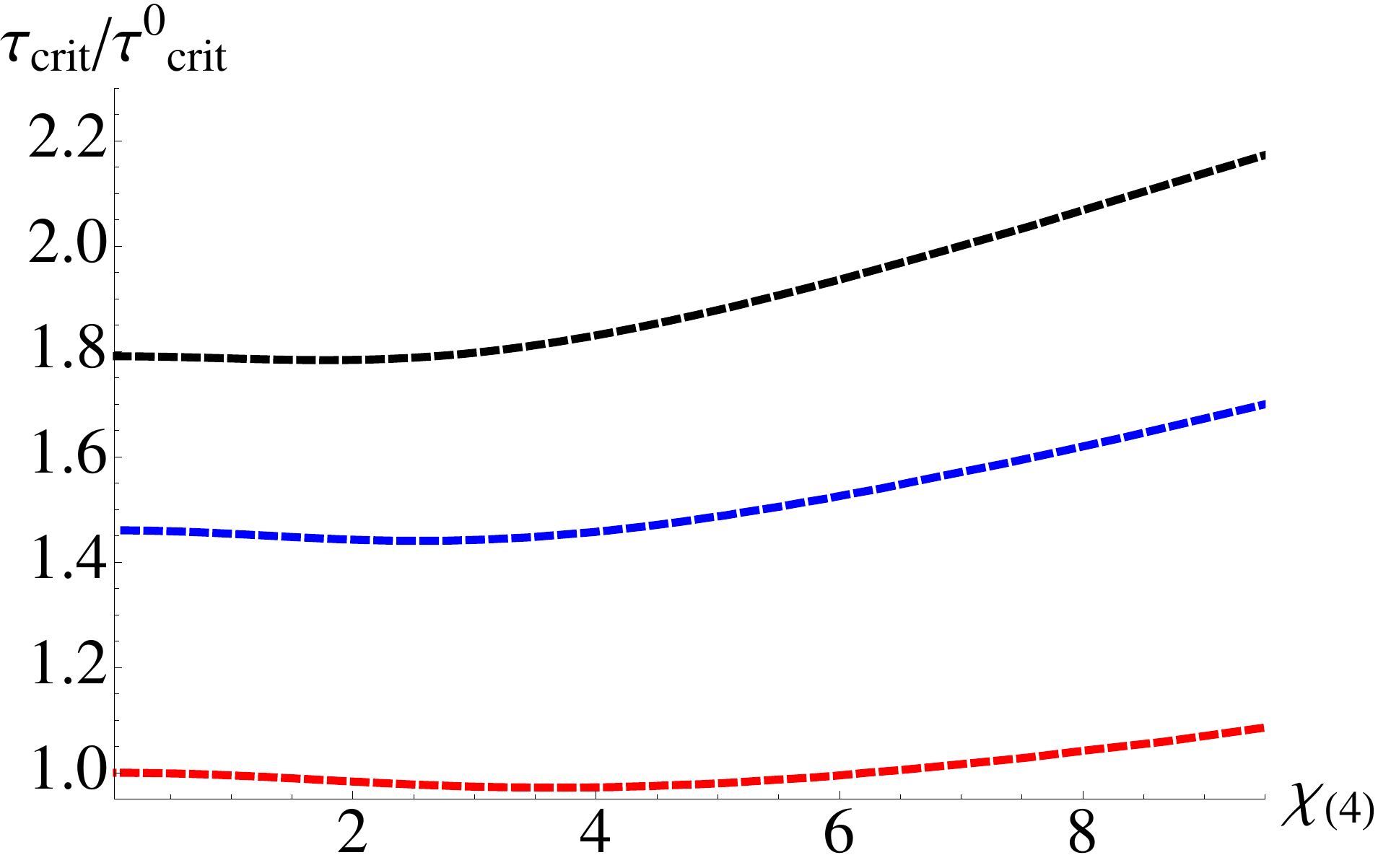} }
\caption{\small For fixed $(4\pi T)\ell=0.7$ we have shown how $\tau_{\rm crit}$ depends on $\chi_{(4)}$ for various probes. The red curve corresponds to the 2-pt function, the blue curve corresponds to the Wilson loop and the black curve corresponds to the entanglement entropy. Here $\tau_{\rm crit}^0$ corresponds to that of the two-point function.}
\label{eerocks}
\end{center}
\end{figure}

Before proceeding further let us offer some comments on the operators that we considered. We have studied three different types of non-local operators as probes of thermalization. It is not {\it a priori} clear which of these operators provides the correct time-scale for thermalization. It can be explicitly checked (like in the case of vanishing chemical potential) that it is actually the entanglement entropy which thermalizes the latest and thereby sets the equilibration time-scale. This is demonstrated in fig.~\ref{eerocks}.

In the presence of a chemical potential the physics becomes richer. First, the relevant quantities to consider here are the following dimensionless combinations: $(T\tau_{\rm crit})$, $(T\ell)$ and $(\mu/T)$ --- where $T$ is the temperature of the thermal background, $\ell$ is the length of the non-local operator, $\mu$ is the chemical potential and $\tau_{\rm crit}$ is the thermalization time. Thus, in general, $(T\tau_{\rm crit})$ is a function of the two independent variables: $(T\ell)$ and $(\mu/T)$ and our goal is to explore this behavior.

Based on general physics intuition, once the thermalization has set in, we can identify the following different regimes
\begin{eqnarray}
T\ell \quad {\rm fixed, \, small:} \quad && \underbrace{\mu/T \gg 1}  \quad {\rm and} \quad \underbrace{\mu/T \ll 1}  \ , \\
&&{\rm quantum} \quad \quad \quad \quad {\rm classical} \nonumber
\end{eqnarray}
and 
\begin{eqnarray}
\mu/T \quad {\rm fixed, \, small:} \quad && \underbrace{T \ell \ll 1}  \quad {\rm and} \quad \underbrace{T \ell \gg 1}  \ . \\
&&{\rm quantum} \quad \quad \quad  {\rm classical} \nonumber
\end{eqnarray}
In the previous subsections, by analyzing the various probes of thermalization, we have seen qualitatively different features in precisely these ``classical" and ``quantum" regimes as defined above. The ubiquitous property that reveals itself through this exercise is that thermalization becomes faster for small values of $\mu/T$, but for large values of $\mu/T$ the thermalization time increases without any upper bound. This non-monotonic behavior in the two regimes are smoothly connected through a minima as depicted in {\it e.g.} figs.~\ref{tcrit_chi_WLrec}(b), \ref{td4eerec}(c). This non-monotonicity is enhanced in higher dimensions.\footnote{Although we have not considered the case of $d=2$, which corresponds to an $(1+1)$-dim dual CFT, we have explicitly checked that the non-monotonic behavior is further suppressed for $d=2$. For small values of $\mu/T$, the variation in $T\tau_{\rm crit}$ is negligible and within the numerical accuracy it is difficult to conclude whether thermalization time increases or decreases. For larger values of $\mu/T$, the dependence is similar to what we see in higher dimensional examples and $T\tau_{\rm crit}$ increases linearly with $\mu/T$.} We should also note that the identification of the ``classical" or the ``quantum" regime is only meaningful when the system has completely thermalized. Although it seems that the thermalization process is sensitive about this equilibrium feature, it is not clear to us whether such a ``classical" or a ``quantum" regime actually exist during the non-equilibrium period.

Let us also briefly comment on the functional dependence of $T\tau_{\rm crit}$ on the dimensionless  combinations $T\ell$ and $\mu/T$. If we take constant $T\ell$-slices, our generic observation suggests a linear relation
\begin{eqnarray}
T\tau_{\rm crit} = A\left(d, T\ell \right) \frac{\mu}{T}  \quad {\rm for} \quad \frac{\mu}{T} \gg 1 \ ,
\end{eqnarray}
where $A(d, T\ell)$ is the slope of the curve which generally depends on the dimensionality of the problem and the value of $T\ell$. On the other hand, if we take constant $(\mu/T)$-slices we get the following polynomial relation\footnote{This can be verified by explicitly fitting the data.}
\begin{eqnarray}
T\tau_{\rm crit} = B\left(d, \mu/T \right) \left(T\ell\right)^2 + C \left(d, \mu/T\right) \left(T \ell\right) \ ,
\end{eqnarray}
where $B$ and $C$ are two constants which depend on the dimensionality of the problem and the value of the chemical potential. Furthermore, it can also be checked explicitly that $B(d, \mu/T)<0$ for $\mu/T \ll 1$ and $B(d, \mu/T) > 0$ for $\mu/T \gg 1$; whereas $C(d, \mu/T)$ is always positive. For small values of the length of the operator, thermalization time behaves linearly with the length of the operator and the slope is given by $C$. For large values of the length, $\tau_{\rm crit}$ is either sub-linear or super-linear depending on the sign of the other constant $B$.

We have  shown that the presence of a chemical potential makes the physics of holographic thermalization richer. Some important qualitative  differences appear, namely,  the non-monotonic behavior of the thermalization time as a function of $\mu/T$ . Some other charcteristics like top-down thermalization  (UV modes thermalize first)  and the fact that it is the entanglement entropy that sets the thermalization scale  are common to the $\mu=0$ case. 

This perhaps is a tip of the iceberg and there is a richer story waiting to be unfolded. Let us point out  some open problems and future directions which would improve the understanding  of holographic thermalization of theories with non-zero chemical potential:

\begin{itemize}
	\item  {\it Fortune favors the brave, or does it:} The time-dependent background we have worked on is more ``phenomenologically motivated" than one obtained from a ``first principle" calculation even from the point of view of classical gravity. One might therefore wonder about the validity or the applicability of the physics we observe within this framework. In this work we have implicitly assumed the existence of at least some approximation in which the physics we see holds true. In \cite{Bhattacharyya:2009uu} the authors showed that a dilaton source of small amplitude and finite duration at the boundary produces a wave that propagates in the bulk and collapses to form an AdS-Schwarzschild geometry. The spacetime for this collapse process was  constructed as an expansion in the amplitude of the dilaton source and it was shown that it takes the Vaidya form at the leading order. A fascinating question on its own right is the study of the  gravitational collapse process that may give rise to the AdS-RN-Vaidya metric. Perhaps a natural point to start from is to consider a generalization of \cite{Bhattacharyya:2009uu} by considering a charged dilaton field. Such an exercise will either establish the AdS-RN-Vaidya background on a firm footing or we will learn some other interesting lesson.

	\item  The fact that thermalization time increases with increasing chemical potential (at least for large enough values of chemical potential as we have observed) can be reconciled with weak-coupling field theory intuition. For a bosonic system, increasing chemical potential ``enhances" the Bose-Einstein condensation and therefore hinders the process of thermalization which will populate excited states. For a system with fermions, increasing chemical potential increases the available states by increasing the Fermi energy. Thus it is intuitive to conclude that a system with bosonic or fermionic degrees of freedom will thermalize slower if the chemical potential is increased.\footnote{We would like to thank Berndt M\"{u}ller for suggesting this possibility.} However, to the best of our knowledge, we are unaware of a field theory (even toy-model) calculation that demonstrates this physics explicitly. On the other hand, for small chemical potential the faster thermalization that we observed does not fit our na\"{i}ve intuition. Thus a field theory computation at weak coupling will shed much light on the underlying physics and we may be able to isolate the features governed by strong coupling and the physics that is governed by the presence of a chemical potential.
	
	\item It would be interesting to study the  spread and evolution of correlations in an out-of-equilibrium system which eventually thermalizes. In \cite{Balasubramanian:2011at} the authors pointed out that mutual and tripartite information are adequate probes to study  this problem and calculated these quantities holographically for a three dimensional bulk theory which is dual to an $(1+1)$-dim CFT. Field theory computations of such observables are so far available only for the $(1+1)$-dim. It will be an interesting exercise to generalize these computations in higher dimensions and analyze the corresponding physics. Furthermore, we have shown that a non-zero chemical potential introduces non-trivial new behavior in the thermalization of entanglement entropy. Thus, we expect that the mutual and tripartite information results will also be  substantially modified when $\mu\ne 0$. 

	\item More phenomenological quantities can also be studied. For example  the stopping distance of a massless particle --- related in the weak coupling to the jet quenching parameter for QCD-like theories --- was calculated in \cite{Arnold:2011qi} for an AdS$_5$-Schwarzchild black hole. An interesting question is  how the stopping distance  is modified in a time dependent background undergoing thermalization  with and without chemical potential\cite{wip}.  

\end{itemize}

We hope to address some of these  problems in the near future.

\section{Acknowledgements}

We would like to thank Willy Fischler and Berndt M\"{u}ller for conversations and encouragement about this work. E.C. acknowledges support of CONACyT grant CB-2008-01-
104649 and CONACyTÕs High Energy Physics Network. AK would like to thank the hospitality of the KITP, Santa Barbara during the workshop ``Novel Numerical Methods for Strongly Coupled Quantum Field Theory and Quantum Gravity" and financial support in part by the National Science Foundation under Grant No. PHY11-25915. AK is supported by the Simons postdoctoral fellowship awarded by the Simons Foundation. This material is based upon work supported by the National Science Foundation under Grant PHY-0969020.

\renewcommand{\theequation}{A.\arabic{equation}}
\setcounter{equation}{0}  
\section*{Appendix A. Geodesics in the background}
\addcontentsline{toc}{section}{Appendix A. Geodesics in the background}

For the sake of completeness, let us first write down the AdS-RN-Vaidya background in general $(d+1)$-bulk dimensions (with $d>2$)
\begin{eqnarray} \label{genvaidya}
&& ds^2  =  \frac{L^2}{z^2} \left( - f(z,v) dv^2 - 2 dv dz + d\vec{x}^2 \right) \ , \quad A_v = q(v) z^{d-2} \ , \\
&& f(z,v) = 1 - m(v) z^d + \frac{(d-2)q(v)^2}{(d-1) L^2} z^{2(d-1)} \ , \quad \Lambda = - \frac{d(d-1)}{2 L^2} \ ,
\end{eqnarray}
which gives
\begin{eqnarray}
2 \kappa T_{\mu\nu}^{\rm ext} = \left( \frac{d-1}{2} z^{d-1} \frac{dm}{dv} - \frac{d-2}{L^2} z^{2d-3} q \frac{dq}{dv} \right) \delta_{\mu v } \delta_{\nu v} \ , \quad \kappa J_{\rm ext}^\mu = (d-2) L^{d-3} \frac{dq}{dv} \delta^{\mu z} \ .
\end{eqnarray}

Analyzing the general behavior of the geodesics or the minimal surfaces of various dimensions in the AdS-RN-Vaidya background is an interesting problem.\footnote{For a recent exhaustive study of such $n$-dimensional extremal surfaces in general static asymptotically AdS background of various dimensions, see {\it e.g.} \cite{Hubeny:2012ry}.} For the rectangular cases we have considered in the main text, the equations of motion of such minimal surfaces take the following form
\begin{eqnarray} \label{genmin}
z'' + v'' f + z' v' \frac{\partial f}{\partial z} + \frac{1}{2} v'^2 \frac{\partial f}{\partial v} & = & 0 \ , \nonumber\\
z v'' + \left(2 p \right) z' v' - p + v'^2 \left(p f - \frac{1}{2} z \frac{\partial f}{\partial z} \right) & = & 0 \ , 
\end{eqnarray}
where
\begin{eqnarray}
p & = & 1 \quad \implies \quad {\rm geodesic} \ , \\
p & = & 2 \quad \implies \quad {\rm Wilson \, loop} \ , \\
p & = & 3 \quad \implies \quad {\rm Entanglement \, entropy} \ .
\end{eqnarray}
On the other hand, the action functional for the minimal surface for the circular region is 
\begin{eqnarray} \label{gencir}
\cS = \int_0^R d\rho \frac{\rho^{p-1}}{z^{p}} \sqrt{1 - f v'^2 - 2 z' v' } \ .
\end{eqnarray}
The equations of motion obtained for the circular region are involved, hence we do not present their explicit form here. We will briefly comment on a few general properties of the geodesics or the minimal surfaces obtained as solutions of (\ref{genmin}).

Before doing so, let us introduce the notion of an apparent horizon which will be relevant for the time-dependent backgrounds we have considered in the main text. A trapped surface $T$ is defined as a co-dimension two spacelike submanifold with the property that the expansion of both ``ingoing" and ``outgoing" future directed null geodesics orthogonal to $T$ is everywhere negative. The boundary of the trapped surfaces associated with a given foliation is defined as the apparent horizon.

In what follows, we will closely follow \cite{Figueras:2009iu}. For the background in (\ref{genvaidya}) the vectors tangent to the ingoing and outgoing null geodesics are given by
\begin{eqnarray}
l_- = - \partial_z \ , \quad l_+ = - \frac{z^2}{L^2} \partial_v + \frac{z^2}{2 L^2} f \partial_z \ 
\end{eqnarray}
such that
\begin{eqnarray}
l_- \cdot l_- = 0 \ , \quad l_+ \cdot l_+ = 0 \ , \quad l_- \cdot l_+ = -1 \ .
\end{eqnarray}
Now the volume element of the co-dimension two spacelike surface (orthogonal to the above null geodesics) is given by
\begin{eqnarray}
\Sigma = \left(\frac{L}{z}\right)^{d-1} \ .
\end{eqnarray}
The expansions are defined to be
\begin{eqnarray}
\theta_{\pm} = \cL_{\pm} \log \Sigma = l_{\pm}^\mu \partial_\mu \left( \log\Sigma \right) \ ,
\end{eqnarray}
where $\cL_{\pm}$ denotes the Lie derivatives along the null vectors $l_{\pm}$. The apparent horizon is then obtained by solving the equation $\Theta = 0$, where $\Theta = \theta_+ \theta_-$ is the invariant quantity. In this case we find
\begin{eqnarray} \label{apphori}
\Theta = \frac{(d-1)^2}{2 L^2} f = 0 
\end{eqnarray}
gives the location of the apparent horizon which we can find numerically.

\begin{figure}[!ht]
\begin{center}
\subfigure[] {\includegraphics[angle=0,
width=0.45\textwidth]{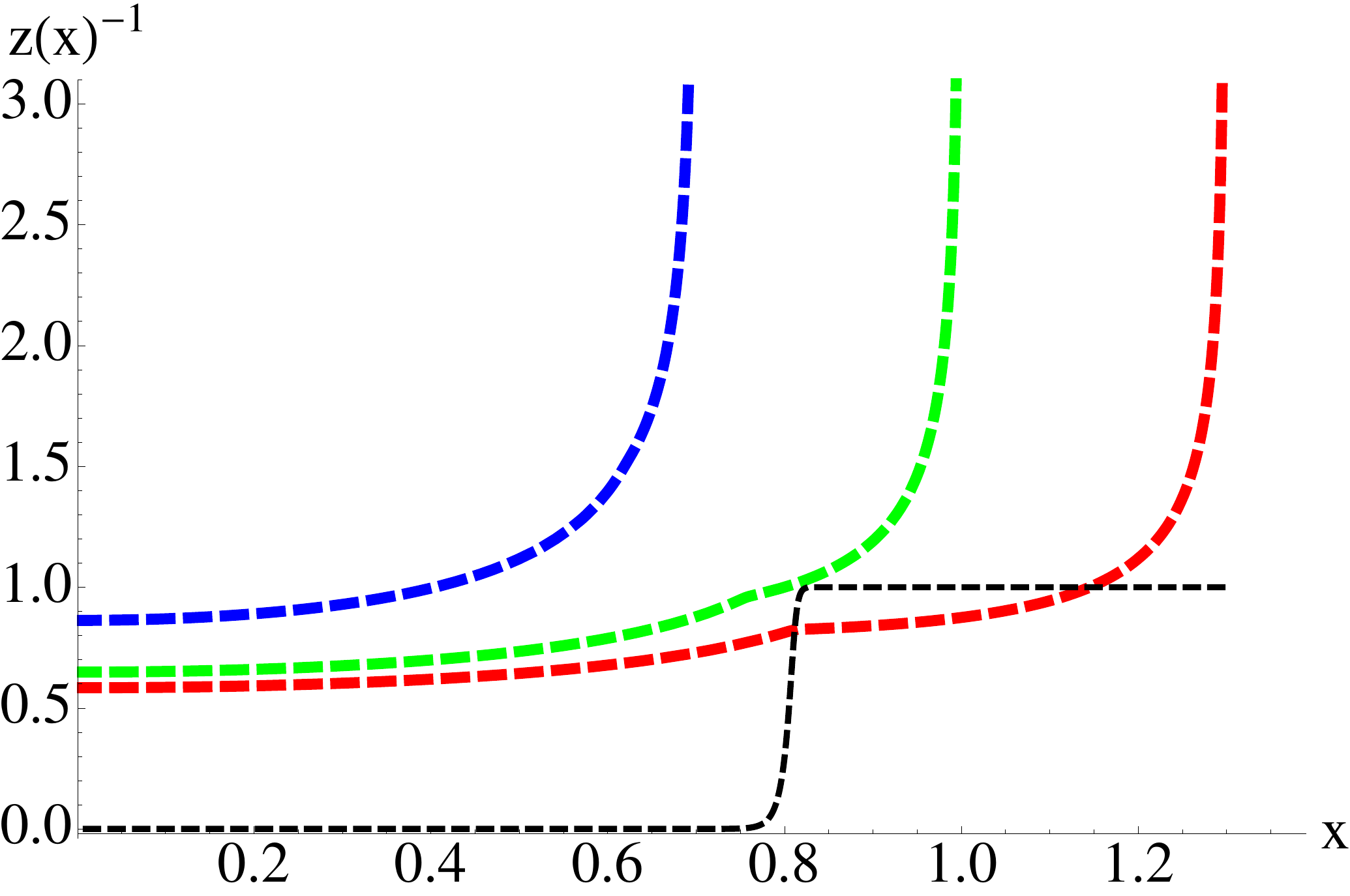} } 
\subfigure[] {\includegraphics[angle=0,
width=0.45\textwidth]{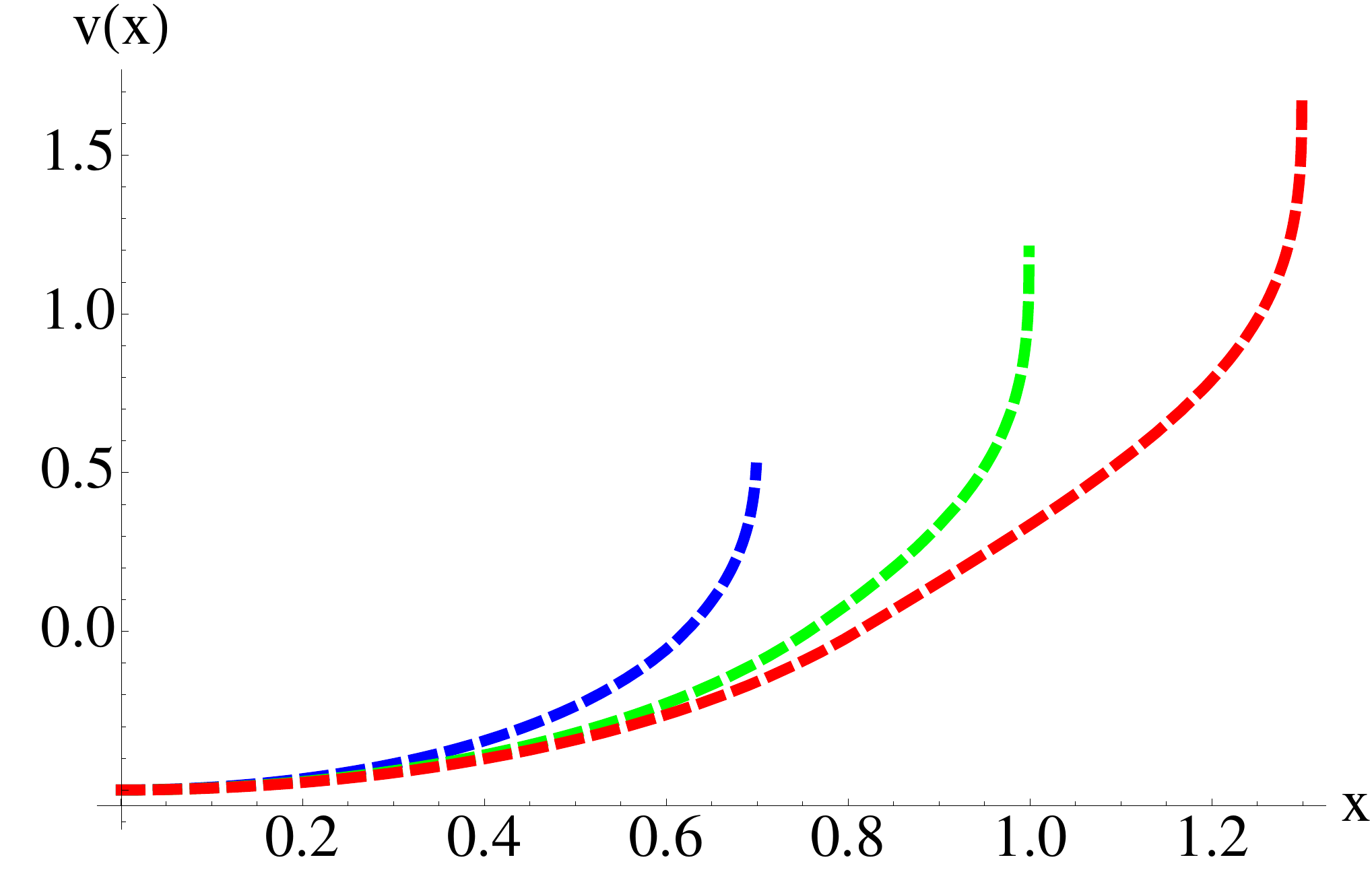} }
\subfigure[] {\includegraphics[angle=0,
width=0.5\textwidth]{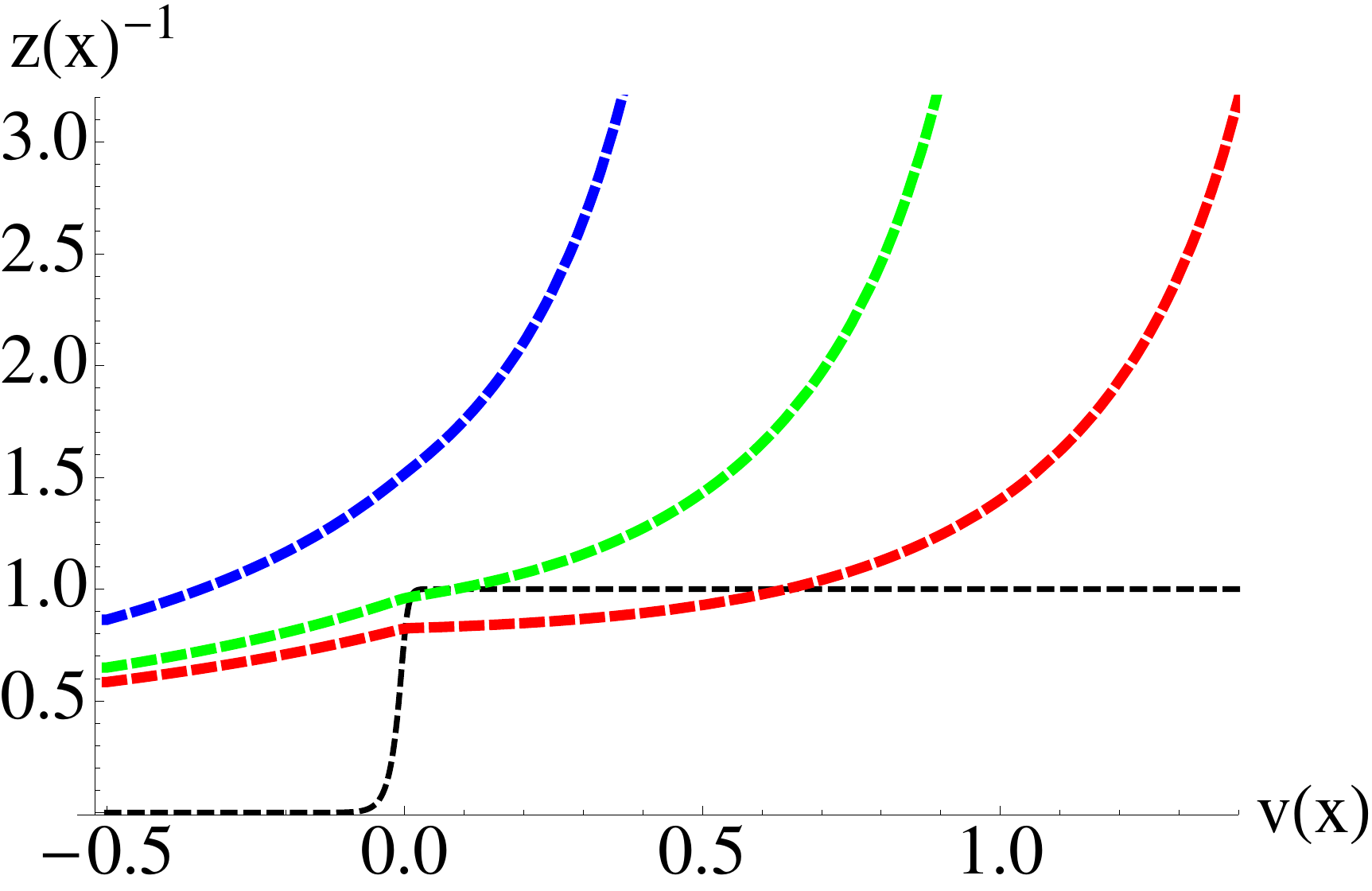} } 
\caption{\small We have presented here the profiles of rectangular Wilson loop for $d=3$. The black dashed line denotes the location of the apparent horizon obtained from solving equation (\ref{apphori}). We have presented here the profiles for only $x>0$ branch since the configuration has an exact symmetry under $x \to -x$. The blue, green and red curves correspond to $(4\pi T)\ell = 5.6, \, 8, \, 10.4$ respectively.}
\label{geotype}
\end{center}
\end{figure}

Now we are ready to present a few representative figures demonstrating how a geodesic or a minimal area surface looks like in our time-dependent backgrounds. Such representative profiles are shown in fig.~\ref{geotype}. It is clear that depending on the boundary separation length $\ell$, the corresponding minimal surface may or may not penetrate the apparent horizon. If the corresponding geodesic or the minimal surface penetrates the apparent horizon, the derivative of the curve becomes discontinuous. The curve consists of two parts, one inside the apparent horizon and the other outside of it. These two sections of the curve are connected according to an analogue of the Snell's law of refraction \cite{Balasubramanian:2011ur}. This discontinuity is visible from the plots in the $\{z(x) - x\}$ and the $\{z(x)- v(x)\}$ planes.

\begin{figure}[!ht]
\begin{center}
 {\includegraphics[angle=0,
width=0.65\textwidth]{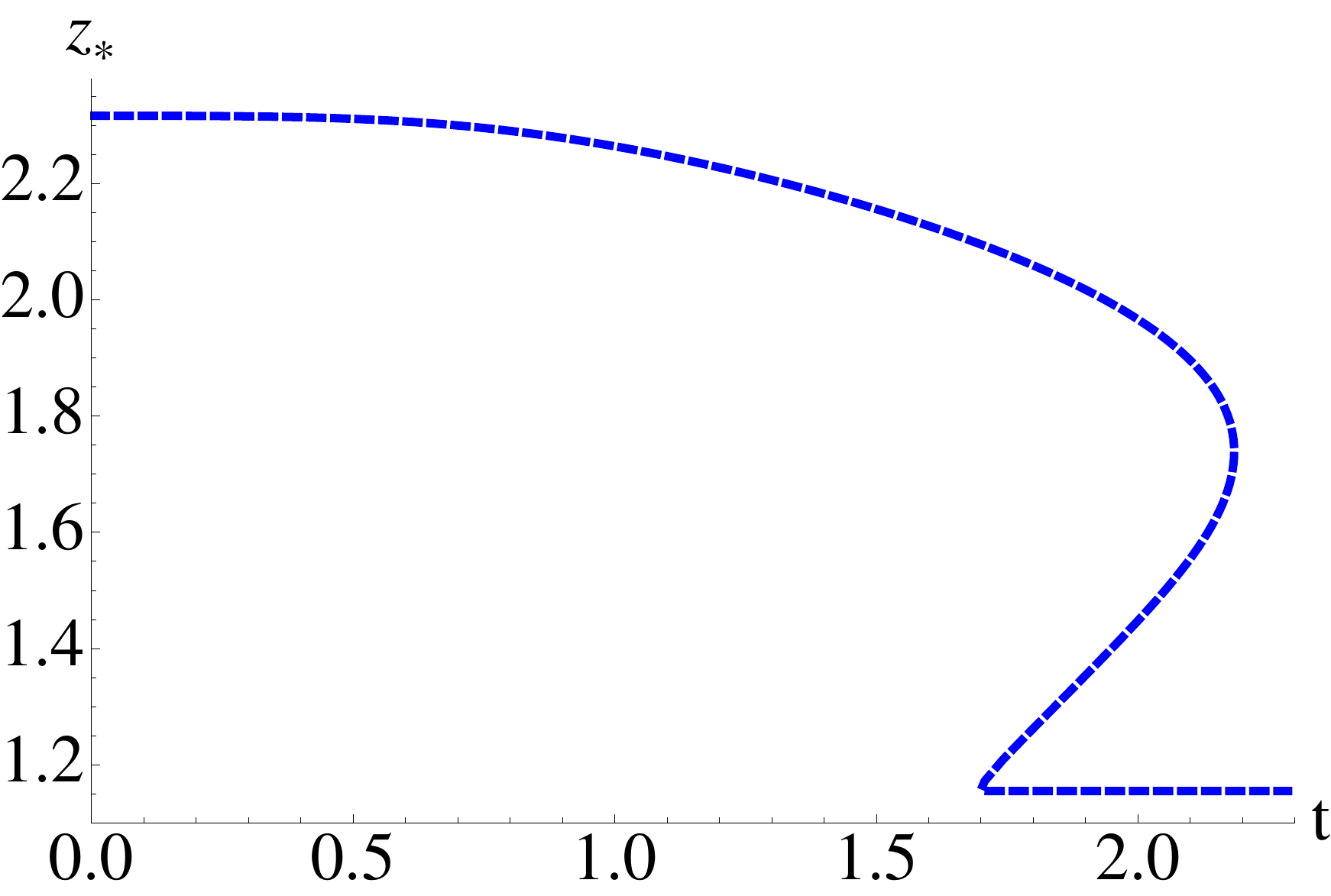} }
\caption{\small The multi-valuedness of $z_*$ with as a function of the boundary time $t$ for $d=4$ rectangular entanglement entropy computation. The boundary separation length is $(4\pi T)\ell \approx 2$.}
\label{multi_demo}
\end{center}
\end{figure}

An interesting feature that we have observed in {\it e.g.} fig.~\ref{rnveerec} and was also pointed out in \cite{Albash:2010mv, Balasubramanian:2011ur} is the swallow-tail behavior of the corresponding evolution function. This is a generic feature that is observed from computing the various probes of thermalization by using solutions obtained from solving the equations in (\ref{genmin}) or the equations obtained from minimizing the action functional in equation (\ref{gencir}) for large enough boundary separation. This swallow-tail behavior results from a multi-valuedness in $z_*$ as a function of the boundary time $t$ as demonstrated in Fig~\ref{multi_demo}. As has been pointed out in \cite{Balasubramanian:2011ur}, this feature does not imply anything unphysical; we just need to be careful and follow the steepest descent procedure.

\newpage

\renewcommand{\theequation}{B.\arabic{equation}}
\setcounter{equation}{0}  
\section*{Appendix B. Charged AdS black holes}
\addcontentsline{toc}{section}{Appendix B. Charged AdS black holes}

Here we will briefly comment on some properties of charged AdS black holes in four and five bulk dimensions and their embedding in $10$ or $11$-dimensional supergravity. Let us begin with the case $d=4$, {\it i.e.} in five bulk dimensions.

The relevant theory here is the $S^5$ reduction of type IIB supergravity \cite{Kim:1985ez, Gunaydin:1984fk} truncated to the $\cN =2$ sector with an $U(1)^3$ symmetry. The corresponding five dimensional theory \cite{Gunaydin:1984ak} includes three $U(1)$ gauge fields, denoted by $A_I$ ($I = 1, 2, 3$), and two scalar fields, denoted by $X_I$ subject the constraint $X_1 X_2 X_3 =1$. This theory has a family of black hole solutions parametrized by three charges (corresponding to the three $U(1)$s) and a non-extremality parameter usually denoted by $\mu$ \cite{Behrndt:1998ns, Behrndt:1998jd}.\footnote{This non-extremality parameter $\mu$ is not to be confused with the chemical potential that we have defined in the main text.} The general form of such black hole solutions take the following form
\begin{eqnarray}
ds^2 & = & \left(H_1 H_2 H_3\right)^{-2/3} f dt^2 + \left(H_1 H_2 H_3 \right)^{1/3} \left( r^2 d\Omega_{3,k}^2 + \frac{dr^2}{f} \right) \ , \\
H_I & = & 1 + \frac{q_I}{r^2} \ , \quad f = k - \frac{\mu}{r^2} + \frac{r^2}{L^2} \left(H_1 H_2 H_3 \right) \ , \quad A_I = \left(H_I^{-1} -1 \right) dt \ , 
\end{eqnarray}
where $k$ takes values $0$ or $1$ depending on whether we work in the Poincar\'{e} patch or the global patch: the line element denoted by $d\Omega_{3,0}^2$ in this case corresponds to that of $\mathbb{R}^3$. The three charges $q_I$, when uplifted to $10$-dim type IIB supergravity, correspond to angular momenta along three Cartan directions on the five-sphere. We have considered only the Poincar\'{e} patch solution in the main text. Furthermore, the AdS-RN solution that we have considered in {\it e.g.} equation (\ref{RN}) corresponds to the case when $q_1 = q_2 = q_3 =q$. In this case, the scalars become trivial and the background metric takes a more familiar form
\begin{eqnarray}
ds^2 & = & - \frac{\rho^2}{L^2} g(\rho) dt^2 + \frac{\rho^2}{L^2} d\vec{x}^2 + \frac{L^2}{\rho^2} \frac{d\rho^2}{g(\rho)} \ , \quad g(\rho) = 1 - \frac{\mu L^2}{\rho^4} + \frac{\mu q L^2}{\rho^6} \ , \\
A & = & \left( \frac{q}{\rho_H^2} - \frac{q}{\rho^2} \right) dt \ ,
\end{eqnarray}
where we have defined $\rho^2 = r^2 + q$ and $\rho_H$ is the location of the event-horizon. We can obtain the background written in equation (\ref{RN}) from the above expression by identifying the following
\begin{eqnarray}
z = \frac{L}{\rho} \ , \quad M = \frac{\mu}{L^2} \ , \quad Q^2 = \frac{3 \mu q}{2 L^2} \ .
\end{eqnarray}

It was noted in \cite{Behrndt:1998jd}, in order to have a horizon the non-extremality parameter $\mu$ must satisfy a lower bound which can be analytically obtained. This means for a given $\mu$, there is an upper bound for the charge beyond which there is no horizon. This fact is revealed in the main text in {\it e.g.} fig.~\ref{Tvsq}. 
If we violate this bound, then a naked singularity appears \cite{Behrndt:1998jd}. As shown in \cite{Myers:2001aq}, this naked singularity has a natural meaning within string theory as an ensemble of giant gravitons which are distributed over the compact $S^5$. For more details, we will refer the reader to \cite{Myers:2001aq}. Note, however, that even within this bound the relevant dimensionless quantity $\chi_{(d)}$, defined in equation (\ref{ratio}), has a range given by $\chi_{(d)} \in [0, \infty]$. Thus the existence of an upper bound in the charge $Q$ does not imply any corresponding bound for the chemical potential measured in units of an appropriate power of the temperature.

\begin{figure}[!ht]
\begin{center}
\subfigure[] {\includegraphics[angle=0,
width=0.65\textwidth]{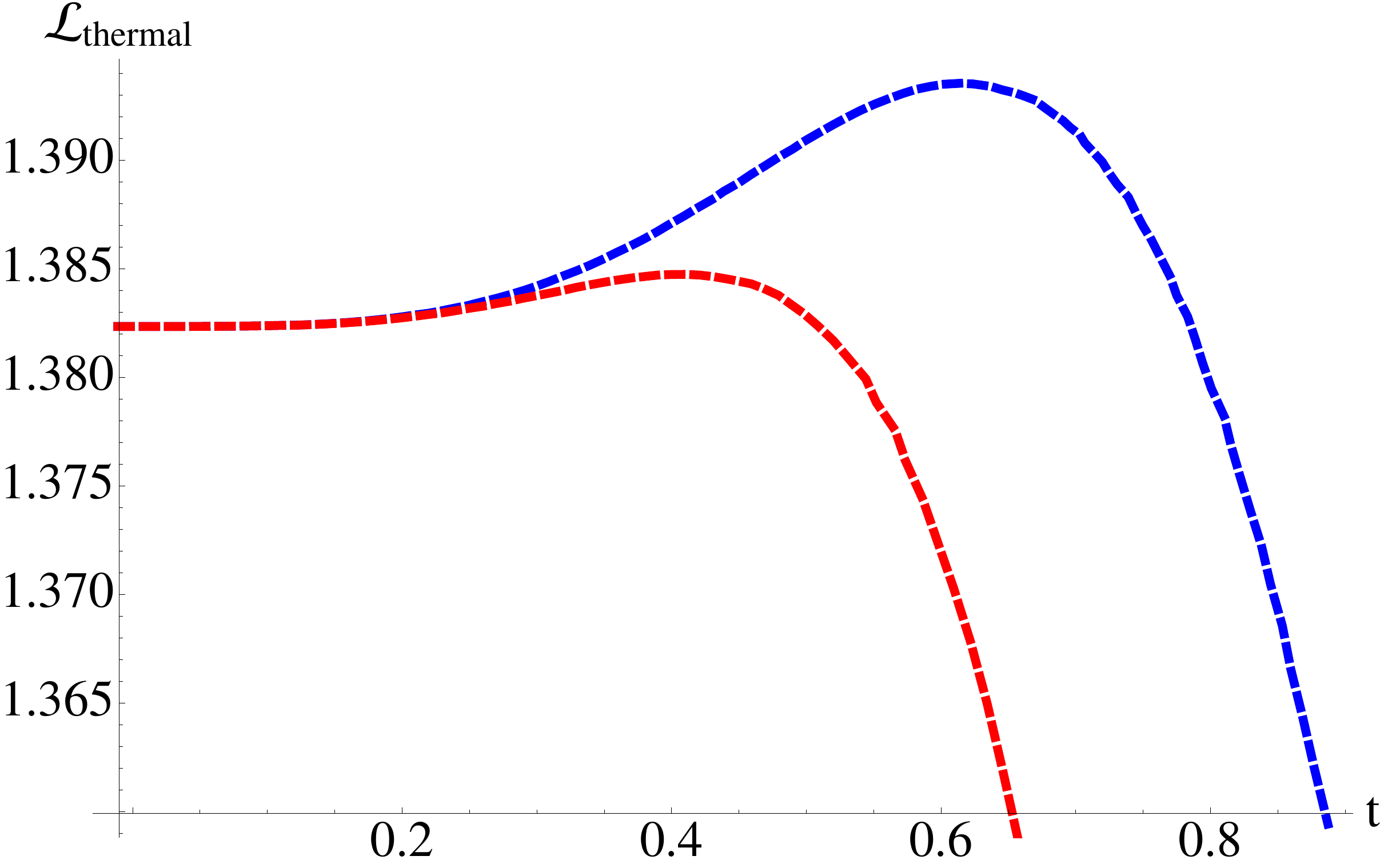} } 
\caption{\small The behavior of $\cL_{\rm thermal}$ (measured in units of the AdS-radius) as a function of boundary time for $d=4$. We have set the boundary separation length $\ell =2$ in units of the black hole mass. The blue curve corresponds to $Q=1$ and the red curve corresponds to $Q=2$ (in units of the black hole mass). Note that with these values of the parameters, the corresponding AdS-RN background does not have an event-horizon; instead it develops a naked singularity.}
\label{notherm}
\end{center}
\end{figure}

It is transparent that if the charge violates this upper bound, then the corresponding physics does not describe a thermal state in the dual field theory. Our primary focus in this article was to study the thermalization process by analyzing various probes of thermalization. It is intriguing to ask what will happen to these probes if we carefully fine-tune the mass parameter $M$ and the charge parameter $Q$ such that the background cannot form a horizon. In this case, the probes should never thermalize. Following the procedure explained in the main text, we can obtain the behavior of probes of thermalization in such a case and one such representative plot is shown in fig.~\ref{notherm}. 

It is clear from fig.~\ref{notherm} that initially $\cL_{\rm therm}$ begins to follow a similar bahviour as in the case when they eventually thermalize; however, after a certain amount of time has elapsed, the curve turns back and never really achieves thermal equilibrium. For larger values of $Q$ (with $M$ fixed), this ``turn over" behavior is further enhanced. Note that in the absence of a horizon it is not entirely clear to us what the corresponding non-local observables correspond to; hence we just present one representative plot and refrain from any further claims.

Similarly, it is also possible to consider the $S^7$-reduction of $11$-dimensional supergravity which gives rise to $SO(8)$ gauged $\cN =8$ supergravity in four dimensions. As in the five dimensional case, it is possible to take an $\cN=2$ consistent truncation which gives rise to a bosonic sector consisting of a graviton, four Abelian gauge potentials, three axions and three dilatons. This theory also has charged black hole solutions parametrized by four charge parameters and is given by
\begin{eqnarray}
ds^2 & = & - \left(H_1H_2H_3H_4\right)^{-1/2} f dt^2 +  \left(H_1H_2H_3H_4\right)^{1/2} \left(r^2 d\Omega_{2,k}^2 + \frac{dr^2}{f} \right) \ , \\
H_I & = & 1 + \frac{q_I}{r} \ , \quad f = k - \frac{\mu}{r} + \frac{r^2}{L^2} \left(H_1 H_2 H_3 H_4\right) \ , \quad A_I = \left(H_I^{-1} -1 \right) dt \ , \\
I & = & 1, \ldots, 4 \ .
\end{eqnarray}
This class of charged black holes can be embedded in M-theory in terms of rotating M2-branes. It is again straightforward to check that the usual AdS-RN background is recovered with four charges set equal.




\begin{thebibliography}{99}


\bibitem{Son:2007vk} 
  D.~T.~Son and A.~O.~Starinets,
  ``Viscosity, Black Holes, and Quantum Field Theory,''
  Ann.\ Rev.\ Nucl.\ Part.\ Sci.\  {\bf 57}, 95 (2007)
  [arXiv:0704.0240 [hep-th]].
  

\bibitem{Hubeny:2010ry} 
  V.~E.~Hubeny and M.~Rangamani,
  ``A Holographic view on physics out of equilibrium,''
  Adv.\ High Energy Phys.\  {\bf 2010}, 297916 (2010)
  [arXiv:1006.3675 [hep-th]].
  
  
\bibitem{Bhattacharyya:2009uu} 
  S.~Bhattacharyya and S.~Minwalla,
  ``Weak Field Black Hole Formation in Asymptotically AdS Spacetimes,''
  JHEP {\bf 0909}, 034 (2009)
  [arXiv:0904.0464 [hep-th]].
  
  
\bibitem{Balasubramanian:2010ce}
  V.~Balasubramanian {\it et al.},
  ``Thermalization of Strongly Coupled Field Theories,''
  Phys.\ Rev.\ Lett.\  {\bf 106}, 191601 (2011)
  [arXiv:1012.4753 [hep-th]].
  

\bibitem{Balasubramanian:2011ur}
  V.~Balasubramanian {\it et al.},
  ``Holographic Thermalization,''
  Phys.\ Rev.\  D {\bf 84}, 026010 (2011)
  [arXiv:1103.2683 [hep-th]].
  

\bibitem{Grumiller:2008va} 
  D.~Grumiller and P.~Romatschke,
  ``On the collision of two shock waves in AdS(5),''
  JHEP {\bf 0808}, 027 (2008)
  [arXiv:0803.3226 [hep-th]].
  
  
\bibitem{Gubser:2008pc} 
  S.~S.~Gubser, S.~S.~Pufu and A.~Yarom,
  ``Entropy production in collisions of gravitational shock waves and of heavy ions,''
  Phys.\ Rev.\ D {\bf 78}, 066014 (2008)
  [arXiv:0805.1551 [hep-th]].
  
  
\bibitem{Albacete:2008vs} 
  J.~L.~Albacete, Y.~V.~Kovchegov and A.~Taliotis,
  ``Modeling Heavy Ion Collisions in AdS/CFT,''
  JHEP {\bf 0807}, 100 (2008)
  [arXiv:0805.2927 [hep-th]].
  
  
\bibitem{AlvarezGaume:2008fx} 
  L.~Alvarez-Gaume, C.~Gomez, A.~Sabio Vera, A.~Tavanfar and M.~A.~Vazquez-Mozo,
  ``Critical formation of trapped surfaces in the collision of gravitational shock waves,''
  JHEP {\bf 0902}, 009 (2009)
  [arXiv:0811.3969 [hep-th]].
  

\bibitem{Lin:2009pn} 
  S.~Lin and E.~Shuryak,
  ``Grazing Collisions of Gravitational Shock Waves and Entropy Production in Heavy Ion Collision,''
  Phys.\ Rev.\ D {\bf 79}, 124015 (2009)
  [arXiv:0902.1508 [hep-th]].
  
  
\bibitem{Albacete:2009ji} 
  J.~L.~Albacete, Y.~V.~Kovchegov and A.~Taliotis,
  ``Asymmetric Collision of Two Shock Waves in AdS(5),''
  JHEP {\bf 0905}, 060 (2009)
  [arXiv:0902.3046 [hep-th]].
  
  
\bibitem{Gubser:2009sx} 
  S.~S.~Gubser, S.~S.~Pufu and A.~Yarom,
  ``Off-center collisions in AdS(5) with applications to multiplicity estimates in heavy-ion collisions,''
  JHEP {\bf 0911}, 050 (2009)
  [arXiv:0902.4062 [hep-th]].
  
  
\bibitem{Kovchegov:2009du} 
  Y.~V.~Kovchegov and S.~Lin,
  ``Toward Thermalization in Heavy Ion Collisions at Strong Coupling,''
  JHEP {\bf 1003}, 057 (2010)
  [arXiv:0911.4707 [hep-th]].
  
\bibitem{Kovchegov:2010zg} 
  Y.~V.~Kovchegov,
  ``Shock Wave Collisions and Thermalization in AdS$_5$,''
  Prog.\ Theor.\ Phys.\ Suppl.\  {\bf 187}, 96 (2011)
  [arXiv:1011.0711 [hep-th]].
    
  
\bibitem{Chesler:2010bi} 
  P.~M.~Chesler and L.~G.~Yaffe,
  ``Holography and colliding gravitational shock waves in asymptotically AdS$_5$ spacetime,''
  Phys.\ Rev.\ Lett.\  {\bf 106}, 021601 (2011)
  [arXiv:1011.3562 [hep-th]].
 
\bibitem{Aref'eva:2009kw}
  I.~Y.~.Aref'eva, A.~A.~Bagrov and L.~V.~Joukovskaya,
  ``Critical Trapped Surfaces Formation in the Collision of Ultrarelativistic Charges in (A)dS,''
  JHEP {\bf 1003} (2010) 002
  [arXiv:0909.1294 [hep-th]].


  \bibitem{Aref'eva:2012ar} 
  I.~Y.~.Aref'eva, A.~A.~Bagrov and E.~O.~Pozdeeva,
  ``Holographic phase diagram of quark-gluon plasma formed in heavy-ions collisions,''
  arXiv:1201.6542 [hep-th].
  
\bibitem{CaronHuot:2011dr} 
  S.~Caron-Huot, P.~M.~Chesler and D.~Teaney,
  ``Fluctuation, dissipation, and thermalization in non-equilibrium AdS$_5$ black hole geometries,''
  Phys.\ Rev.\ D {\bf 84}, 026012 (2011)
  [arXiv:1102.1073 [hep-th]].
  
  
\bibitem{GalanteSchvellinger} 
  D.~Galante and M.~Schvellinger,
  ``Thermalization with a chemical potential from AdS spaces,''
  arXiv:1205.1548 [hep-th].
 
 
\bibitem{Behrndt:1998jd}
  K.~Behrndt, M.~Cvetic and W.~A.~Sabra,
  ``Nonextreme black holes of five-dimensional N=2 AdS supergravity,''
  Nucl.\ Phys.\  B {\bf 553}, 317 (1999)
  [arXiv:hep-th/9810227].
    

\bibitem{Jensen:2010em} 
  K.~Jensen,
  ``Chiral anomalies and AdS/CMT in two dimensions,''
  JHEP {\bf 1101}, 109 (2011)
  [arXiv:1012.4831 [hep-th]].


\bibitem{Behrndt:1998ns}
  K.~Behrndt, A.~H.~Chamseddine, W.~A.~Sabra,
  ``BPS black holes in N=2 five-dimensional AdS supergravity,''
  Phys.\ Lett.\  {\bf B442}, 97-101 (1998).
  [hep-th/9807187].
  
  
\bibitem{Cvetic:1999ne} 
  M.~Cvetic and S.~S.~Gubser,
  ``Phases of R charged black holes, spinning branes and strongly coupled gauge theories,''
  JHEP {\bf 9904}, 024 (1999)
  [hep-th/9902195].
  
  
\bibitem{Chamblin:1999tk} 
  A.~Chamblin, R.~Emparan, C.~V.~Johnson and R.~C.~Myers,
  ``Charged AdS black holes and catastrophic holography,''
  Phys.\ Rev.\ D {\bf 60}, 064018 (1999)
  [hep-th/9902170].
  
  
\bibitem{Hartnoll:2009sz} 
  S.~A.~Hartnoll,
  ``Lectures on holographic methods for condensed matter physics,''
  Class.\ Quant.\ Grav.\  {\bf 26}, 224002 (2009)
  [arXiv:0903.3246 [hep-th]].

  
  
\bibitem{Balasubramanian:1999zv}
  V.~Balasubramanian and S.~F.~Ross,
  ``Holographic particle detection,''
  Phys.\ Rev.\  D {\bf 61}, 044007 (2000)
  [arXiv:hep-th/9906226].
  
  
\bibitem{Maldacena:1998im}
  J.~M.~Maldacena,
  ``Wilson loops in large N field theories,''
  Phys.\ Rev.\ Lett.\  {\bf 80}, 4859 (1998)
  [arXiv:hep-th/9803002].
  
  
\bibitem{Berenstein:1998ij} 
  D.~E.~Berenstein, R.~Corrado, W.~Fischler and J.~M.~Maldacena,
  ``The Operator product expansion for Wilson loops and surfaces in the large N limit,''
  Phys.\ Rev.\ D {\bf 59}, 105023 (1999)
  [hep-th/9809188].
  
  
\bibitem{Ryu:2006bv} 
  S.~Ryu and T.~Takayanagi,
  ``Holographic derivation of entanglement entropy from AdS/CFT,''
  Phys.\ Rev.\ Lett.\  {\bf 96}, 181602 (2006)
  [hep-th/0603001].

\bibitem{Hubeny:2007xt} 
  V.~E.~Hubeny, M.~Rangamani and T.~Takayanagi,
  ``A Covariant holographic entanglement entropy proposal,''
  JHEP {\bf 0707}, 062 (2007)
  [arXiv:0705.0016 [hep-th]].

\bibitem{Nishioka:2009un} 
  T.~Nishioka, S.~Ryu and T.~Takayanagi,
  ``Holographic Entanglement Entropy: An Overview,''
  J.\ Phys.\ A A {\bf 42}, 504008 (2009)
  [arXiv:0905.0932 [hep-th]].
  
  
\bibitem{AbajoArrastia:2010yt}
  J.~Abajo-Arrastia, J.~Aparicio and E.~Lopez,
  ``Holographic Evolution of Entanglement Entropy,''
  JHEP {\bf 1011}, 149 (2010)
  [arXiv:1006.4090 [hep-th]].
  
  
\bibitem{Albash:2010mv} 
  T.~Albash and C.~V.~Johnson,
  ``Evolution of Holographic Entanglement Entropy after Thermal and Electromagnetic Quenches,''
  New J.\ Phys.\  {\bf 13}, 045017 (2011)
  [arXiv:1008.3027 [hep-th]].
  
  
\bibitem{Balasubramanian:2011at} 
  V.~Balasubramanian, A.~Bernamonti, N.~Copland, B.~Craps and F.~Galli,
  ``Thermalization of mutual and tripartite information in strongly coupled two dimensional conformal field theories,''
  Phys.\ Rev.\ D {\bf 84}, 105017 (2011)
  [arXiv:1110.0488 [hep-th]].

  \bibitem{Arnold:2011qi} 
  P.~Arnold and D.~Vaman,
  ``Jet quenching in hot strongly coupled gauge theories simplified,''
  JHEP {\bf 1104}, 027 (2011)
  [arXiv:1101.2689 [hep-th]].
  

\bibitem{wip} 
Work in progress.


\bibitem{Hubeny:2012ry} 
  V.~E.~Hubeny,
  ``Extremal surfaces as bulk probes in AdS/CFT,''
  arXiv:1203.1044 [hep-th].
  

\bibitem{Figueras:2009iu} 
  P.~Figueras, V.~E.~Hubeny, M.~Rangamani and S.~F.~Ross,
  ``Dynamical black holes and expanding plasmas,''
  JHEP {\bf 0904}, 137 (2009)
  [arXiv:0902.4696 [hep-th]].
  
  
\bibitem{Kim:1985ez} 
  H.~J.~Kim, L.~J.~Romans and P.~van Nieuwenhuizen,
  ``The Mass Spectrum of Chiral N=2 D=10 Supergravity on S**5,''
  Phys.\ Rev.\ D {\bf 32}, 389 (1985).
  

\bibitem{Gunaydin:1984fk} 
  M.~Gunaydin and N.~Marcus,
  ``The Spectrum of the s**5 Compactification of the Chiral N=2, D=10 Supergravity and the Unitary Supermultiplets of U(2, 2/4),''
  Class.\ Quant.\ Grav.\  {\bf 2}, L11 (1985).
  
  
\bibitem{Gunaydin:1984ak} 
  M.~Gunaydin, G.~Sierra and P.~K.~Townsend,
  ``Gauging the d = 5 Maxwell-Einstein Supergravity Theories: More on Jordan Algebras,''
  Nucl.\ Phys.\ B {\bf 253}, 573 (1985).
  
   
\bibitem{Myers:2001aq} 
  R.~C.~Myers and O.~Tafjord,
  ``Superstars and giant gravitons,''
  JHEP {\bf 0111}, 009 (2001)
  [hep-th/0109127].

  

\end{thebibliography}
\end{document}